\documentclass[11pt,reqno]{amsproc}
\usepackage[colorinlistoftodos,shadow]{todonotes}
\usepackage[T1]{fontenc}
\usepackage{amsmath,amscd,amssymb}
\usepackage{cite}
\usepackage{color}
\usepackage{graphicx}
\usepackage{psfrag,epsfig}
\usepackage{multirow}
\usepackage{rotating}
\usepackage{fullpage}
\usepackage{subfigure}
\usepackage{enumerate}
\usepackage{comment}
\usepackage{lineno}
\usepackage{algorithmic}
\usepackage{algorithm}
\usepackage[small]{caption}
\usepackage[debug=false, colorlinks=true, pdfstartview=FitV, 
linkcolor=blue, citecolor=blue, urlcolor=blue]{hyperref}
\numberwithin{equation}{section}
\usepackage{chemarr}
\usepackage[version=3]{mhchem}
\newcommand*\patchAmsMathEnvironmentForLineno[1]{%
  \expandafter\let\csname old#1\expandafter\endcsname\csname #1\endcsname
  \expandafter\let\csname oldend#1\expandafter\endcsname\csname end#1\endcsname
  \renewenvironment{#1}%
     {\linenomath\csname old#1\endcsname}%
     {\csname oldend#1\endcsname\endlinenomath}}%
\newcommand*\patchBothAmsMathEnvironmentsForLineno[1]{%
  \patchAmsMathEnvironmentForLineno{#1}%
  \patchAmsMathEnvironmentForLineno{#1*}}%
\AtBeginDocument{%
\patchBothAmsMathEnvironmentsForLineno{equation}%
\patchBothAmsMathEnvironmentsForLineno{align}%
\patchBothAmsMathEnvironmentsForLineno{flalign}%
\patchBothAmsMathEnvironmentsForLineno{alignat}%
\patchBothAmsMathEnvironmentsForLineno{gather}%
\patchBothAmsMathEnvironmentsForLineno{multline}%
}

\title[Non-Negative Tensor Factorization for Reactive-Mixing]{Unsupervised Machine Learning Based on Non-Negative Tensor Factorization for Analyzing Reactive-Mixing}
\author[V.~V.~Vesselinov, et.al.]{V.~V.~Vesselinov$^{1,*}$,
M.~K.~Mudunuru$^{1}$,
S.~Karra$^{1}$,
D.~O'Malley$^{1,2}$, and 
B.~S.~Alexandrov$^{3}$ \\
{\scriptsize $^{1}$Computational Earth Science Group,
Earth and Environmental Sciences Division, 
Los Alamos National Laboratory, Los Alamos, NM 87545. \\
$^{2}$Department of Computer Science and Electrical Engineering,
University of Maryland, Baltimore County, 
Baltimore, MD 21250. \\
$^{3}$Physics and Chemistry of Materials Group,
Theoretical Division, Los Alamos National Laboratory, 
Los Alamos, NM 87545. \\
}
}
\address{\small $^*$Corresponding author: Dr.~Velimir 
Valentinov Vesselinov, Computational Earth Science Group 
(EES-16), MS T003 Earth and Environmental Sciences Division, 
Los Alamos National Laboratory, Los Alamos, NM 87545. \\
\textbf{E-mail address:}~vvv@lanl.gov \newline \newline
}
\date{\today}
\begin{document}
\maketitle
%
%
\section*{ABSTRACT} 
Analysis of reactive-diffusion simulations representing complex mixing processes requires a large number of independent model runs.
For each high-fidelity model simulation, the model inputs are varied and the predicted mixing behavior is represented by temporal and spatial changes in species concentration.
It is then required to discern how the model inputs (such as diffusivity, dispersion, anisotropy, and velocity field properties) impact the mixing process. 
This task is challenging and typically involves interpretation of large model outputs representing temporal and spatial changes of species concentration within the model domain.
However, the task can be automated and substantially simplified by applying Machine Learning (ML) methods.
In this paper, we present an application of an unsupervised ML method (called NTF$k$) using Non-negative Tensor Factorization (NTF) coupled with a custom clustering procedure based on $k$-means to reveal the temporal and spatial features in product concentrations.
An attractive and unique aspect of the proposed ML method is that it ensures the extracted features are non-negative, which are important to obtain a meaningful deconstruction of the mixing processes.
The ML methodology is applied to a large set of high-resolution finite-element model simulations representing anisotropic reaction-diffusion processes in perturbed vortex-based velocity fields.
The applied finite-element method ensures that spatial and temporal species concentration are always non-negative, even in the case of high anisotropic contrasts.
The simulated reaction is a fast irreversible bimolecular reaction $A + B \rightarrow C$, where species $A$ and $B$ react to form species $C$.
The reactive-diffusion model input parameters that control mixing include properties of velocity field (such as vortex structures), anisotropic dispersion, and molecular diffusion.
We demonstrate the applicability of the ML feature extraction method to produce a meaningful deconstruction of model outputs to discriminate between different physical processes impacting the reactants, their mixing, and the spatial distribution of the product $C$.
The presented ML analysis allowed us to identify additive temporal and spatial features that characterize mixing behavior.
The application of the proposed NTF$k$ approach is not limited to reactive-mixing.
NTF$k$ can be readily applied to any observed or simulated datasets that can be represented as tensors and have separable latent signatures or features.
\\
\\
\textbf{Keywords:}~Non-negative tensor factorization;
Tucker decompositions;
Structure-preserving feature extraction;
Unsupervised machine learning;
Reaction-diffusion equations;
Anisotropic dispersion;
Reactive-mixing;
Non-negative solutions;
Fast irreversible bimolecular reactions


\section*{HIGHLIGHTS OF PAPER}
\begin{itemize}
  \item NTF$k$ provide an elegant way to extract hidden features in irreversible fast bimolecular reaction-diffusion systems.
  \item Features extracted using NTF$k$ are non-negative and physically meaningful, which are an important constraint to understand mixing process.
  \item Another added benefit of our proposed NTF$k$ method is data compression, which is achieved with minimal loss of information. 
  In our case, the compression ratio is in the $\mathcal{O}(10^{-4})$
\end{itemize}

\section{INTRODUCTION}
\label{sec:intro}
Mixing and reaction of species is an important physical phenomena crucial in many applied research areas related not only to flow in porous/fractured media but also to turbulent flows at various scales in groundwater, rivers, lakes, oceans, and the atmosphere \cite{Donaldson1972,Ottino1990,Ivey1991,Imboden1995,Ismagilov2000,2015_Mudunuru_etal_ASME,2017_Mudunuru_Nakshatrala_MAMS,Hessel2005,Willingham2008,Stumm2012}.
A problem of predominant interest is groundwater contaminant remediation.
For several decades in the past and for the foreseeable future, aquifer contamination has been and is expected to be one of the most important application in the hydrogeological sciences \cite{gelhar1993stochastic,fetter1999contaminant,vengosh2014critical}.
Research has been driven by substantial scientific and engineering challenges associated with prediction and remediation of contaminant plumes in natural environments.
Many of these challenges are due to geochemical complexities associated with contaminant fate and transport in the subsurface, which is affected by numerous physical processes.
One of the most important factors is fluid flow.
The flow controls how the geochemical species of interest (e.g., contaminants and remedial agents) move and interact in the subsurface porous/fractured media.
Typically, water in an aquifer is a mixture of different groundwater types with different origins and geochemical signatures \cite{deutsch1997groundwater}.
For example, groundwater might be originating from different recharge sources with contrasting geochemical properties.
Also, groundwater may have been flowing through different rock types which may have altered the composition by means of geochemical reactions.
Furthermore, some of the groundwater recharge sources might be associated with contamination sources with different geochemical signatures.
In the case of active contaminant remediation, additional complexity occurs due to injection of water with species which are expected to alter the groundwater composition and eliminate the contaminant species (for example, through reduction and/or precipitation).
Understanding how different groundwater types mix and how some of the species in these mixtures react is a challenging but very important task.
This task is typically performed using complex inverse models representing groundwater mixing processes in the aquifer \cite{wagner1992simultaneous,neupauer2000comparison,atmadja2001pollution,michalak2004estimation,guan2006identification,mamonov2013point,hamdi2013inverse,murray2014spatio,borukhov2015identification}.

Typically, mixing analyses are performed in the laboratory based on microfluidics experiments \cite{ottino2004introduction}.
Another option is to perform high-resolution numerical simulations which are capable of representing the governing processes.  
These simulations usually involve solving anisotropic reaction-diffusion equations using numerical methods like the pseudo-spectral and finite element methods \cite{2013_Nakshatrala_Mudunuru_Valocchi_JCP_v253_p278_p307,chang2017large,2002_Adrover_etal_CCE_v26_p125_p139, 2009_Tsang_PRE_v80_p026305}.
From these simulations, one then infers scaling laws related to quantities of interest such as species decay rates, product yield, degree of mixing, extent of spreading, etc \cite{2016_Mudunuru_Nakshatrala_JCP_v305_p448_p493,2002_Adrover_etal_CCE_v26_p125_p139, 2009_Tsang_PRE_v80_p026305}. 
Since the reactive-mixing simulations are dependent on a series of model input parameters and produce large model outputs of state variables, the interpretation of the subsequent modeling results is challenging.
For instance, one can be interested in how the model input parameters impact the aforementioned quantities of interest over time.
This interpretation can be automated and substantially simplified by applying unsupervised Machine Learning (ML) methods such as Non-negative Matrix Factorization (NMF) and Non-negative Tensor Factorization (NTF) \cite{cichocki2009nonnegative}.
For example, NMF/NTF-based ML methods have been successfully used for analysis of Monte Carlo simulated fission chamber's output signals \cite{laassiri2017application}, for compression of scientific simulation data \cite{austin2016parallel}, and for a variety of other applications \cite{cichocki2009nonnegative}. 
To avoid confusion, we should emphasize that in this paper the term tensor is used to define two different types of mathematical objects.
We use tensors to define multi-dimensional arrays as commonly used in the unsupervised machine learning literature \cite{cichocki2009nonnegative}.
We also use the term tensor as a geometric object applied to characterize properties of physical parameters such as dispersivity; the dispersivity tensor is formed as a cross product of two vectors.

The primary goal of these unsupervised ML methods (such as NMF and NTF) is to separate the dependence of a quantity of interest into independent physical/chemical processes.
Through such a decomposition, NMF/NTF methods not only provide rapid data analysis and data reduction but also provide insight into the physics of the underlying system.
For instance, physical insight can be:~How does reactive-mixing progress under low and highly anisotropic conditions?
Which processes (e.g., longitudinal dispersion, transverse dispersion, molecular dispersion) contribute to the formation of product at early and later stages of time?
NMF/NTF methods provide an elegant approach to answer these questions, which are addressed in this paper.
It should be noted that the hidden features extracted with the proposed ML methods can also be used to construct reduced-order models.
In addition, the proposed ML models can also be used for extrapolation of the concentration beyond the simulation time.
However, the development of reduced-order models is not the main intention of this paper.

Conventional numerical formulations for anisotropic reaction-diffusion equations can produce negative and unphysical solutions for concentration of chemical species \cite{Roos_Stynes_Tobiska,2013_Stynes_arXiv,Gresho_Sani_v1,Mei}, and this is exacerbated when anisotropy dominates \cite{2013_Nakshatrala_Mudunuru_Valocchi_JCP_v253_p278_p307,2015_Mudunuru_etal_ASME,chang2017large}.
However, it is physically nonsensical for the concentration of a species to be negative.
To overcome this first challenge, we employ a novel non-negative finite element method \cite{2013_Nakshatrala_Mudunuru_Valocchi_JCP_v253_p278_p307,2016_Mudunuru_Nakshatrala_JCP_v305_p448_p493,chang2017large} to produce high-fidelity non-negative numerical solutions.
This non-negative finite element method provides physically meaningful concentrations for reaction-diffusion equations even under high anisotropy.
Then, in order to discover hidden features in the high-fidelity data, we present a new unsupervised ML methodology by combining NTF and $k$-means clustering.
The proposed ML method is called NTF$k$. 
NTF$k$ ensures that the discovered features are additive and are in accordance with the non-negativity of underlying numerical solutions.
For this reason, NTF$k$ is very attractive for our problem as it is a structure-preserving feature extraction approach, which other unsupervised ML methods that does not include non-negativity constraints do not guarantee.
Note that the proposed method is robust and can be applied to a wide variety of reactive-transport applications. 
For example, the robustness of the proposed method is studied in a similar context for contaminant source identification \cite{iliev2018nonnegative,stanev2018identification,vesselinov2018contaminant} using NMF$k$.
These references highlight the fact that the proposed method can be used to study data from a variety of simulation sources \cite{iliev2018nonnegative,stanev2018identification} and from observational field data \cite{vesselinov2018contaminant}.
While these analyses were performed using matrix versions of these methods, similar analyses can be performed using NTF$k$.

The paper is organized as follows. 
In Section ~\ref{sec:methodology}, we describe briefly a mathematical model for reaction-diffusion mixing simulations.
Then, we present our NTF$k$ method with the goal of discovering hidden non-negative features present in the high-fidelity model outputs.
High-fidelity datasets for NTF$k$ are generated by performing a series of model runs using a parallel non-negative finite element solver with different model input parameters representing uncertainties in the underlying reaction-diffusion processes.
The obtained results are presented in Section~\ref{sec:results} followed by conclusions in Section~\ref{sec:conclusions}.
%

\section{METHODOLOGY}
\label{sec:methodology}

\subsection{Reaction-Diffusion Mixing Simulations}
\label{sec:simulation}
Let $\Omega \subset \mathbb{R}^{n}$ be a domain with $n$  spatial dimensions and a piecewise smooth boundary $\partial \Omega$. 
Let $\overline{\Omega}$ be the set closure of $\Omega$ and $\boldsymbol{x} \in \overline{\Omega}$ be a spatial point.
The divergence and gradient operators with respect to $\boldsymbol{x}$ are denoted by $\mathrm{div}[\bullet]$ and $\mathrm{grad}[\bullet]$, respectively.
Let $\boldsymbol{n}(\boldsymbol{x})$ be the unit outward normal to $\partial \Omega$.
Let $t  \in \, (0, \mathcal{I})$ denote the time, where $\mathcal{I}$  is the length of time of interest.
Consider a bimolecular reaction where species $A$ and $B$ react irreversibly to produce species $C$:
\begin{linenomath}
\begin{align}
\label{eqn:Bimolecular_fast_reaction}
n_{A} \, A \, + \, n_{B} \, B \rightarrow n_{C} \, C
\end{align}
\end{linenomath}
where $n_A$, $n_B$, and $n_C$ are (positive) stoichiometric coefficients.

The governing equations for the reaction in Eq.~\eqref{eqn:Bimolecular_fast_reaction} without any sources/sinks are given by: 
\begin{linenomath}
\begin{subequations}
	\label{eqn:DRs_for_A_B_C}
	\begin{align}
	\label{eqn:DRs_for_A}
	&\frac{\partial c_A}{\partial t} - \mathrm{div}[\boldsymbol{D}
	(\boldsymbol{x},t) \, \mathrm{grad}[c_A]] =  - 
	n_{\small{A}} \, k_{AB} c_A c_B \quad \mathrm{in} \; \Omega \times ]0, 
	\mathcal{I}[ \\
	\label{eqn:DRs_for_B} 
	&\frac{\partial c_B}{\partial t} - \mathrm{div}[\boldsymbol{D}
	(\boldsymbol{x},t) \, \mathrm{grad}[c_B]] =  - 
	n_{\small{B}} \, k_{AB} c_A c_B \quad \mathrm{in} \; \Omega \times ]0, 
	\mathcal{I}[ \\
	\label{eqn:DRs_for_C} 
	&\frac{\partial c_C}{\partial t} - \mathrm{div}[\boldsymbol{D}
	(\boldsymbol{x},t) \, \mathrm{grad}[c_C]] = 
	+n_{\small{C}} \, k_{AB} c_A c_B \quad \mathrm{in} \; \Omega \times ]0, 
	\mathcal{I}[ \\
	\label{eqn:DRs_for_Dirchlet}
	&c_i(\boldsymbol{x},t) = c^{\mathrm{p}}_i(\boldsymbol{x},t) \quad 
	\mathrm{on} \; \Gamma^{\mathrm{D}}_{i} \times ]0, 
	\mathcal{I}[ \quad (i = A, \, B, \, C) \\
	\label{eqn:DRs_for_Neumann}
	& \left(-\boldsymbol{D} (\boldsymbol{x},t) \, \mathrm{grad}
	[c_i] \right) \bullet \boldsymbol{n}(\boldsymbol{x}) = h^
	{\mathrm{p}}_i(\boldsymbol{x},t) \quad \mathrm{on} \; 
	\Gamma^{\mathrm{N}}_{i} \times ]0, \mathcal{I}[ 
	\quad (i = A, \, B, \, C) \\
	\label{eqn:DRs_for_IC}
	&c_i(\boldsymbol{x},t=0) = c^{0}_i(\boldsymbol{x}) \quad 
	\mathrm{in} \; \Omega \quad (i = A, \, B, \, C)
	\end{align}
\end{subequations}
\end{linenomath}
where $c_i$ is the molar concentration of $i$-th chemical species, $\boldsymbol{D}(\boldsymbol{x},t)$ is the anisotropic dispersion tensor, and $k_{AB}$ is the bilinear reaction rate coefficient. 
$c^{\mathrm{p}}_i(\boldsymbol{x},t)$  and $h^{\mathrm{p}}_i(\boldsymbol{x},t)$ are the prescribed molar concentration and flux on Dirichlet and Neumann boundaries $\Gamma^{\mathrm{D}}_{i}$ and $\Gamma^{\mathrm{N}}_{i}$, respectively. 
Additionally, $c^{0}_i(\boldsymbol{x})$ is the initial concentration of $i$-th chemical species.

Using the non-negative linear transformations \cite{2013_Nakshatrala_Mudunuru_Valocchi_JCP_v253_p278_p307,2016_Mudunuru_Nakshatrala_JCP_v305_p448_p493},
\begin{subequations}
	\label{eqn:Definitions_of_F_G}
	\begin{align}
	\label{eqn:Definitions_of_F}
	c_F &:= c_A + \left( \frac{n_A}{n_C} \right) c_C \\ 
	\label{eqn:Definitions_of_G}
	c_G &:= c_B + \left( \frac{n_B}{n_C} \right) c_C 
	\end{align}
\end{subequations}
Eqs.~\eqref{eqn:DRs_for_A}--\eqref{eqn:DRs_for_IC} can be transformed to
\begin{subequations}
	\begin{align}
	\label{eqn:Diffusion_for_F}
	&\frac{\partial c_F}{\partial t} - \mathrm{div}[\boldsymbol{D}
	(\boldsymbol{x},t) \, \mathrm{grad}[c_F]] = 0
	\quad \mathrm{in} \; \Omega \times ]0, \mathcal{I}[ \\
	\label{eqn:Diffusion_for_Dirchlet_F}
	&c_F(\boldsymbol{x},t) = c_F^{\mathrm{p}}(\boldsymbol{x},t) := 
	c^{\mathrm{p}}_A(\boldsymbol{x},t) + \left( \frac{n_A}{n_C} 
	\right) c^{\mathrm{p}}_C(\boldsymbol{x},t) \quad \mathrm{on} 
	\; \Gamma^{\mathrm{D}} \times ]0, \mathcal{I}[ \\
	\label{eqn:Diffusion_for_Neumann_F}
	& \left(-\boldsymbol{D} (\boldsymbol{x},t) \, \mathrm{grad}[c_F]
	\right) \bullet \boldsymbol{n}(\boldsymbol{x}) =  h^{\mathrm{p}}
	_F(\boldsymbol{x},t) := h^{\mathrm{p}}_A(\boldsymbol{x},t) + 
	\left( \frac{n_A}{n_C} \right) h^{\mathrm{p}}_C(\boldsymbol{x},
	t) \quad \mathrm{on} \; \Gamma^{\mathrm{N}} \times ]0, 
	\mathcal{I}[ \\
	\label{eqn:Diffusion_for_IC_F}
	&c_F(\boldsymbol{x},t=0) = c^{0}_F(\boldsymbol{x}) := 
	c^{0}_A(\boldsymbol{x}) + \left( \frac{n_A}{n_C} \right) 
	c^{0}_C(\boldsymbol{x}) \quad \mathrm{in} \; \Omega
	\end{align}
\end{subequations} 
and 
\begin{subequations}
	\begin{align}
	\label{eqn:Diffusion_for_G} 
	&\frac{\partial c_G}{\partial t} - \mathrm{div}[\boldsymbol{D}
	(\boldsymbol{x},t) \, \mathrm{grad}[c_G]] = 0
	\quad \mathrm{in} \; \Omega \times ]0, \mathcal{I}[ \\
	\label{eqn:Diffusion_for_Dirchlet_G}
	&c_G(\boldsymbol{x},t) = c^{\mathrm{p}}_G(\boldsymbol{x},t) := 
	c^{\mathrm{p}}_B (\boldsymbol{x},t) + \left( \frac{n_B}{n_C} 
	\right) c^{\mathrm{p}}_C (\boldsymbol{x},t) \quad \mathrm{on} 
	\; \Gamma^{\mathrm{D}} \times ]0, \mathcal{I}[ \\
	\label{eqn:Diffusion_for_Neumann_G}
	& \left(-\boldsymbol{D} (\boldsymbol{x},t) \, \mathrm{grad}[c_G] 
	\right) \bullet \boldsymbol{n}(\boldsymbol{x}) = h^{\mathrm{p}}_
	G(\boldsymbol{x},t) := h^{\mathrm{p}}_B(\boldsymbol{x},t) + 
	\left( \frac{n_B}{n_C} \right) h^{\mathrm{p}}_C(\boldsymbol{x},
	t) \quad \mathrm{on} \; \Gamma^{\mathrm{N}} \times ]0, 
	\mathcal{I}[ \\
	\label{eqn:Diffusion_for_IC_G}
	&c_G(\boldsymbol{x},t=0) = c^{0}_G(\boldsymbol{x}) := c^{0}_B
	(\boldsymbol{x}) + \left( \frac{n_B}{n_C} \right) c^{0}_
	C(\boldsymbol{x}) \quad \mathrm{in} \; \Omega 
	\end{align}
\end{subequations}

Since fast bimolecular reactions are considered, it is assumed that the reactants $A$ and $B$ cannot co-exist at the same location $\boldsymbol{x}$ at any instant of time.
Based on this assumption, quantities $c_F(\boldsymbol{x},t)$ and $c_G(\boldsymbol{x},t)$ can be used to evaluate $c_A(\boldsymbol{x},t)$, $c_B(\boldsymbol{x},t)$ and, $c_C(\boldsymbol{x},t)$ via:
\begin{subequations}
	\label{eqn:Fast_A_B_C}
	\begin{align}
	\label{eqn:Fast_A}
	&c_A(\boldsymbol{x},t) = \mathrm{max} \left[c_F(\boldsymbol{x},t) 
	- \left(\frac{n_A}{n_B}\right) c_G(\boldsymbol{x},t), \, 0 
	\right] \\
	\label{eqn:Fast_B}
	&c_B(\boldsymbol{x},t) = \left( \frac{n_B}{n_A} \right) \; 
	\mathrm{max}\left[- c_F(\boldsymbol{x},t) + 
	\left(\frac{n_A}{n_B} \right) c_G(\boldsymbol{x},t), \, 0 \right] \\
	\label{eqn:Fast_C}
	&c_C(\boldsymbol{x},t) = \left( \frac{n_C}{n_A} \right) \; 
	\left(c_F(\boldsymbol{x},t) - c_A(\boldsymbol{x},t) \right)
	\end{align}
\end{subequations}
The uncoupled governing equations given by Eqs.~\eqref{eqn:Diffusion_for_F}--\eqref{eqn:Diffusion_for_IC_G} are solved using the lower-order non-negative finite element methods \cite{2013_Nakshatrala_Mudunuru_Valocchi_JCP_v253_p278_p307,2016_Mudunuru_Nakshatrala_JCP_v305_p448_p493,chang2017large} to produce non-negative solutions. 
Once we obtain the non-negative quantities $c_F$ and $c_G$, Eqs.~\eqref{eqn:Fast_A}--\eqref{eqn:Fast_C} provide the concentrations of $A$, $B$, and $C$.
It should be noted that a simpler version of this problem related to contaminant transport was investigated by Vesselinov and co-authors in their previous works using the NMF method \cite{vesselinov2018contaminant,stanev2018identification,iliev2018nonnegative}.
NMF can extract features only from two-dimensional datasets \cite{cichocki2009nonnegative}.
As a result, it cannot extract hidden features from datasets with a temporal and multiple spatial dimensions such as the ones considered here.
Fast bimolecular reactions and other complex reactions fall under this category \cite{kim2018computationally}.
In order to extract hidden features from spatio-temporal data, we need a method which decomposes tensors.
Our proposed NTF$k$ method overcomes the limitation of the NMF-based methods.

\begin{figure}
	\centering
	{\includegraphics[width = 0.4\textwidth]{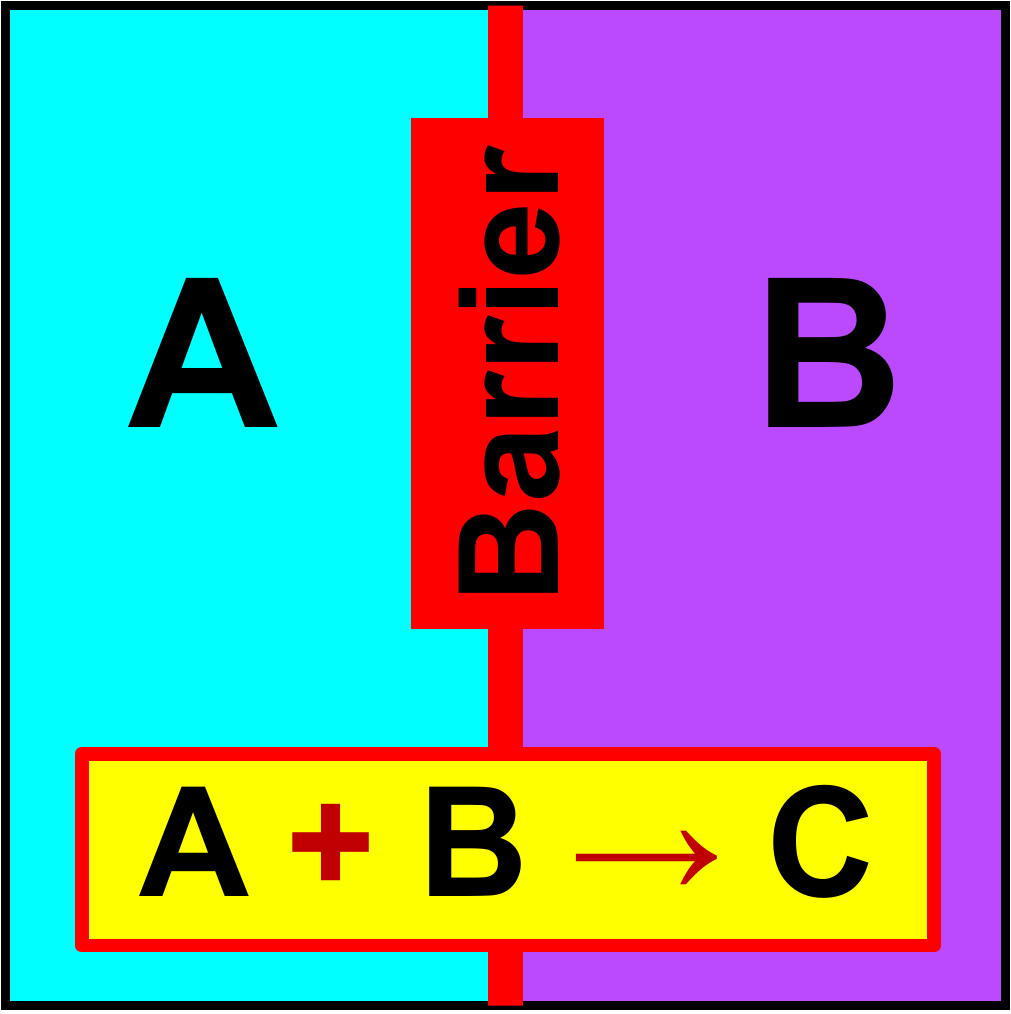}}
	\caption{Schematic representation of the initial boundary value problem with zero flux boundary conditions along the domain boundaries. 
	Initially species $A$ and $B$ are in the left and right part of the domain, respectively, and they are separated by a barrier at $t = 0$; $A$ and $B$ are allowed to mix for time $t > 0$ and form $C$ driven by synthetic vortex-based velocity field defined in Eq.~\eqref{eqn:Vel_x}--\eqref{eqn:Vel_y}. 
	The initial concentrations of $A$ and $B$ are equal to $1.0$.
	The initial concentration of $C$ is $0.0$.
	The maximum plausible concentration of $C$ is $0.5$.
	There are no sources/sinks.
	\label{Fig:ROM_BVPs}}
\end{figure}

Fig.~\ref{Fig:ROM_BVPs} provides a pictorial description of the initial boundary value problem.
The computational domain is a square with $L = 1$ and zero flux boundary conditions are assumed on all sides of the domain.
For all the chemical species, the non-reactive volumetric source $f_i(\boldsymbol{x}, t)$ is equal to zero.
Initially, species $A$ and $B$ are separated; $A$ is placed in the left half of the domain and $B$ is placed in the right half.
The stoichiometric coefficients are taken as $n_A = 1$, $n_B = 1$, and $n_C = 1$.
The total time of interest is taken as $\mathcal{I} = 1$.
The dispersion tensor follows the formulation of \cite{Pinder_Celia}:
\begin{linenomath}
\begin{align}
\label{Eqn:Anisotropic_Dispersion_Tensor}
\boldsymbol{D}_{\mathrm{subsurface}}
(\boldsymbol{x}) = D_{m} \boldsymbol{I} + 
\alpha_{T} \|\boldsymbol{v}\| \boldsymbol{I} + 
\frac{\alpha_L - \alpha_T}{\|\boldsymbol{v}\|} 
\boldsymbol{v} \otimes \boldsymbol{v}
\end{align}
\end{linenomath}
where $D_m$ is the molecular diffusivity, $\alpha_L$ is the longitudinal dispersivity, $\alpha_T$ is the transverse dispersivity, $\boldsymbol{I}$ is the identity tensor, $\otimes$ is the tensor product, $\boldsymbol{v}$ is the velocity vector field, and $\| \bullet \|$ is the Frobenius norm.
Note that the governing equations given by Eqs.~\ref{eqn:DRs_for_A}--\ref{eqn:DRs_for_IC} have no advection term.
The parameterization of dispersion given in Eq.~\ref{eqn:DRs_for_A_B_C} and Eq.~\ref{Eqn:Anisotropic_Dispersion_Tensor} is for porous media where the anisotropic mechanical dispersion term is a direct consequence of fluid flow.
Since there is not really any fluid flow, we assume that Eq.~\ref{Eqn:Anisotropic_Dispersion_Tensor} provides a physics-based formula to compute a spatially and temporally varying `diffusion/dispersion' tensor.
As a result, the velocity field assumed in Eq.~\ref{Eqn:Anisotropic_Dispersion_Tensor} need not be divergence free.
However, we choose $\boldsymbol{v}$ to be divergence free.

Synthetic model velocity field is used to define the anisotropic dispersion tensor through the following stream function \cite{2002_Adrover_etal_CCE_v26_p125_p139,2009_Tsang_PRE_v80_p026305,2016_Mudunuru_Nakshatrala_JCP_v305_p448_p493}:
\begin{align}
\label{eqn:Div_Free_Stream_Function}
\psi(\boldsymbol{x},t) = 
\begin{cases}
\frac{1}{2 \pi \kappa_f} \left( \sin(2 \pi \kappa_f x) 
- \sin(2 \pi \kappa_f y) + v_o \cos(2 \pi \kappa_f y)
\right) &\; \; \mathrm{if} \; \nu T \leq t < \left( \nu 
+ \frac{1}{2} \right) T  \\
\frac{1}{2 \pi \kappa_f} \left( \sin(2 \pi \kappa_f x) 
- \sin(2 \pi \kappa_f y) - v_o \cos(2 \pi \kappa_f x)
\right) &\; \; \mathrm{if} \; \left( \nu + \frac{1}{2} 
\right) T \leq t < \left( \nu + 1 \right) T
\end{cases}
\end{align}
where $\nu = 0, 1, 2, 3, \cdots$ is an integer and $v_0$ is the perturbation parameter. 
$\kappa_fL$ and $T$ are characteristic spatial and time-scales of the vortex-based velocity field.
Units of $\kappa_f$ is $[\mathrm{m}^{-1}]$.
Using Eq.~\eqref{eqn:Div_Free_Stream_Function}, the divergence free velocity field components are given as follows:
\begin{align}
\label{eqn:Vel_x}
\mathrm{v}_{x}(\boldsymbol{x},t) = -\frac{\partial \psi}{\partial \mathrm{y}} = 
\begin{cases}
\cos\left(2 \pi \left(\kappa_fL \right) \left(\frac{y}{L} \right) \right) + v_o \sin\left(2 \pi \left(\kappa_fL \right) \left(\frac{y}{L} \right) \right) 
&\quad \mathrm{if} \; \nu T \leq t < \left( \nu + 
\frac{1}{2} \right) T  \\
\cos\left(2 \pi \left(\kappa_fL \right) \left(\frac{y}{L} \right) \right) &\quad \mathrm{if} \; 
\left( \nu + \frac{1}{2} \right) T \leq t < 
\left( \nu + 1 \right) T
\end{cases}
\end{align}
\begin{align}
\label{eqn:Vel_y}
\mathrm{v}_{y}(\boldsymbol{x},t) = +\frac{\partial \psi}{\partial \mathrm{x}} = 
\begin{cases}
\cos\left(2 \pi \left(\kappa_fL \right) \left(\frac{x}{L} \right) \right) &\quad \mathrm{if} \; 
\nu T \leq t < \left( \nu + \frac{1}{2} \right) T \\
\cos\left(2 \pi \left(\kappa_fL \right) \left(\frac{x}{L} \right) \right) + v_o \sin\left(2 \pi \left(\kappa_fL \right) \left(\frac{x}{L} \right) \right) 
&\quad \mathrm{if} \; \left( \nu + \frac{1}{2} \right) 
T \leq t < \left( \nu + 1 \right) T
\end{cases}
\end{align}

\begin{figure}
	\centering
	{\includegraphics[width = 0.85\textwidth]{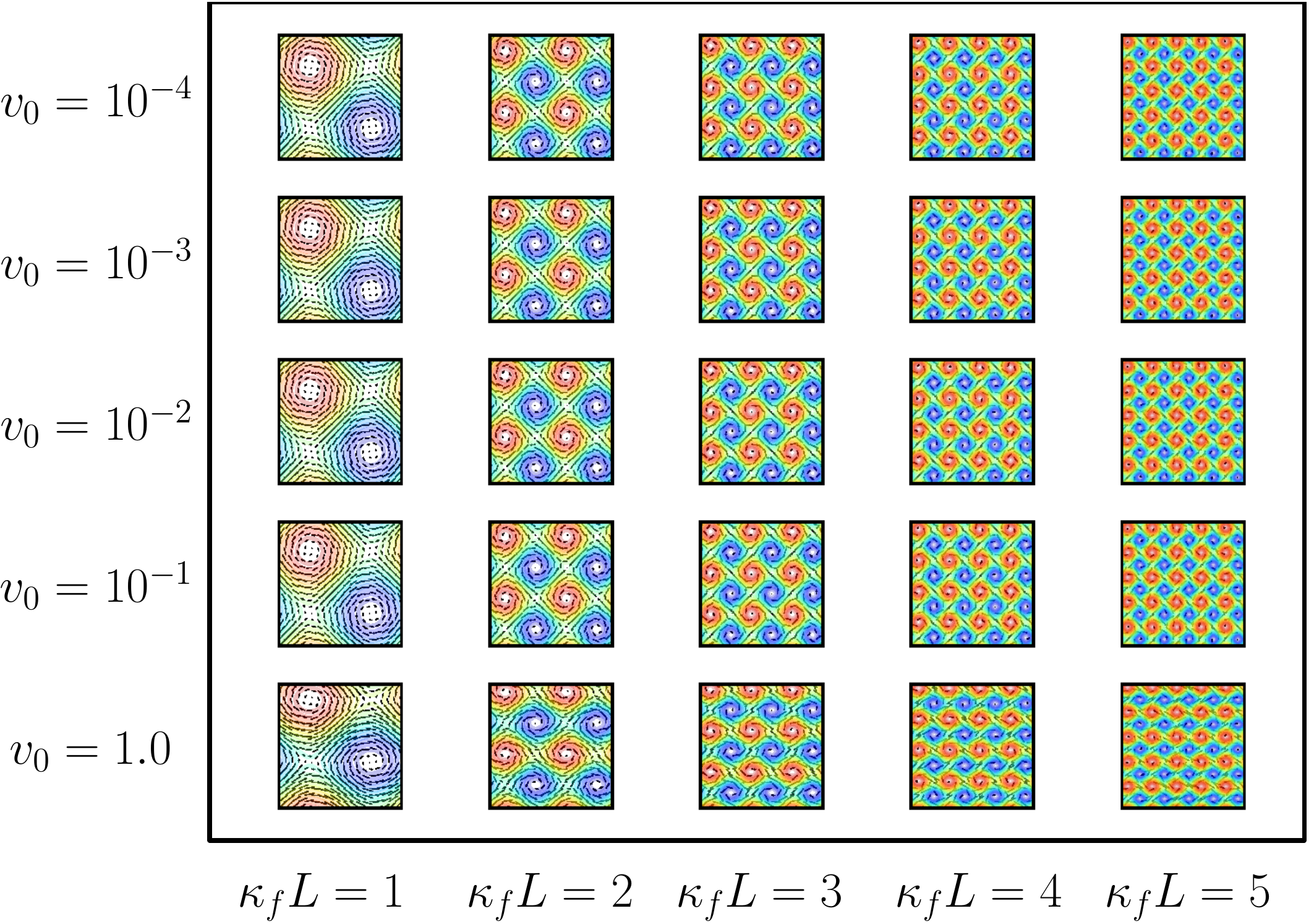}}
	\caption{Streamlines and associated velocity vectors for different values of $v_0$ and $\kappa_fL$ representing large-scale and small-scale vortex structures present in the velocity field for $\nu T \leq t < \left( \nu + \frac{1}{2} \right) T$.
	The number of vortices increases for larger values of $\kappa_fL$ and as a result there are more small-scale features present in the velocity field.
	Note that the parameter $v_0$ does not substantially impact the location of vortices.
	\label{Fig:Large_Small_Scales_VelField}}
\end{figure}

\begin{figure}
	\centering
	{\includegraphics[width = 0.85\textwidth]{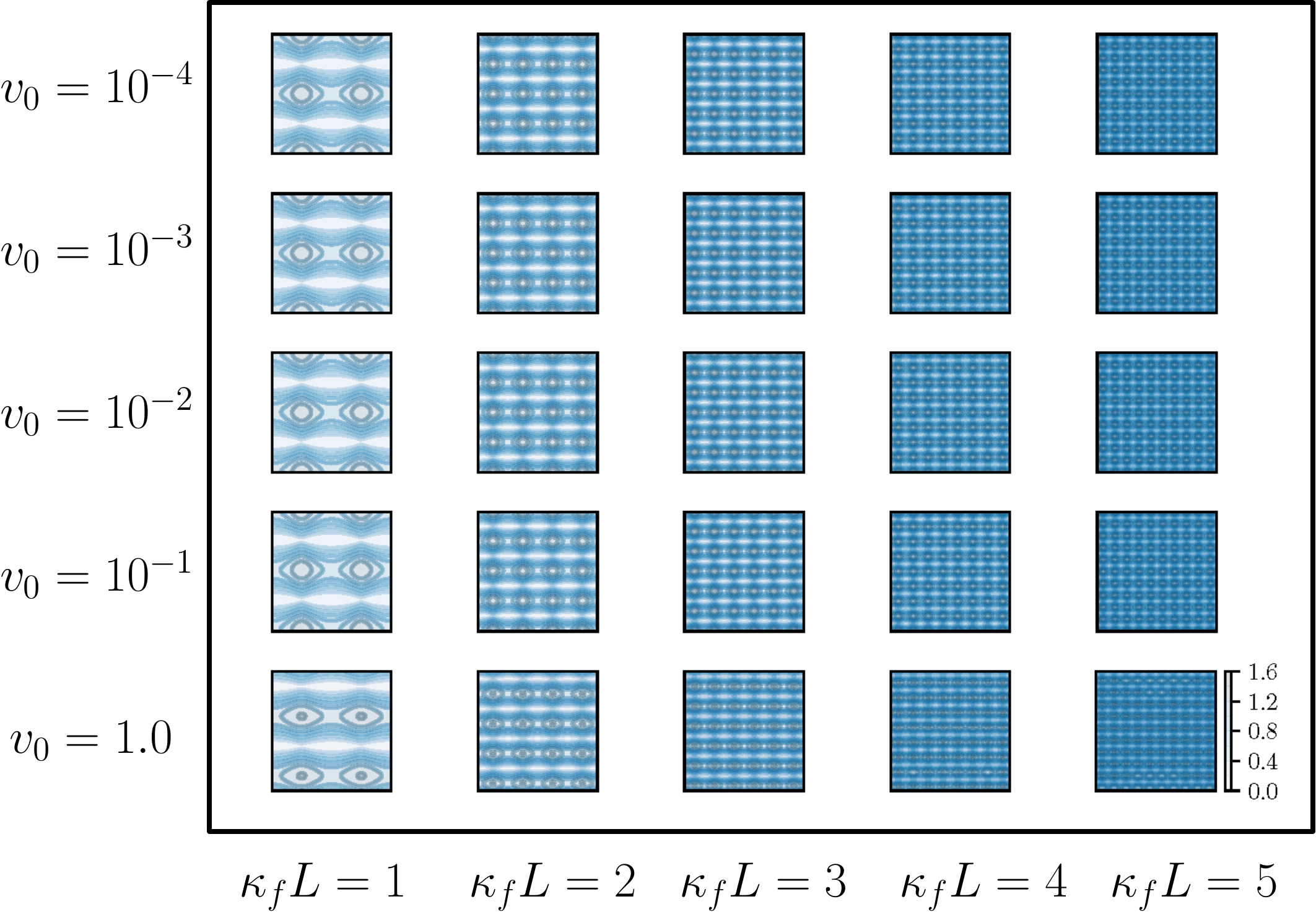}}
	\caption{Graphic representation of the $ij$-th components of the anisotropic dispersion tensor, where $i = 1$ and $j = 1$.
	Analyses are performed for different values of $v_0$ and $\kappa_fL$ for $\nu T \leq t < \left( \nu + \frac{1}{2} \right) T$.
	\label{Fig:D11_Dispersion}}
\end{figure}

\begin{figure}
	\centering
    {\includegraphics[width = 0.85\textwidth]{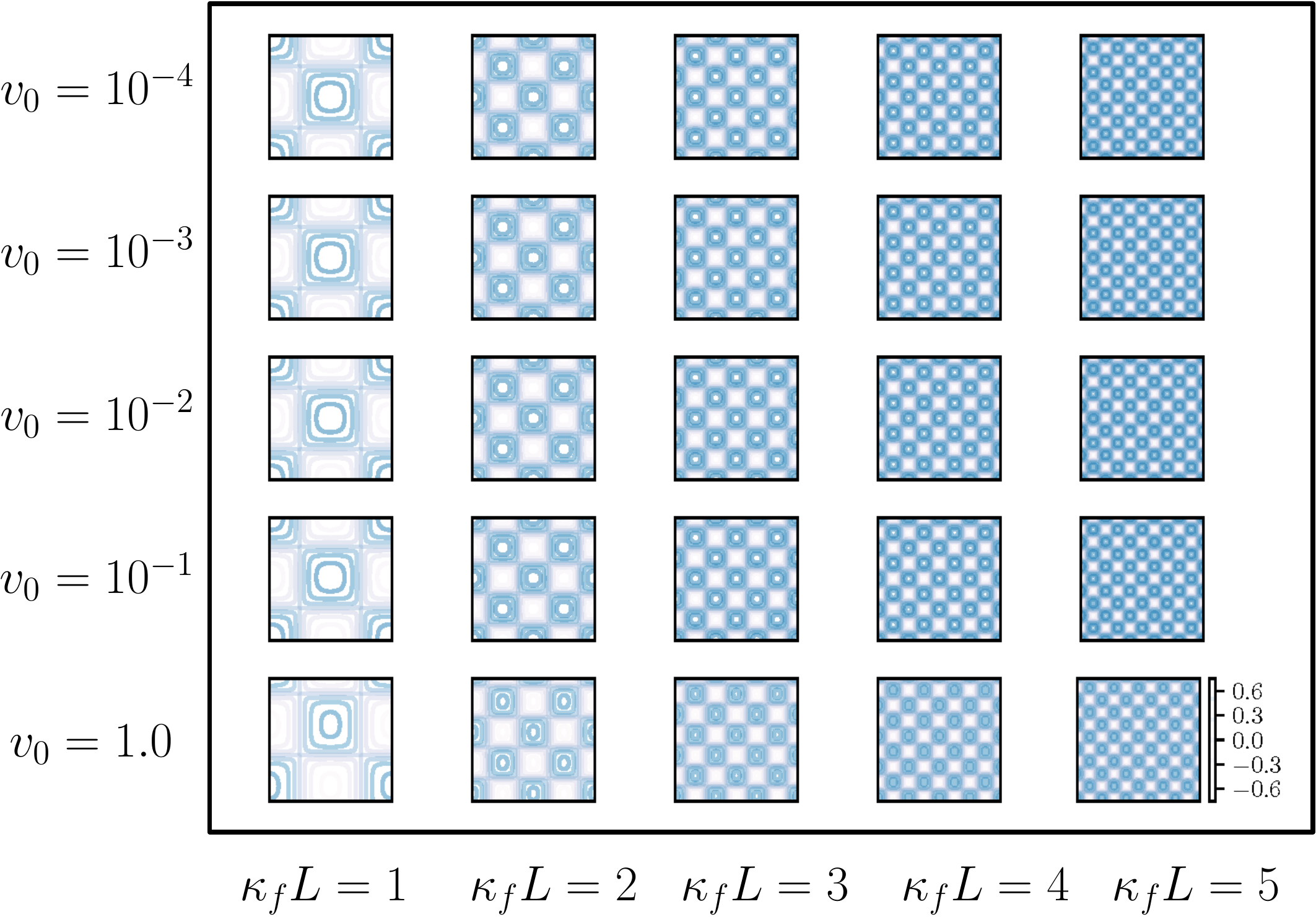}}
	\caption{Graphic representation of the $ij$-th components of the anisotropic dispersion tensor, where $i = 1$ and $j = 2$.
	Analyses are performed for different values of $v_0$ and $\kappa_fL$ for $\nu T \leq t < \left( \nu + \frac{1}{2} \right) T$.
	\label{Fig:D12_Dispersion}}
\end{figure}

\begin{figure}
	\centering
	{\includegraphics[width = 0.85\textwidth]{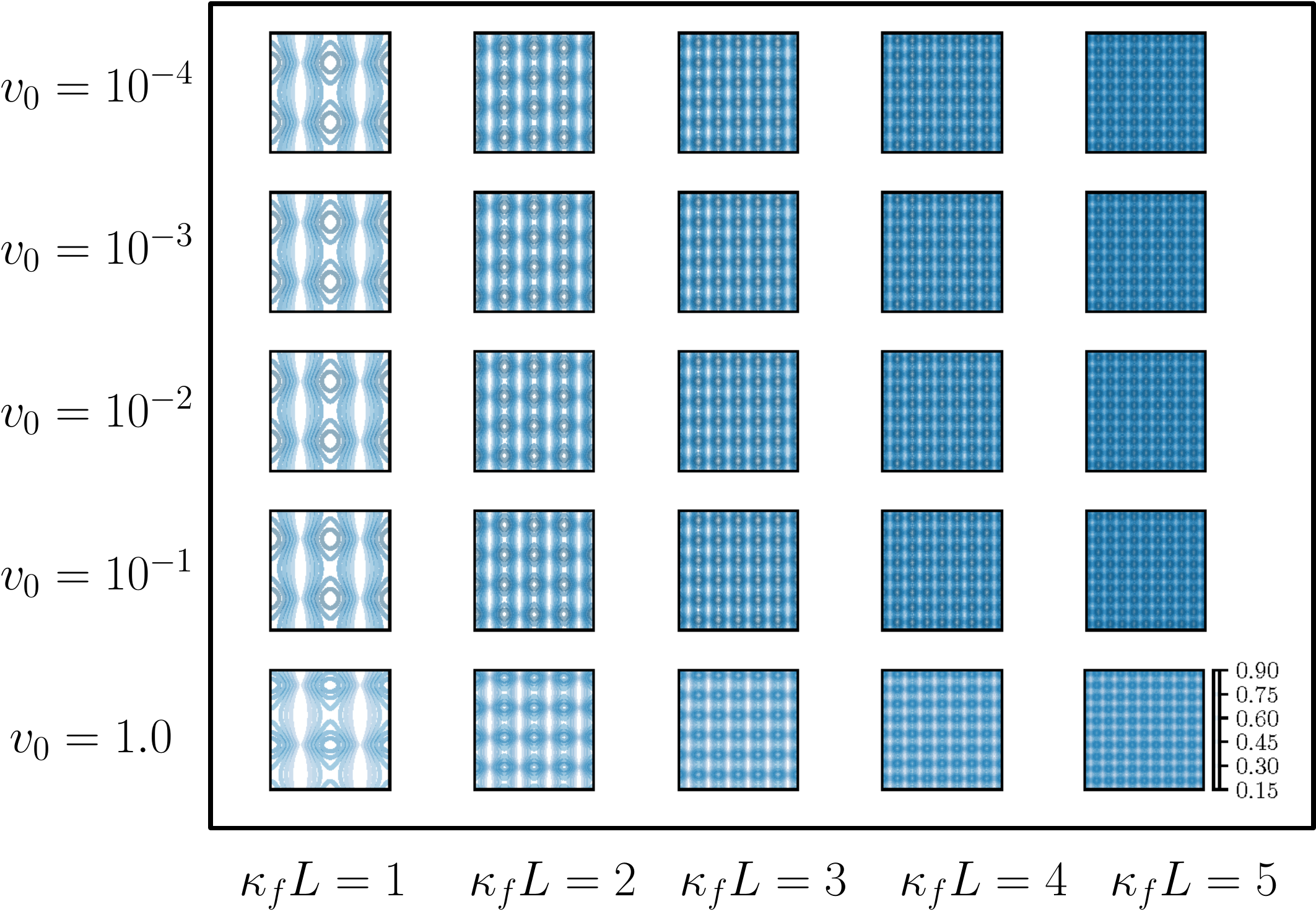}}
	\caption{Graphic representation of the $ij$-th components of the anisotropic dispersion tensor, where $i = 2$ and $j = 2$.
	Analyses is performed for different values of $v_0$ and $\kappa_fL$ for $\nu T \leq t < \left( \nu + \frac{1}{2} \right) T$.
	\label{Fig:D22_Dispersion}}
\end{figure}

Fig.~\ref{Fig:Large_Small_Scales_VelField} shows the streamlines of the vortex-based velocity field given by Eqs.~\eqref{eqn:Vel_x}--\eqref{eqn:Vel_y}.
Small-scale and large-scale vortices are observed when $\kappa_fL$ is high and low, respectively.
The number of vortices increases for larger values of $\kappa_fL$ and as a result there are more small-scale features in the velocity field.
Fig.~\ref{Fig:Large_Small_Scales_VelField} also shows that the perturbation parameter $v_0$ does not substantially impact the location of the vortices.
Figs.~\ref{Fig:D11_Dispersion}--\ref{Fig:D22_Dispersion} provide the contours for different components of anisotropic dispersion.
The analyses are performed by varying $v_0$ and $\kappa_fL$ for $\nu T \leq t < \left( \nu + \frac{1}{2} \right) T$.
Other model parameters are kept fixed.
These include $\alpha_L = 1$, $\alpha_T = 10^{-4}$, and $D_m = 10^{-8}$, which correspond to high anisotropic contrast and low molecular diffusion.
In Figs.~\ref{Fig:D11_Dispersion}--\ref{Fig:D22_Dispersion}, light blue and dark blue colors represent low and high values of dispersion.
For low values of $\kappa_fL$, we can see a clear separation between the low and high values of dispersion.
For example, this can be observed by comparing the plots for $\kappa_fL$ is equal to 1 and 2 related to dispersion components $D_{11}$, $D_{12}$, and $D_{22}$ in Figs.~\ref{Fig:D11_Dispersion}, \ref{Fig:D12_Dispersion}, and \ref{Fig:D22_Dispersion}, respectively.
As the value of $\kappa_fL$ increases, this separation decreases only for diagonal components ($D_{11}$ and $D_{22}$) of dispersion (see Fig.~\ref{Fig:D11_Dispersion} and Fig.~\ref{Fig:D22_Dispersion}). 
However, the off-diagonal terms of the dispersion tensor still have distinct islands of low and high values (see Fig.~\ref{Fig:D12_Dispersion}) for all $\kappa_fL$ values.

\subsection{Non-negative Tensor Factorization}
\label{sec:NTF}
In this subsection, we present a novel structure-preserving feature extraction method to extract hidden features from high-resolution reaction-diffusion model outputs.
These outputs are generated by varying input parameters in Eqs.~\eqref{Eqn:Anisotropic_Dispersion_Tensor}--\eqref{eqn:Vel_y}.
The proposed methodology is based on NTF, which is an emerging research field in the area of data analytics and data compression. 
In addition to extracting hidden features that are buried in large high-dimensional datasets \cite{cichocki2009nonnegative}, NTF-based methods are also used in source separation.
For example, Blind Source Separation (BSS) techniques based on matrix factorization methods such as Principle Component Analysis (PCA) \cite{jolliffe1986principal}, Independent Component Analysis (ICA) \cite{amari1996new}, and Non-negative Matrix/Tensor Factorization (NMF/NTF) \cite{paatero1994positive}, form a class of unsupervised machine learning (ML) methods that are instrumental in model-free feature extraction and dimensionality reduction.
When a BSS technique is applied in signal processing, the extracted features are the unique original signals represented in the mixtures of signals that are recorded by a set of spatially distributed sensors (e.g., the voices of several speakers recorded by multiple microphones placed at different locations in a ball room \cite{haykin2005cocktail}).
However, the matrix factorization methods are inherently deficient for examining multidimensional datasets (i.e., tensor datasets), which are the natural extensions of the matrix datasets.
Numerical model outputs are typically multidimensional and discrete.
They often represent one or more state variables at a discrete set of locations in space and time, and, as a result, are ideal for NTF analyses.
There are two main tensor factorization methods:~Canonical Polyadic (CP) decomposition (CANDECOMP/PARAFAC) \cite{hitchcock1927expression, harshman1994parafac, de2000multilinear} and Tucker decomposition \cite{tucker1966some, andersson2000n}.
Both can be computationally effective for interpreting multidimensional datasets \cite{gredilla2013unsupervised}.

\begin{figure}
	\centering
	{\includegraphics[clip=true,width = 0.985\textwidth]{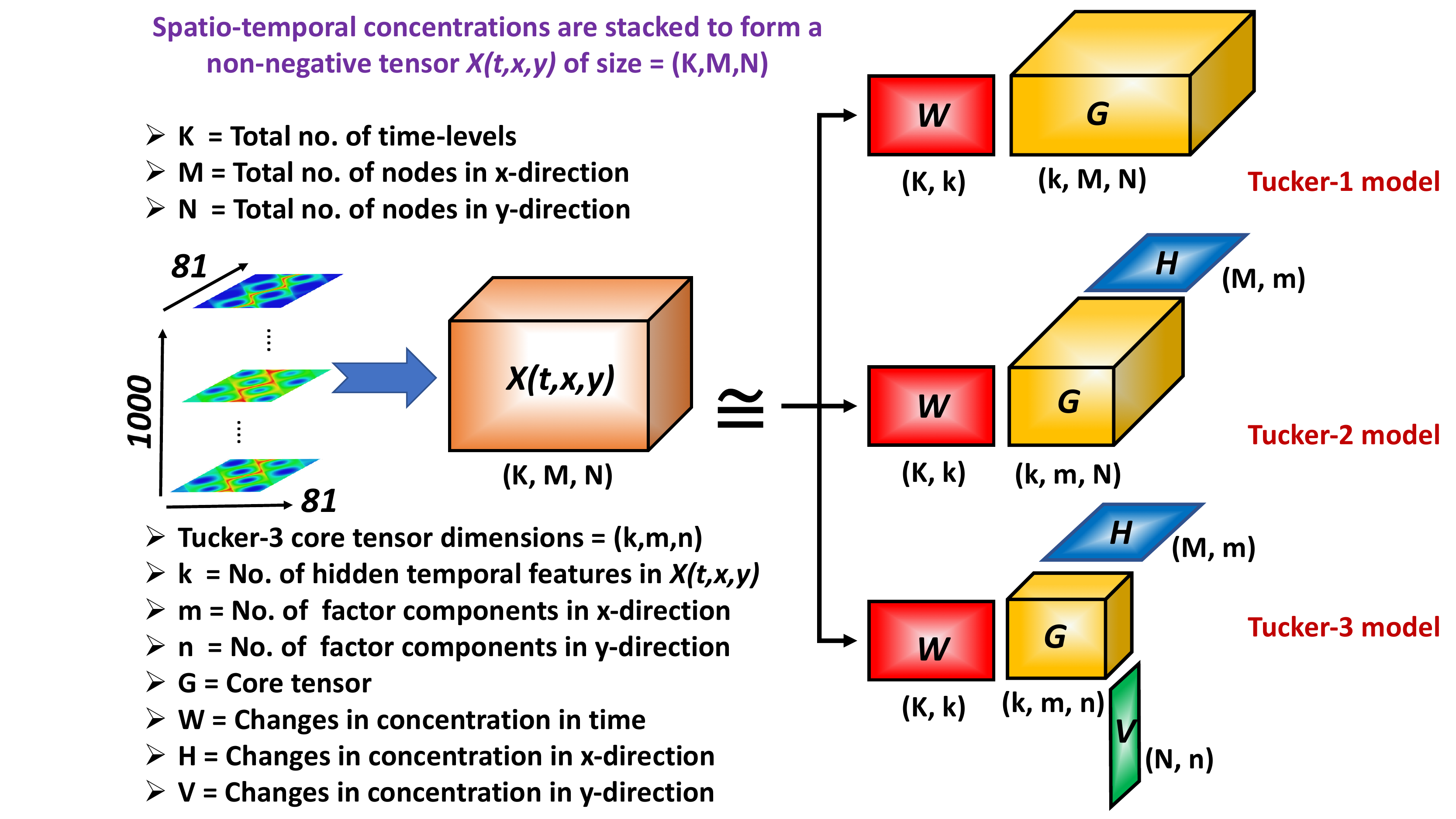}}
	\caption{Schematic representation of various Tucker-based tensor factorization models. 
	Herein, we employ Tucker-3 model to decompose the non-negative concentration tensor $X(t,x,y)$ into a core tensor $G$ and factors $W$, $V$, and $H$.
	\label{Fig:Tucker_Models}}
\end{figure}

Herein, we analyze three-dimensional data that present the evolution in time and space of concentration of product $C(t, x, y)$.
These concentrations are generated by non-negative finite element numerical simulations of the reaction-diffusion equation described in Section \ref{sec:simulation}.
Fig.~\ref{Fig:Tucker_Models} and Algorithm \ref{Algo:NTFk_Transient_Bimolecular} summarizes the proposed tensor-factorization methodology.
For a given set of parameters, the output of our simulations is a data-tensor $X_{ijl}$ in three dimensions: $ (t, x, y) \Rightarrow ( 1000  \times 81  \times 81 )$.
$X(t,x,y)$ corresponds to concentrations of non-negative invariants, which are $c_F$ and $c_G$.
In each cell of this non-negative tensor, we have the concentration $X(t_i, x_j, y_l)$ at a specific time $t_i$ and at a spatial point whose coordinates are $(x_j, y_l)$. 
Due to inherent non-linearity in the problem, first, hidden features are extracted from each non-negative invariant.
Then, using the non-linear transformations given by Eqs.~\eqref{eqn:Fast_A}--\eqref{eqn:Fast_C}, we obtain the corresponding hidden features of product $C$.

To analyze such a three-dimensional simulation dataset, we utilize a sparse non-negative Tucker-3 decomposition model \cite{morup2008algorithms}.
Our choice for non-negative constraints is motivated by (i) the fact that the considered data-tensors are inherently non-negative and (ii) our goal to relate the extracted features to easily interpretable quantities without introducing any a priori assumptions.
Indeed, a meaningful interpretation of the obtained results requires the extracted features to be \emph{parts} of the original data \cite{lee1999learning}.
The non-negative constraints in the tensor factorization process lead to extraction of strictly additive components, which are parts of the original data \cite{ross2006learning}.
Thus, the non-negative tensor factorization has the ability to identify readily understandable structure-preserving features that enable the discovery of new causal structures and unknown mechanisms hidden in the data \cite{cichocki2009nonnegative}. 

The non-negative Tucker-3 decomposition of the three dimensional data-tensor representing the temporal evolution of $X(t, x, y)$ on a two-dimensional $(x,y)$ non-negative finite element grid is,
\begin{align}
X(t, x, y) = G \otimes  W(t) \otimes H(x) \otimes V(y) + \epsilon(t, x, y)
\end{align}
where $\otimes$ denotes the tensor product.
The decomposition of the non-negative tensor $X(t, x, y)$ ($X \in {\mathbb{R}_{\geq 0}^{K \times M \times N}}$) can be expressed by components:
\begin{align}
X_{ijl} =  \sum_{p=1}^{k}\sum_{q=1}^{m}\sum_{r=1}^{n}{ {G_{pqr}} {W_{ip}} {H_{jq}} {V_{lr}} } + {\epsilon_{ijl}} \quad \forall i, j, l
\label{eqn:tucker3}
\end{align}
where all the elements of $X$, $G$, $W$, $H$, and $V$ are non-negative,
\begin{align}
X_{ijl}, G_{pqr}, W_{ip}, H_{jq} , V_{lr} \geq 0 \quad \forall i, j, l, p, q, r
\end{align}
Eq.~\ref{eqn:tucker3} can be also represented as:
\begin{align}
X_{ijl} =  \widetilde{X}_{ijl} + {\epsilon_{ijl}}
\label{eqn:tucker3a}
\end{align}
where $\widetilde{X}$ ($\widetilde{X} \in {\mathbb{R}_{\geq 0}^{K \times M \times N}}$) is the Tucker-3 estimate of $X$:
\begin{align}
\widetilde{X}_{ijl} = \sum_{p=1}^{k}\sum_{q=1}^{m}\sum_{r=1}^{n}{ {G_{pqr}} {W_{ip}} {H_{jq}} {V_{lr}} }
\end{align}

The Tucker-3 decomposition includes
(i) an unknown core-tensor $G$ ($G \in {\mathbb{R}_{\geq 0}^{k \times m \times n}}) $ that represents the interactions between the $t$, $x$, and $y$ components of $W(t), H(x),$ and $V(y)$;
(ii) an unknown matrix $W$ ($W \in \mathbb{R}_{\geq 0}^{K \times k}$) representing the changes of $X$ in time (the time-component);
(iii) an unknown matrix $H$ ($H \in \mathbb{R}_{\geq 0}^{M \times m}$) representing the changes of $X$ in the x-direction (the x-component), and
(iv) an unknown matrix $V$ ($V \in \mathbb{R}_{\geq 0}^{N \times n}$) representing the changes of $X$ in the y-direction (the y-component).
Here, $\mathbb{R}_{\geq 0}$ denotes the set of non-negative real numbers $\mathbb{R}_{\geq 0} = \left\{ x \in \mathbb{R} \mid x \geq 0 \right\}$. 
Additionally, $\epsilon$ (${\epsilon} \in {\mathbb{R}^{K \times M \times N}}$) denotes the unknown discrepancy between the original data $X(t,x,y)$ and estimate $\widetilde{X}(t,x,y)$;
the discrepancy is caused by presence of a noise or errors in the measurements.
If the Tucker decomposition is successful, $\epsilon$ should represent a white noise.
An important part of our NTF$k$ reconstruction analyses is to make sure that $\epsilon$ characterizes a white noise and there additional signals that are not extracted by our procedure.

Mathematically, the solution of the non-negative Tucker-3 tensor decomposition is a solution of a complex multi-dimensional optimization with non-negative constraints given by:
\begin{align}
\min\limits_{G, W, H, V \geq 0}  \| X - G \otimes  W \otimes H \otimes V \|^{2}
\label{eqn:min}
\end{align}

To extract the unknown core-tensor $G$, and the factor matrices $W$, $H$, and $V$, we utilize the block coordinate descent method for regularized multi-convex optimization algorithms introduced by Xu et.al. \cite{Xu2013, Xu2015} and programmed in \texttt{Julia} \cite{bezanson2012julia}. 
In the analyses below, we focus on the temporal features $W$, i.e., we estimate the robustness of the factorization related to the temporal factor $W$.
To identify the optimal number of features (i.e. the optimal number of columns of $W$), we perform `$N$' NTF runs with random initial guesses for the unknown parameters, and then cluster the resulting sets of columns of the factors $W^k_1, W^k_2, ...,W^k_N$, where $k$ is the number of features used in this set of simulations.
We use a custom clustering algorithm for this purpose, which is based on $k$-means clustering and described by \cite{alexandrov2014blind} and \cite{vesselinov2018contaminant}.
$k$-means is  one of  the simplest unsupervised  learning  algorithms  that  solve  the  clustering problem.
The procedure classifies every single dataset entry to belong to a certain number of clusters fixed apriori based on minimization of a user-defined objective function (in our case, the objective functions is based on cosine distances; \cite{alexandrov2014blind},  \cite{vesselinov2018contaminant}).  
The optimal number of features is evaluated by comparison of the quality of the reconstruction, $R$, based on the Frobenius norm (Eq.~\ref{eqn:min}):  $R = ||X - \widetilde{X}||/||X||$ and the quality of the derived clusters for different number of features, $k$, estimated by their average Silhouettes \cite{rousseeuw1987Silhouettes}.
Silhouette  is a measure of how similar a given single dataset entry is to its own cluster compared to other clusters.
The silhouette ranges from $-1$ to $+1$, where values close to $+1$ indicate that the dataset entry is well matched to its own cluster and poorly matched to neighboring clusters.
High average Silhouette values suggest that the clustering configuration is appropriate.
Low average Silhouette values  suggest that the clustering configuration is not optimal; potentially, there are too many or too few clusters.

\begin{algorithm}
  \caption{{\small Overview of NTF$k$-based structure-preserving feature extraction framework for reactive-mixing}}
  \label{Algo:NTFk_Transient_Bimolecular}
  \begin{algorithmic}[1]
    \STATE Input:~Time step $\Delta t$; total time of interest $\mathcal{I}$; stoichiometric coefficients; initial and boundary conditions for the chemical species $A$, $B$, and $C$; finite element mesh parameters; model parameters for anisotropic reaction-diffusion equation.
    %
    \STATE Calculate initial and boundary conditions for the non-negative invariants $c_F$ and $c_G$ using Eqs.~\eqref{eqn:Definitions_of_F}--\eqref{eqn:Definitions_of_G}.
    %
    \STATE Solve for invariant concentrations $c_F$ and $c_G$ for all times.
    
    \FOR{$n = 0, 1, \cdots, \left(\texttt{NumTimeSteps} - 1 \right)$}
    %
    \STATE Call optimization-based diffusion with decay solver to obtain $\boldsymbol{c}_F^{(n+1)}$:    
    \begin{align*}
      \mathop{\mbox{minimize}}_{\boldsymbol{c}^{(n + 1)}_F \in 
        \mathbb{R}^{ndofs}} & \quad \frac{1}{2}  \left \langle 
      \boldsymbol{c}^{(n + 1)}_F; \boldsymbol{K} 
      \boldsymbol{c}^{(n + 1)}_F \right \rangle - \left \langle 
      \boldsymbol{c}^{(n + 1)}_F; \boldsymbol{f}^{(n + 1)}_F  
      \right \rangle - \frac{1}{\Delta t}  \left \langle 
      \boldsymbol{c}^{(n + 1)}_F; \boldsymbol{c}^{(n)}_F
      \right \rangle \\
      \mbox{subject to} & \quad c_F^{\mathrm{min}} \boldsymbol{1} \preceq 
      \boldsymbol{c}^{(n + 1)}_F \preceq c_F^{\mathrm{max}} \boldsymbol{1} 
    \end{align*}
    \STATE Call optimization-based diffusion with decay solver to obtain $\boldsymbol{c}_G^{(n+1)}$:    
    \begin{align*}
      \mathop{\mbox{minimize}}_{\boldsymbol{c}^{(n + 1)}_G \in 
        \mathbb{R}^{ndofs}} & \quad \frac{1}{2}  \left \langle 
      \boldsymbol{c}^{(n + 1)}_G; \boldsymbol{K} 
      \boldsymbol{c}^{(n + 1)}_G \right \rangle - \left \langle 
      \boldsymbol{c}^{(n + 1)}_G; \boldsymbol{f}^{(n + 1)}_G  
      \right \rangle - \frac{1}{\Delta t}  \left \langle 
      \boldsymbol{c}^{(n + 1)}_G; \boldsymbol{c}^{(n)}_G
      \right \rangle \\
      \mbox{subject to} & \quad c_G^{\mathrm{min}} \boldsymbol{1} \preceq 
      \boldsymbol{c}^{(n + 1)}_G \preceq c_G^{\mathrm{max}} \boldsymbol{1} 
    \end{align*}
    \STATE where $\boldsymbol{K}$ is the symmetric positive definite coefficient matrix, $ndofs$ is the number of degrees of freedoom, $\langle \cdot;\cdot \rangle$ represents the standard inner-product on Euclidean spaces, $\boldsymbol{1}$ denotes a vector of ones of size $ndofs \times 1$, and the symbol $\preceq$ represents the component-wise inequality for vectors. 
    \ENDFOR
    %
    \STATE Get non-negative solutions for $c_A$, $c_B$, and $c_C$ using Eqs.~\eqref{eqn:Fast_A}--\eqref{eqn:Fast_C}.    
    %
    \STATE Decompose $c_F$ and $c_G$ using Eq.~\eqref{eqn:min}. 
    \begin{itemize}
      \item $c_F(\mathbf{x},t) = c_F(\mathbf{x},t)|_{\mathrm{T1}} + c_F(\mathbf{x},t)|_{\mathrm{T2}} + c_F(\mathbf{x},t)|_{\mathrm{T3}} \quad \forall (\mathbf{x}, t) \in \overline{\Omega} \times [0, \mathcal{I}]$
      \item $c_G(\mathbf{x},t) = c_G(\mathbf{x},t)|_{\mathrm{T1}} + c_G(\mathbf{x},t)|_{\mathrm{T2}} + c_G(\mathbf{x},t)|_{\mathrm{T3}} \quad \forall (\mathbf{x}, t) \in \overline{\Omega} \times [0, \mathcal{I}]$
    \end{itemize}
    where T1, T2, and T3 are temporal features identified by NTF$k$ method.
    %
    \STATE Using non-linear transformations given by Eqs.~\eqref{eqn:Fast_A}--\eqref{eqn:Fast_C}, obtain the hidden features of $c_C$.
    Based on Eq.~\eqref{eqn:Fast_A}, as $c_A \leq c_F \quad \forall t \in [0, \mathcal{I}]$, we have
    \begin{itemize}
      \item \mbox{$c_A(\mathbf{x},t) = \underbrace{\mathrm{max} \left[c_F - \left(\frac{n_A}{n_B}\right) c_G, \, 0 
	    \right]|_{\mathrm{T1}}}_{c_A|_{\mathrm{T1}}} + \underbrace{\mathrm{max} \left[c_F - \left(\frac{n_A}{n_B}\right) c_G, \, 0 
	    \right]|_{\mathrm{T2}}}_{c_A|_{\mathrm{T2}}} + \underbrace{\mathrm{max} \left[c_F - \left(\frac{n_A}{n_B}\right) c_G, \, 0 
	    \right]|_{\mathrm{T3}}}_{c_A|_{\mathrm{T3}}}$}
      \item $c_C(\mathbf{x},t) = \left( \frac{n_C}{n_A} \right) \; \left( \left(c_F - c_A\right)|_{\mathrm{T1}} 
        + \left(c_F - c_A\right)|_{\mathrm{T2}} + \left(c_F - c_A\right)|_{\mathrm{T3}} \right) \quad \forall (\mathbf{x}, t) 
        \in \overline{\Omega} \times [0, \mathcal{I}]$
    \end{itemize}
    %
    \STATE Relate the extracted hidden features to physical processes based on the impact of model input parameters on the features.
    That is,
    \begin{itemize}
      \item T1 $\Rightarrow$ mixing of reactants due to longitudinal dispersion
      \item T2 $\Rightarrow$ mixing of reactants due to transverse dispersion
      \item T3 $\Rightarrow$ mixing of reactants due to molecular diffusivity
    \end{itemize}
    %
    \STATE Compare the decomposed features with the ground truth (for example, see Fig.~\ref{Fig:Truth_vs_NTFk}).
  \end{algorithmic}
\end{algorithm}
%

\section{RESULTS}
\label{sec:results}
The NTF$k$ methodology was applied to analyze more than 2000 simulations predicting the concentrations of product $C$ using the methodology outlined in Section \ref{sec:simulation}.
In our case, $K = 1000$, $M = 81$, and $N = 81$. 
For $\epsilon \approx \mathcal{O}(10^{-9})$, the size of the obtained core tensor $G$, which is given in terms of $k$, $m$, and $n$, vary from 2 to 4, 2 to 14, and 3 to 15, respectively.
Each numerical simulation is obtained for a different set of reaction-diffusion model input parameters.
The varied input parameters are:~perturbation parameter of the underlying vortex-based velocity field $v_0$, longitudinal-to-transverse anisotropic dispersion ratio $\frac{\alpha_L}{\alpha_T}$, molecular diffusivity $D_m$, and velocity field characteristic scales $\kappa_fL$ and $T$.
$\alpha_L$ is taken to be equal to 1 and $\alpha_T$ is varied accordingly.
Note that simulations will also depend upon the selected values of $\alpha_L$.
The corresponding values for the varied input parameters are: $v_o = \left[1, 10^{-1}, 10^{-2}, 10^{-3}, 10^{-4} \right]$, $\frac{\alpha_L}{\alpha_T} = \left[1, 10^{1}, 10^{2}, 10^{3}, 10^{4} \right]$, $D_m = \left[10^{-8}, 10^{-1}, 10^{-2}, 10^{-3} \right]$, $\kappa_fL = \left[2, 3, 4, 5 \right]$, and $T = \left[1 \times 10^{-4}, 2 \times 10^{-4}, 3 \times 10^{-4}, 4 \times 10^{-4}, 5 \times 10^{-4} \right]$.
In our simulations, $\alpha_L = 1$ and $\alpha_T$ is varied accordingly.
There is no existing methodology or closed-form mathematical solutions that can predict how the variation of these input parameters impacts the concentration of product $C$.
The goal of the proposed unsupervised ML analyses of the simulation outputs is to provide insights about the influences of the input parameters on species mixing, decay of reactants, and formation of the reaction product. 

\begin{figure}
	\centering
	\subfigure[$\kappa_fL = 2$ and $t = 0.1$]
	{\includegraphics[clip=true,width = 0.37\textwidth]{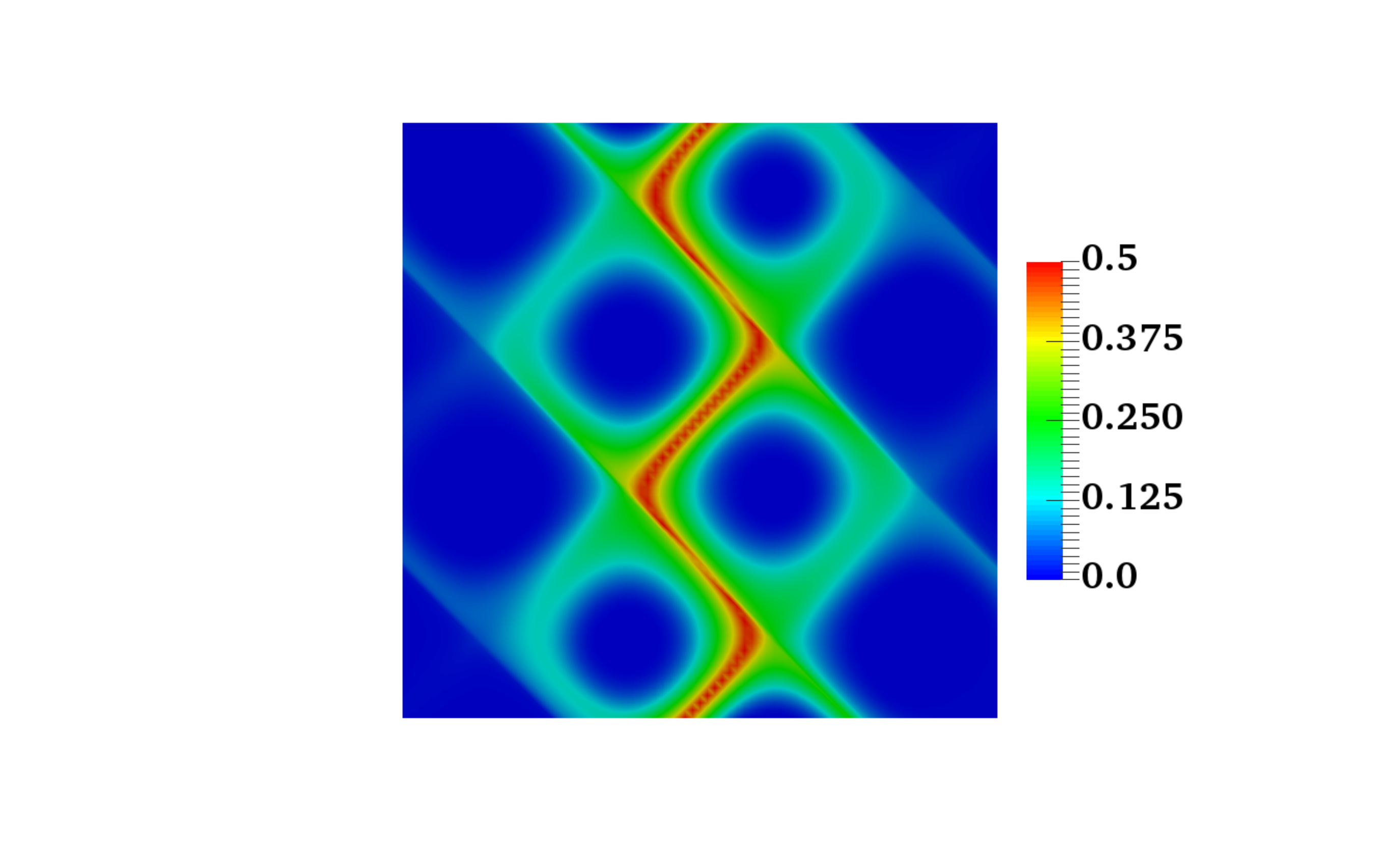}}
	\hspace{-0.5in}
	\subfigure[$\kappa_fL = 2$ and $t = 0.5$]
	{\includegraphics[clip=true,width = 0.37\textwidth]{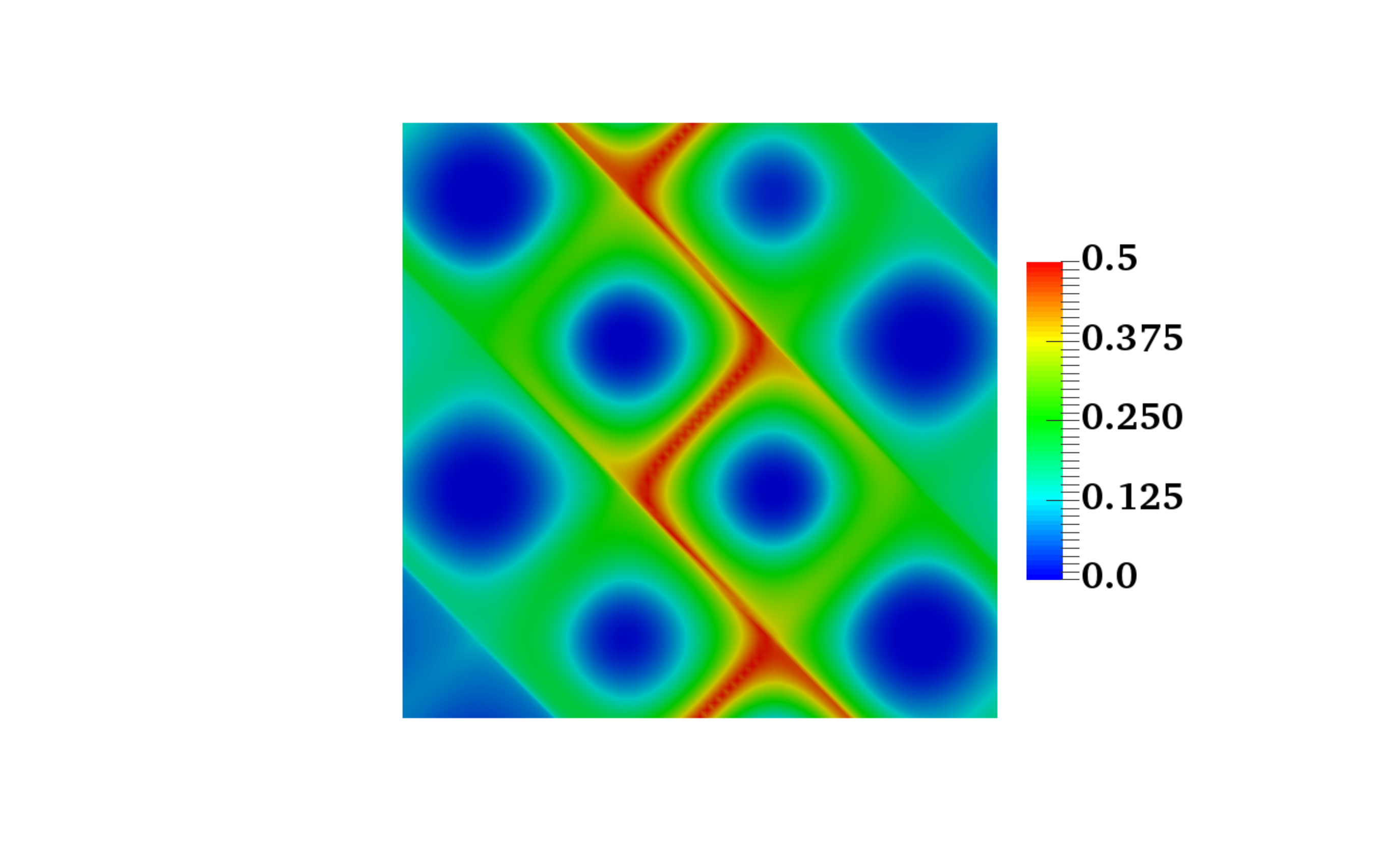}}
	\hspace{-0.5in}
	\subfigure[$\kappa_fL = 2$ and $t = 1.0$]
	{\includegraphics[clip=true,width = 0.37\textwidth]{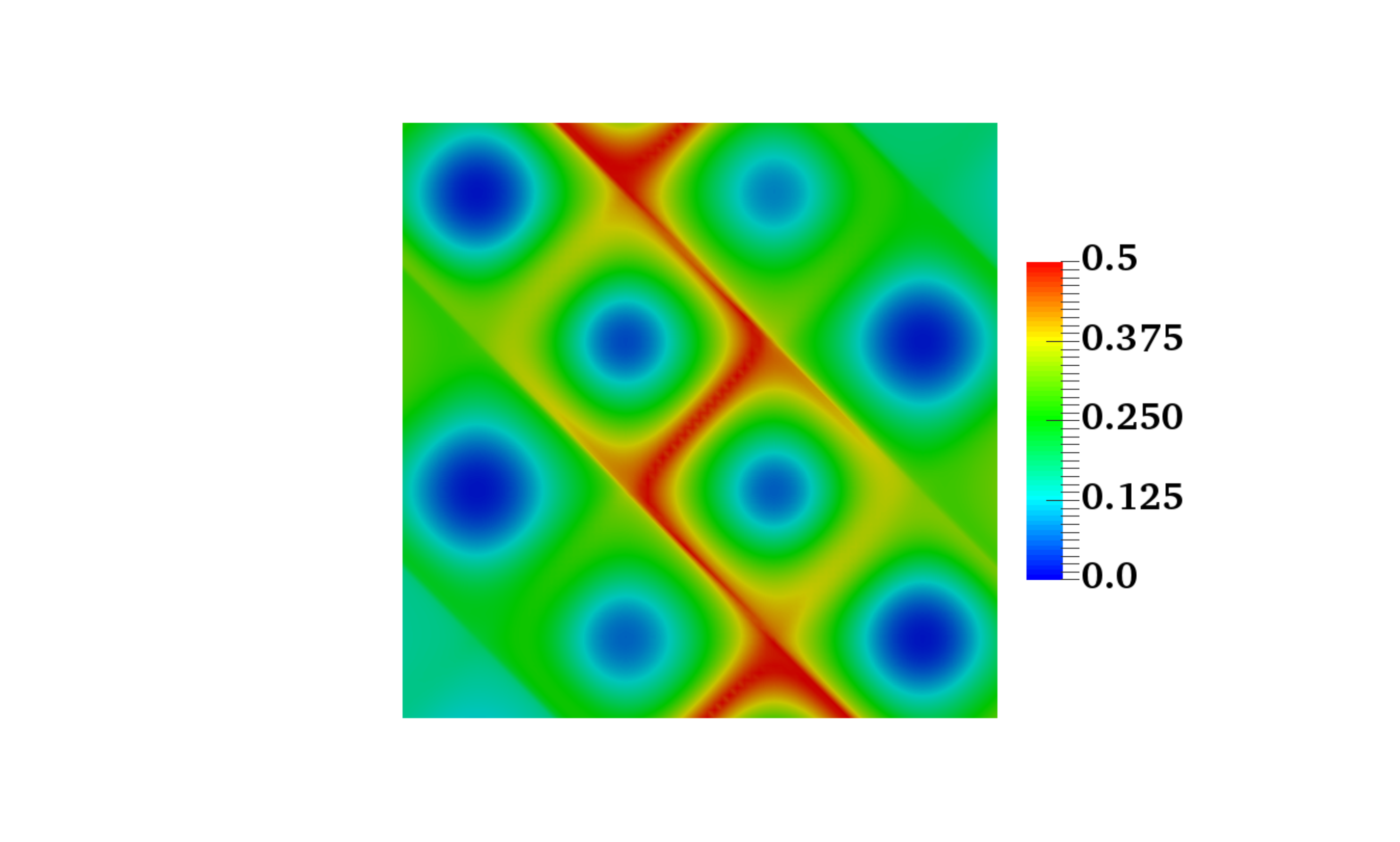}}
	\subfigure[$\kappa_fL = 3$ and $t = 0.1$]
	{\includegraphics[clip=true,width = 0.37\textwidth]{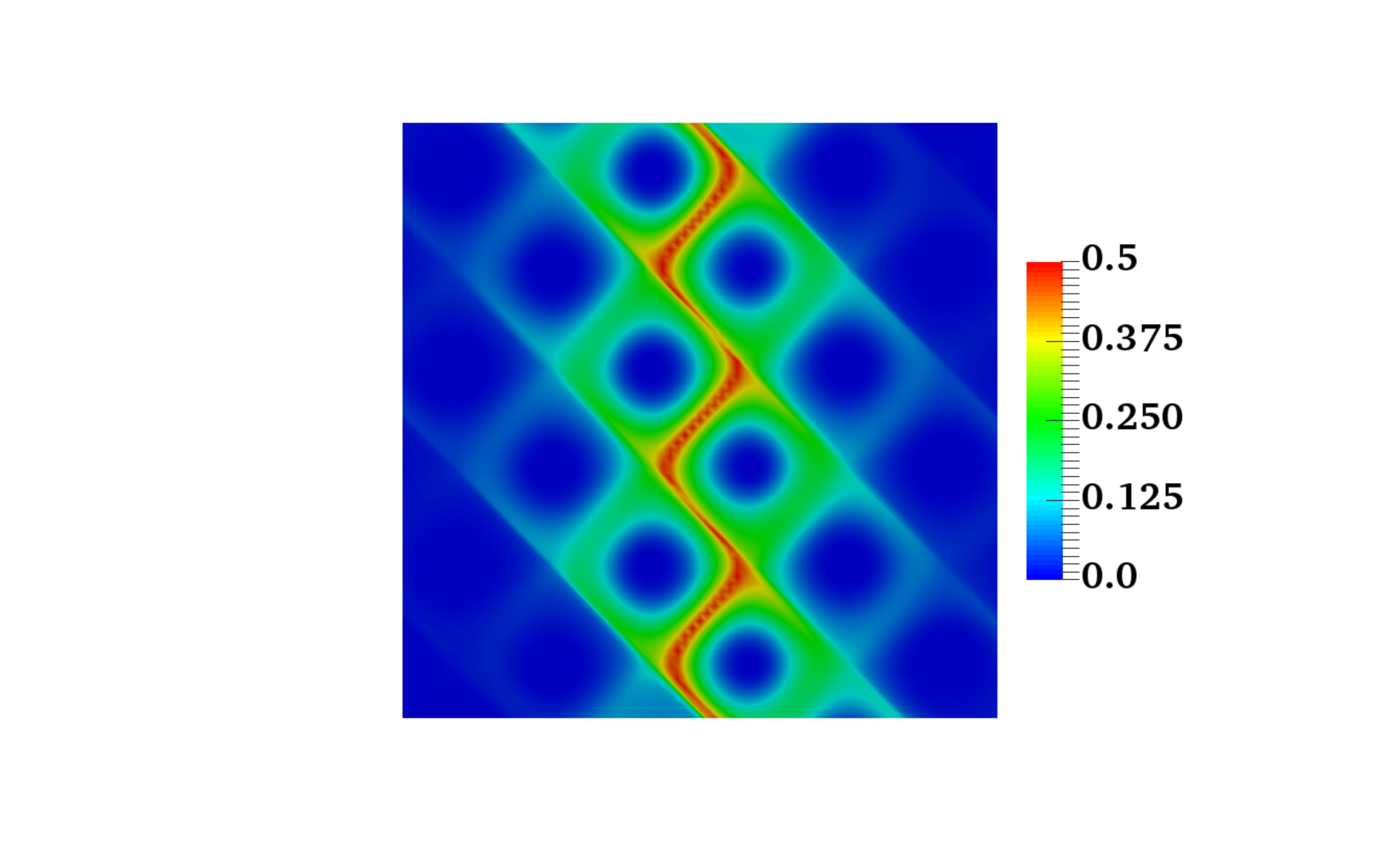}}
	\hspace{-0.5in}
	\subfigure[$\kappa_fL = 3$ and $t = 0.5$]
	{\includegraphics[clip=true,width = 0.37\textwidth]{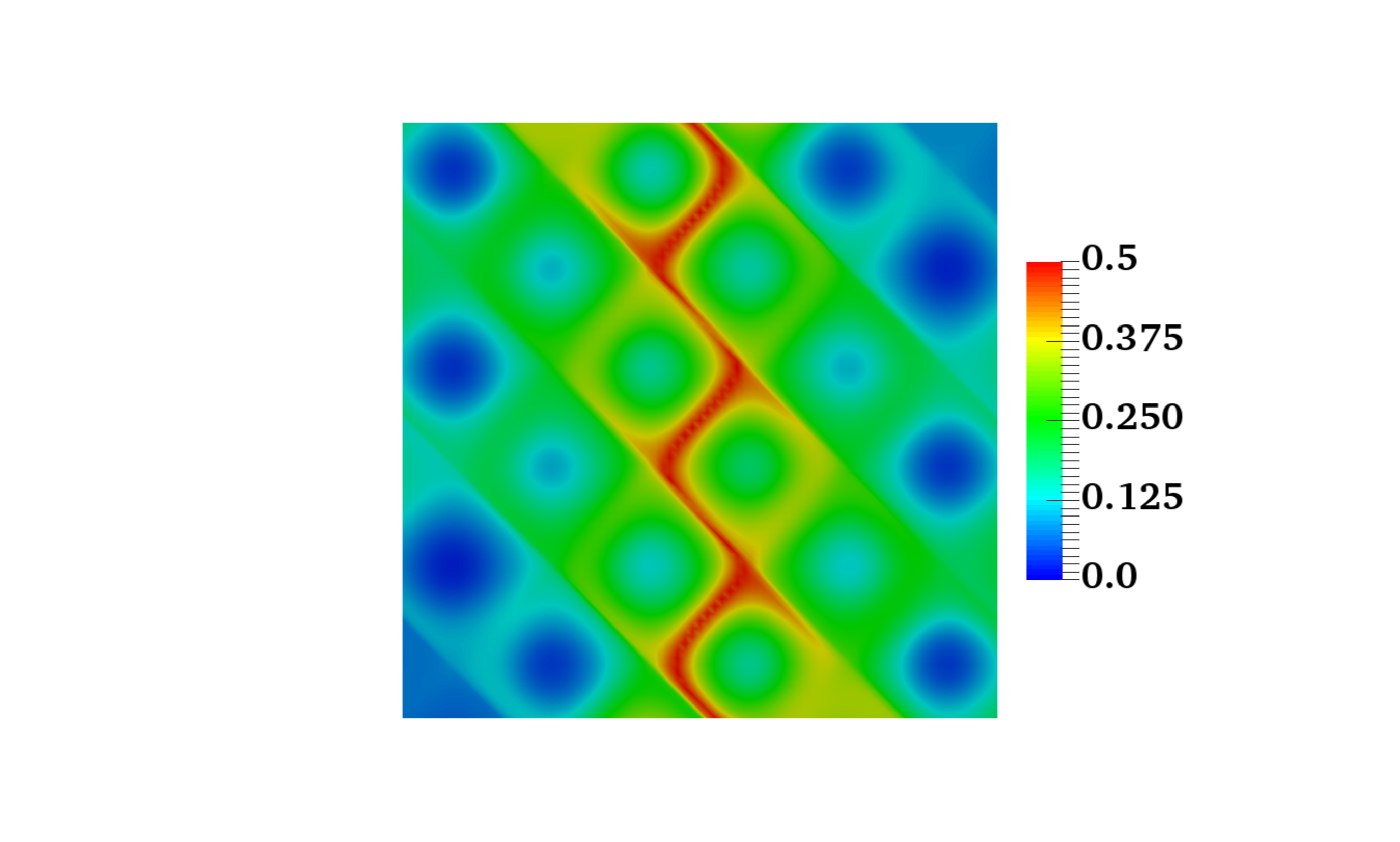}}
	\hspace{-0.5in}
	\subfigure[$\kappa_fL = 3$ and $t = 1.0$]
	{\includegraphics[clip=true,width = 0.37\textwidth]{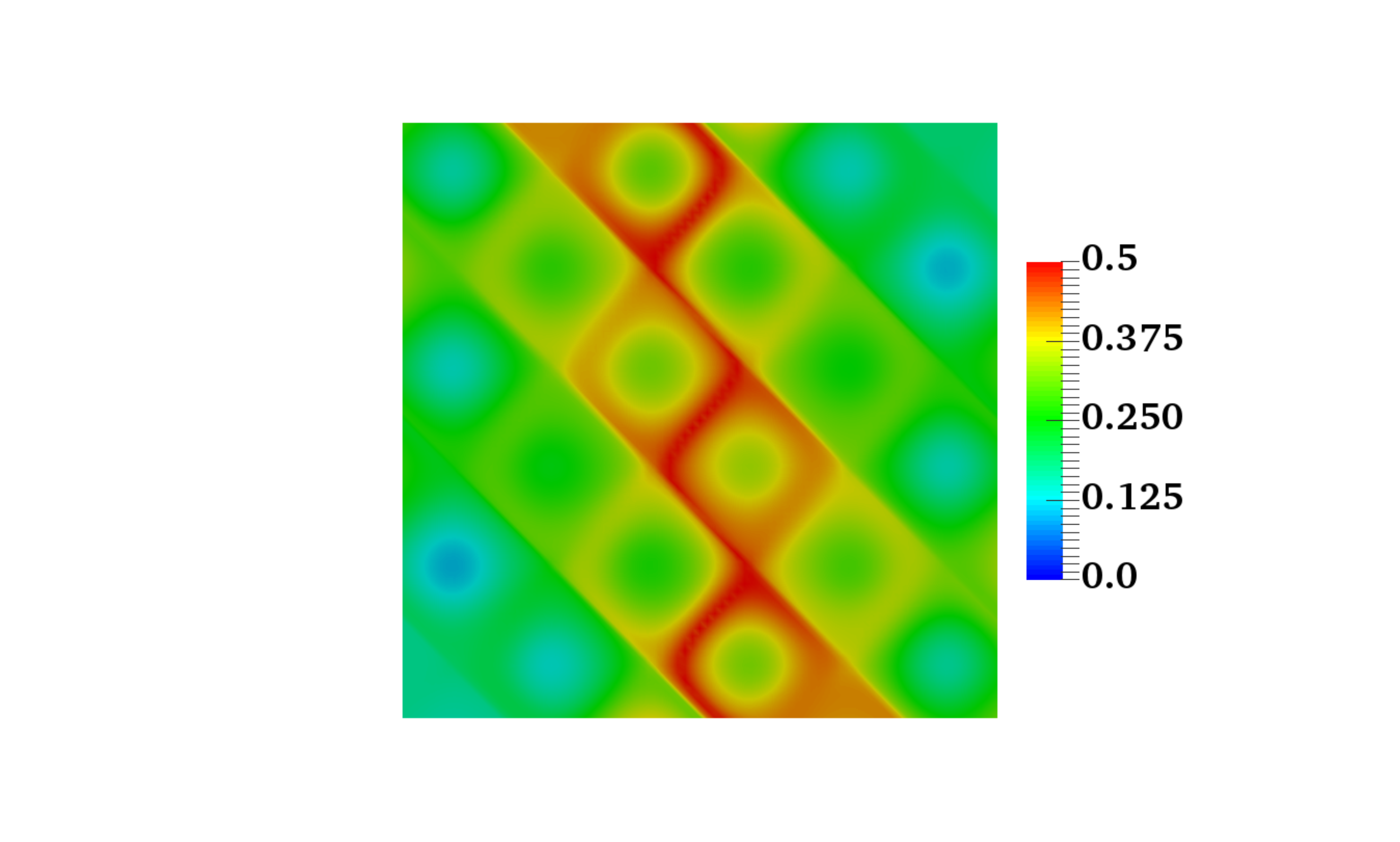}}
	\subfigure[$\kappa_fL = 4$ and $t = 0.1$]
	{\includegraphics[clip=true,width = 0.37\textwidth]{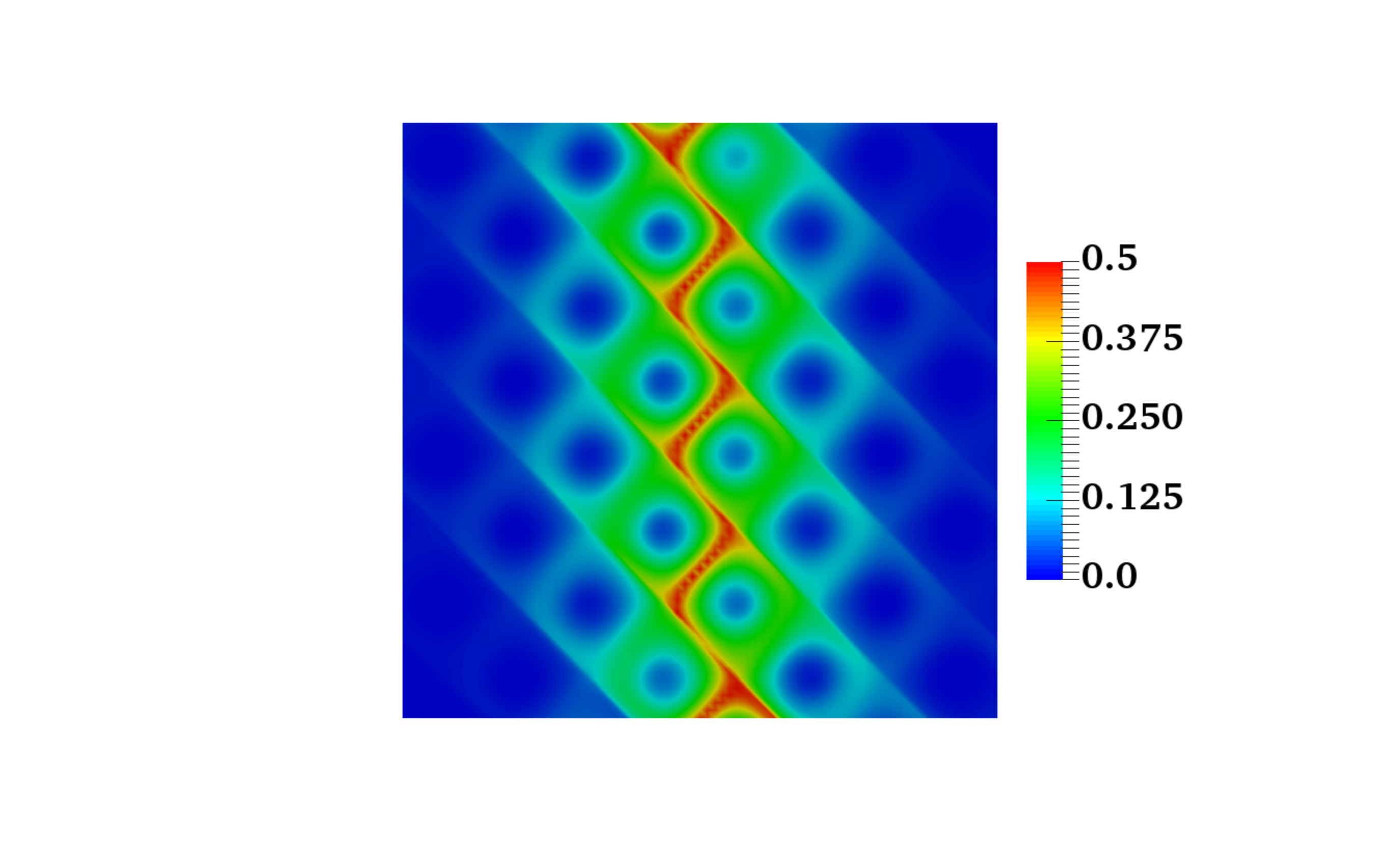}}
	\hspace{-0.5in}
	\subfigure[$\kappa_fL = 4$ and $t = 0.5$]
	{\includegraphics[clip=true,width = 0.37\textwidth]{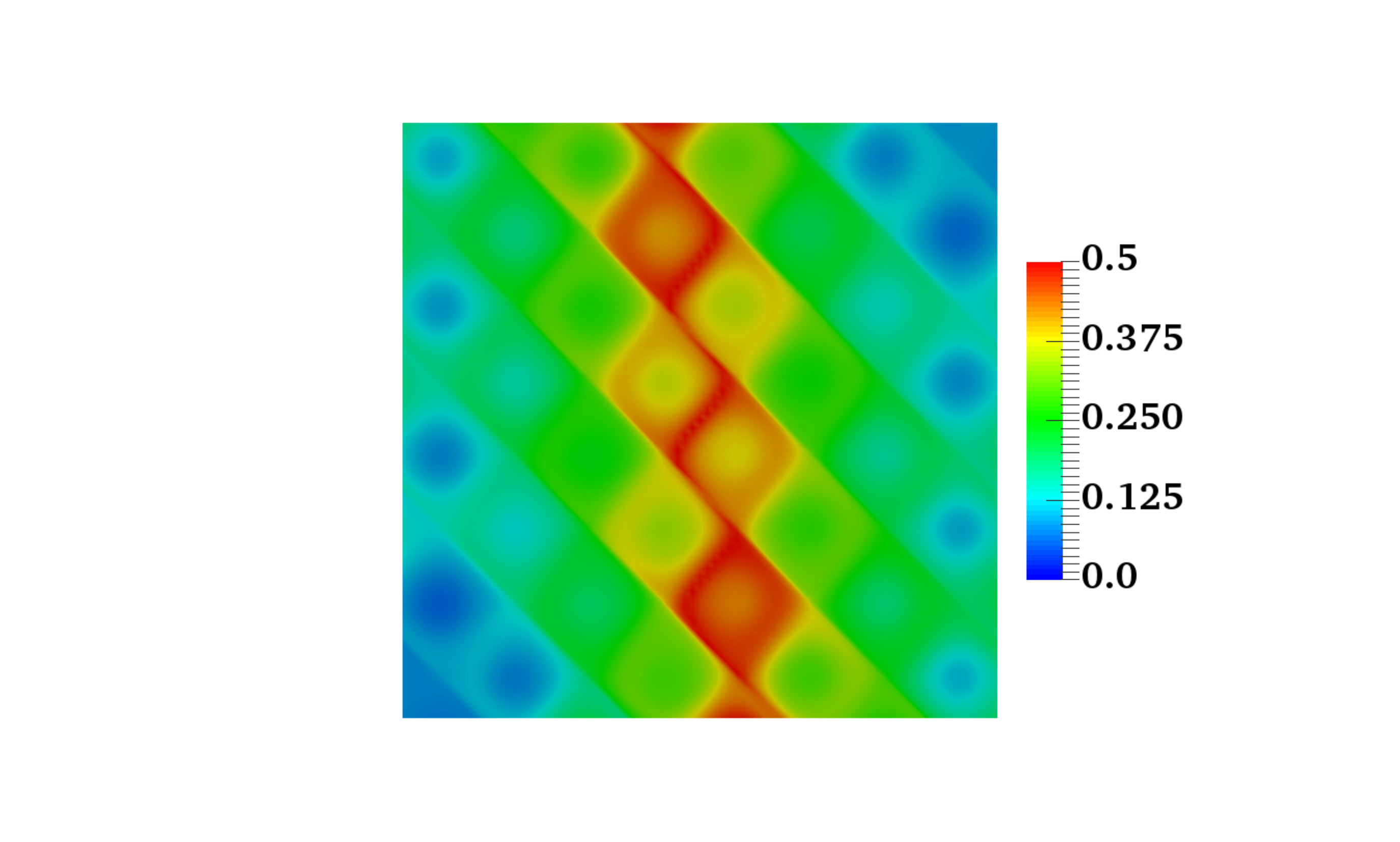}}
	\hspace{-0.5in}
	\subfigure[$\kappa_fL = 4$ and $t = 1.0$]
	{\includegraphics[clip=true,width = 0.37\textwidth]{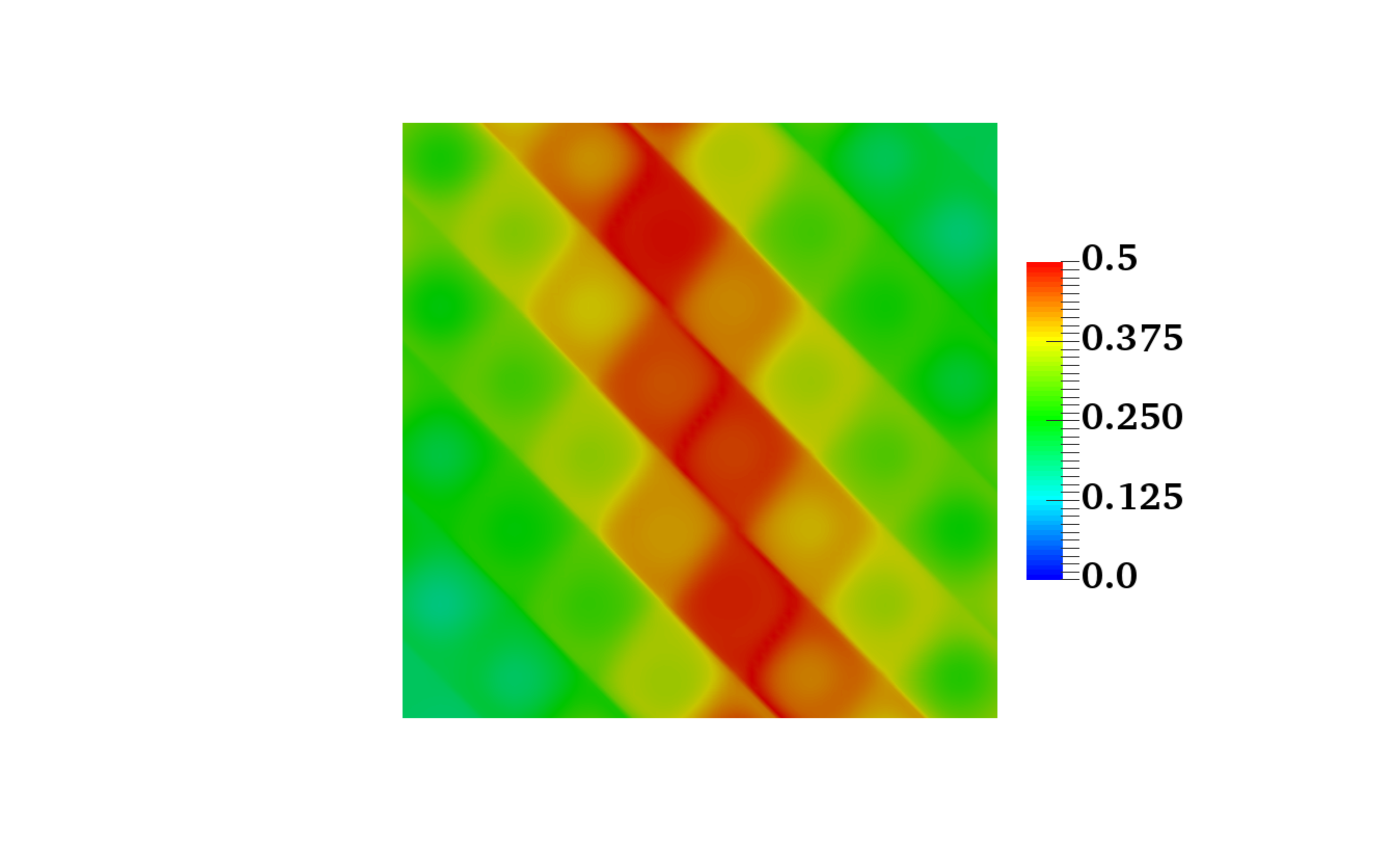}}
	\subfigure[$\kappa_fL = 5$ and $t = 0.1$]
	{\includegraphics[clip=true,width = 0.37\textwidth]{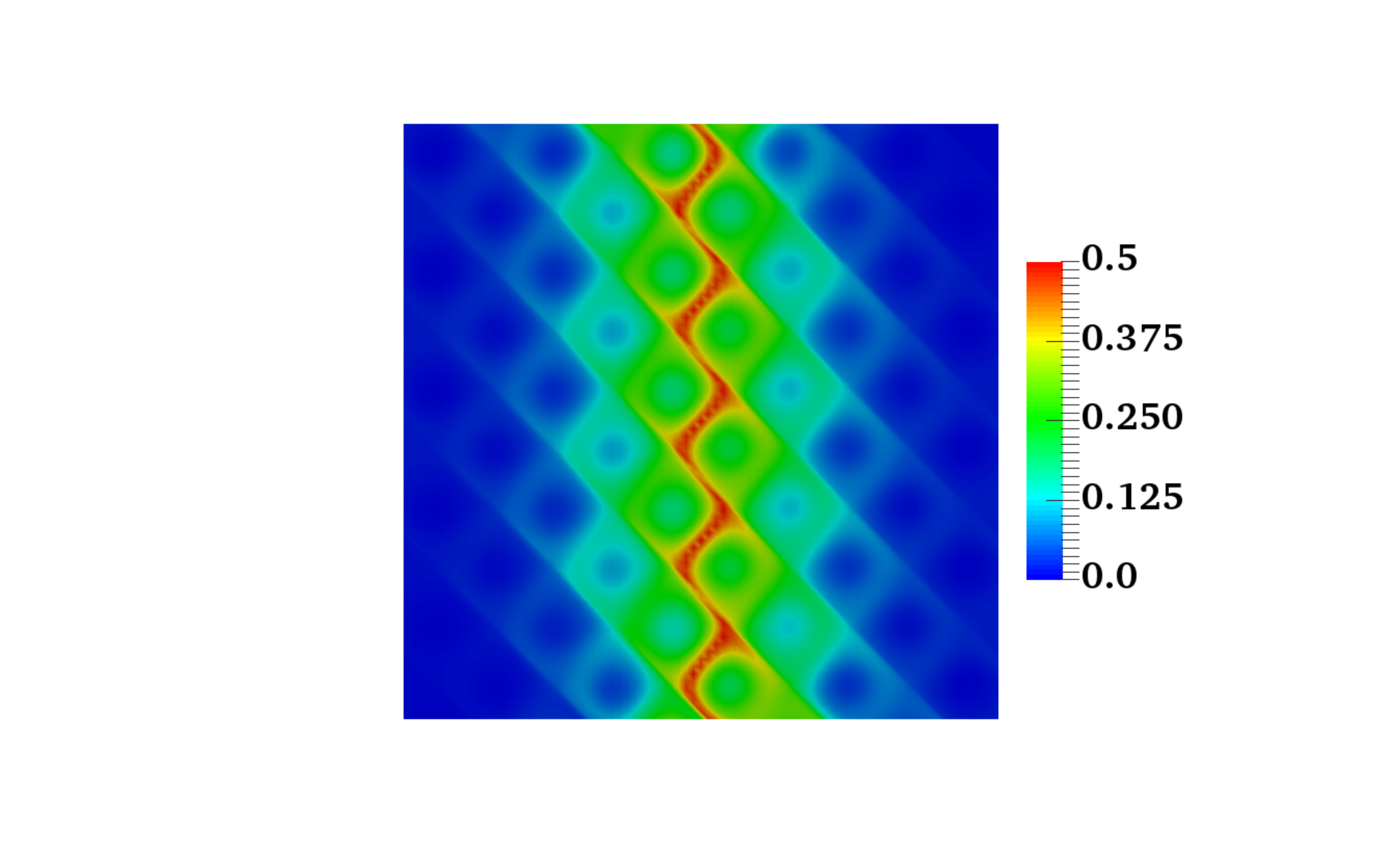}}
	\hspace{-0.5in}
	\subfigure[$\kappa_fL = 5$ and $t = 0.5$]
	{\includegraphics[clip=true,width = 0.37\textwidth]{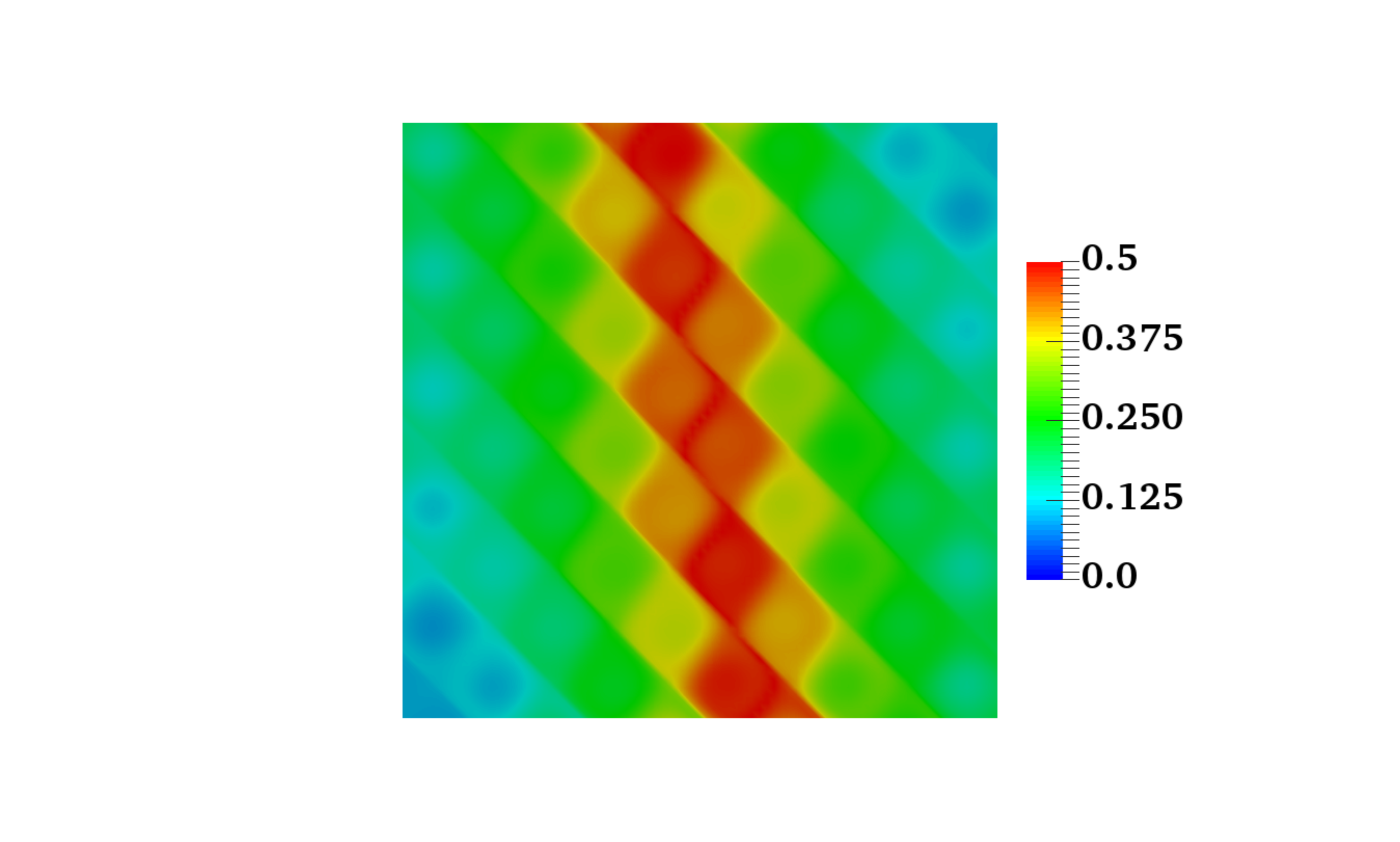}}
	\hspace{-0.5in}
	\subfigure[$\kappa_fL = 5$ and $t = 1.0$]
	{\includegraphics[clip=true,width = 0.37\textwidth]{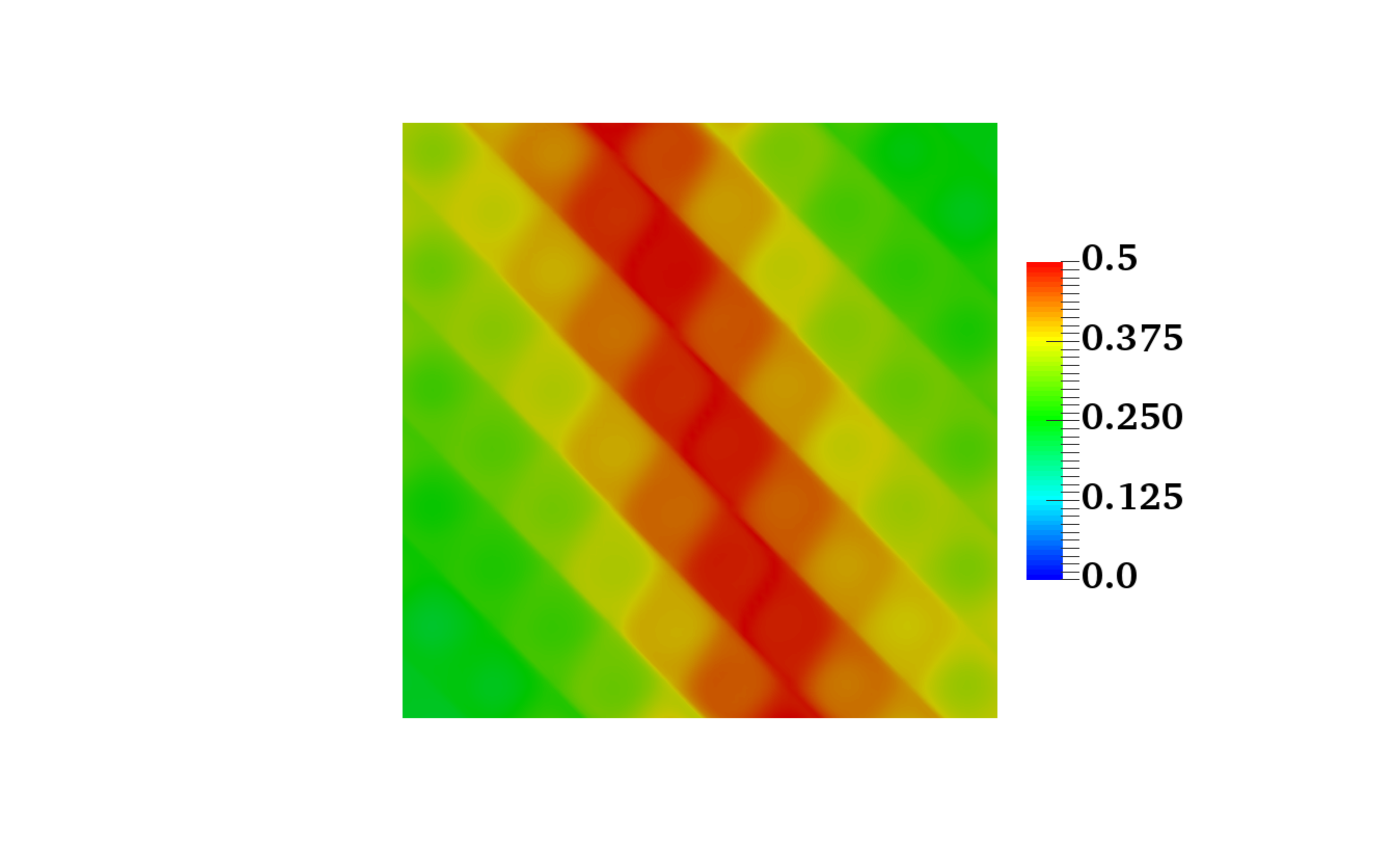}}
	\caption{Concentration contours of product $C$ at times $t = 0.1$, $0.5$, and $1.0$ for different $\kappa_fL$ values.
	Other input parameters are $\frac{\alpha_L}{\alpha_T} = 10^{4}$, $v_o = 10^{-1}$, $T = 1 \times 10^{-4}$, and $D_m = 10^{-3}$.
	At early times and for smaller values of $\kappa_fL$, concentration of product $C$ is zero in the vortex centers.
	At late times and as $\kappa_fL$ increases, the regions with zero concentration of product $C$ decrease. 
	\label{Fig:Contours_C_Difftimes}}
\end{figure}

The output from each numerical simulation is a tensor with dimensions $1000 \times 81 \times 81$ representing the concentration of product $C$ in time and space.
That is, there are 1000 time steps for computational times ranging from 0.001 to 1.0 with a uniform time step equal to 0.001.
The two-dimensional computational domain is discretized by low-order structured triangular finite elements 
with 81 nodes on each side.
The resulting finite element mesh is of size $81 \times 81$.
Mudunuru and co-authors \cite{2013_Nakshatrala_Mudunuru_Valocchi_JCP_v253_p278_p307,2016_Mudunuru_Nakshatrala_JCP_v305_p448_p493,2017_Mudunuru_Nakshatrala_MAMS} have performed numerical $h$-convergence studies in their previous works.
Through grid refinement studies, they showed that $81 \times 81$ is sufficient to get accurate non-negative FEM solution for fine spatial scale variation in dispersion.
Fig.~\ref{Fig:Contours_C_Difftimes} shows the concentration contours of product $C$ at various times for a case when anisotropic dispersion is very high and molecular diffusivity is very low.
The input parameter that is varied is $\kappa_fL$ and other parameters such as $\frac{\alpha_L}{\alpha_T}$, $v_0$, $T$, and $D_m$ are kept constant.
Small values of $\kappa_fL$ correspond to large-scale vortex structures and large values of $\kappa_fL$ correspond to small-scale vortex structures in the velocity field.
From this figure, one can infer that the small-scale vortex structures present in the velocity field enhance mixing even when the value of molecular diffusivity is small and anisotropic dispersivity is large.

\begin{figure}
	\centering
	{\includegraphics[width = 1.05\textwidth]{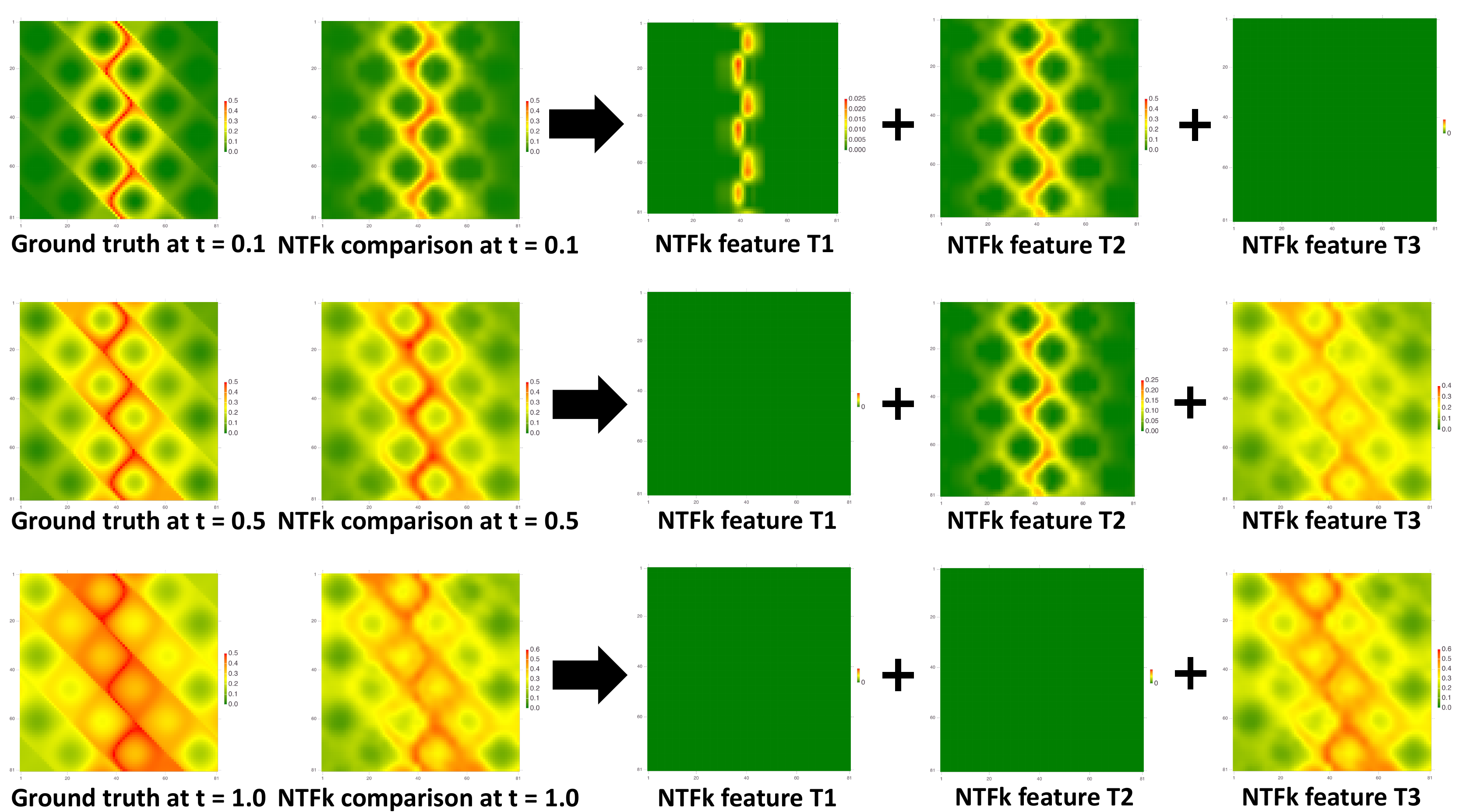}}
	\caption{Comparison of ground truth and NTF$k$ results at various times for product $C$.
	This figure also shows the decomposition of NTF$k$ result into features T1, T2, and T3, which represent reactive-mixing due to longitudinal dispersion, transverse dispersion, and molecular diffusion, respectively. 
	The input parameters corresponding to this particular case are:~$v_o = 10^{-3}$, $\frac{\alpha_L}{\alpha_T} = 10^{4}$, $D_m = 10^{-3}$, $\kappa_fL = 3$, and $T = 1 \times 10^{-4}$.
	$v_o = 10^{-3}$ corresponds to small perturbations in underlying vortex-based velocity field.
	$T = 1 \times 10^{-4}$ corresponds to fast flipping of vortex structures from clockwise direction to anti-clockwise direction.
	$\frac{\alpha_L}{\alpha_T} = 10^{4}$ and $D_m = 10^{-3}$ corresponds to high anisotropic dispersion and low molecular diffusivity.
	$\kappa_fL = 3$ corresponds to medium-scale vortex structures present in the velocity field.
	The numerical errors ($L_2$-norm) at times $t = 0.1$, $t = 0.5$, and $t = 1.0$ are equal to 1.954, 1.285, and 1.587.
	\label{Fig:Truth_vs_NTFk}}
\end{figure}

\begin{figure}
	\centering
	\subfigure[$v_o = 10^{-4}$]
	{\includegraphics[clip=true,width = 0.49\textwidth]{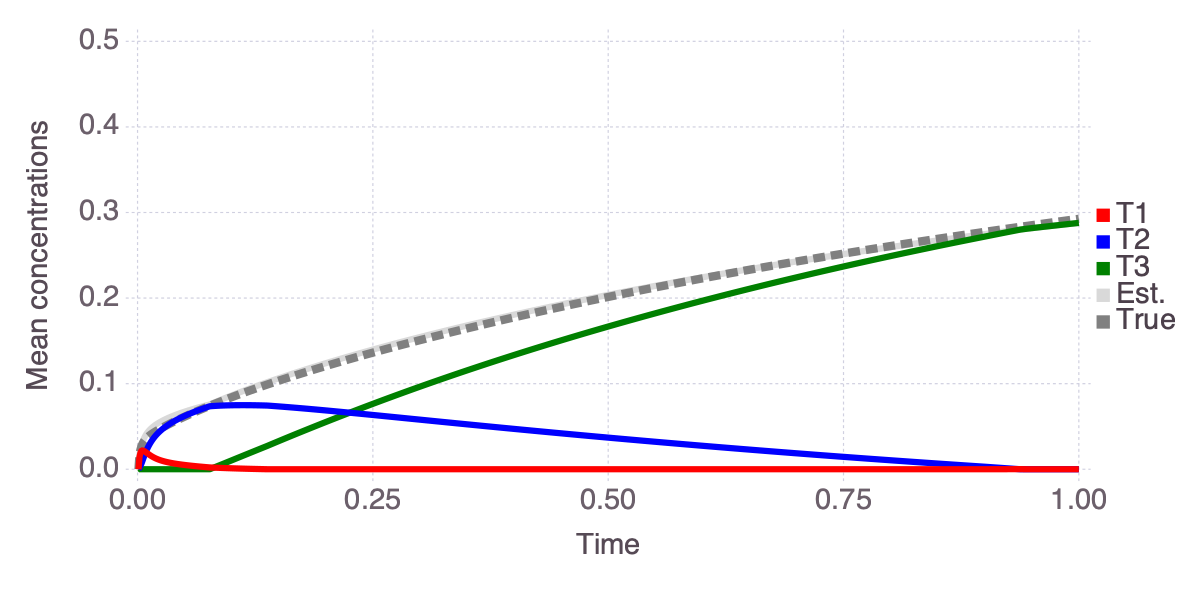}}
	\subfigure[$v_o = 10^{-3}$]
	{\includegraphics[clip=true,width = 0.49\textwidth]{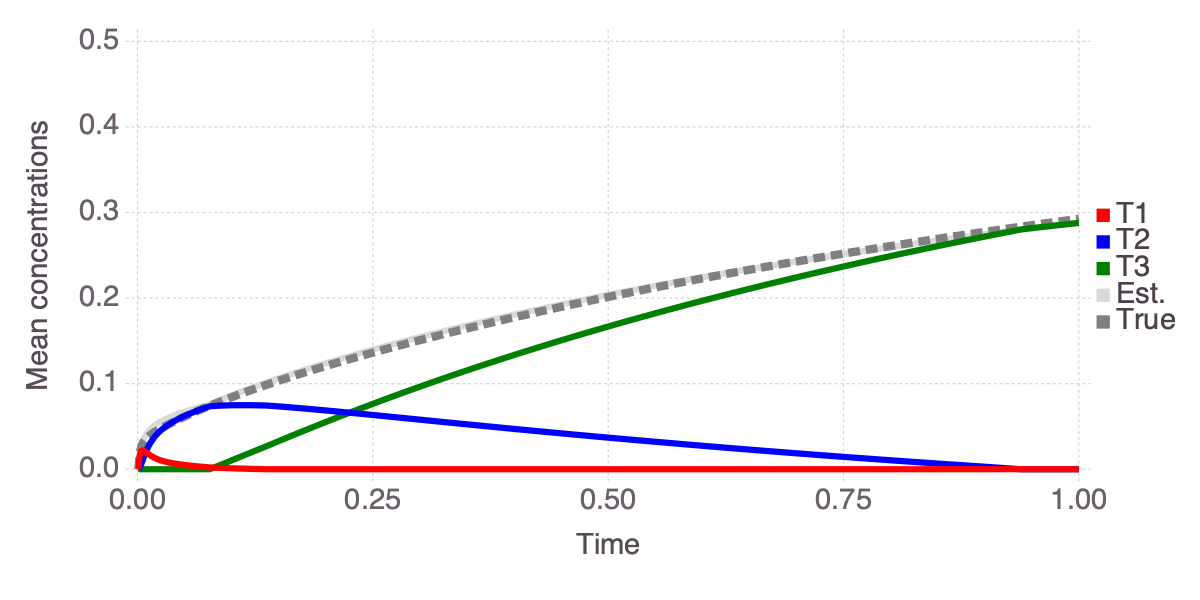}}\\
	\subfigure[$v_o = 10^{-2}$]
	{\includegraphics[clip=true,width = 0.49\textwidth]{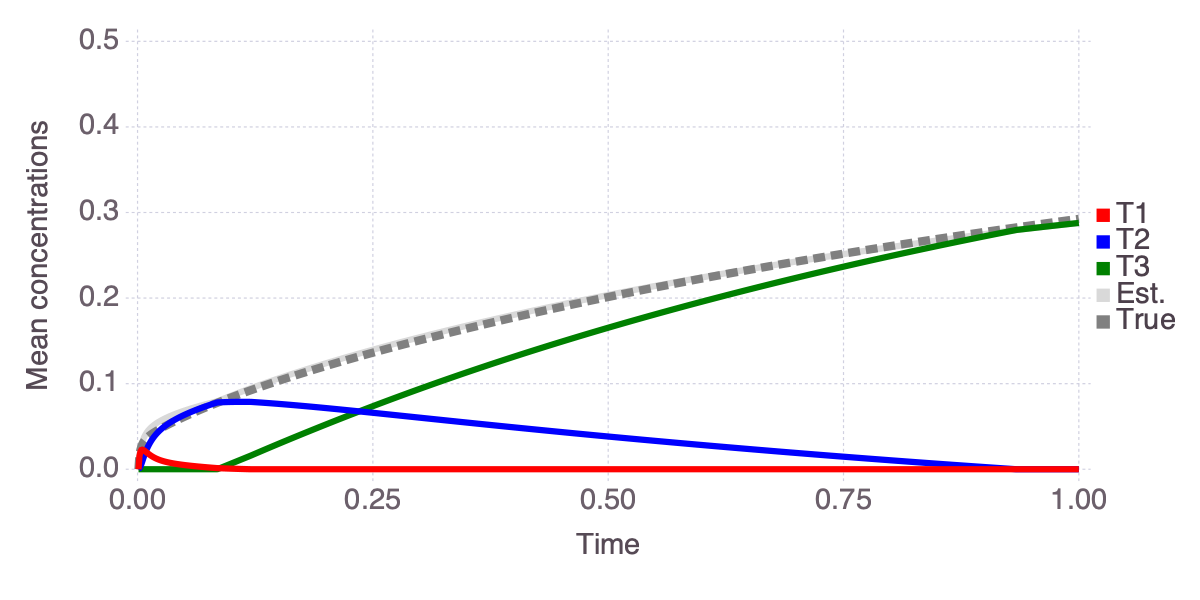}}
	\subfigure[$v_o= 10^{-1}$]
	{\includegraphics[clip=true,width = 0.49\textwidth]{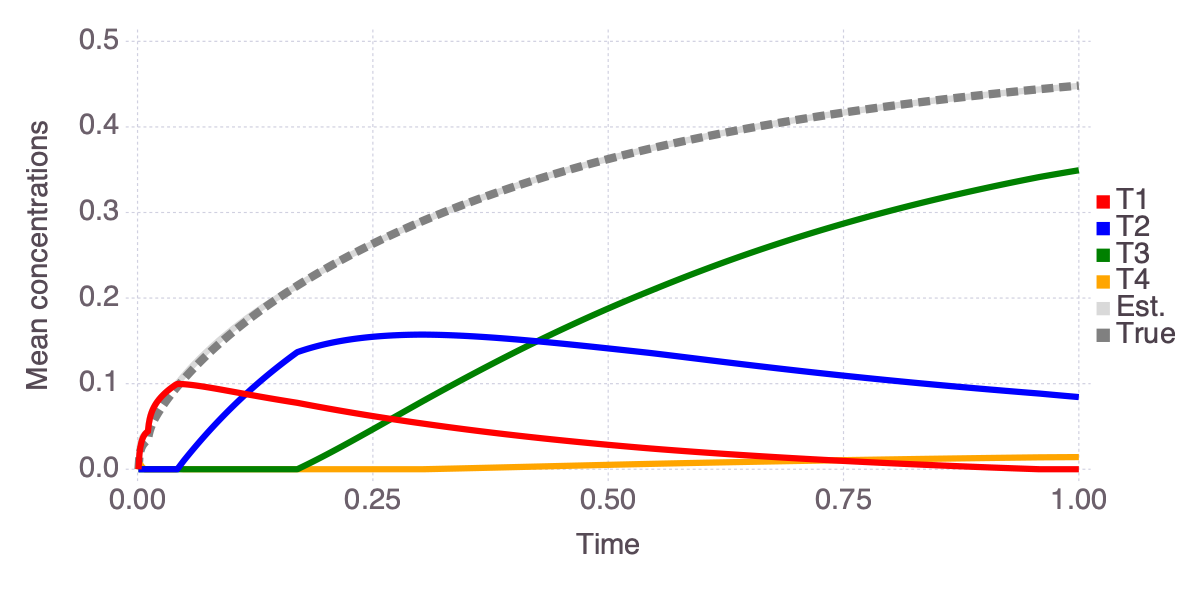}}
	\caption{Transients in the estimated mean concentrations of product $C$ associated with the identified temporal features (T1, T2, T3 and T4) for a series of simulations representing varying $v_0$ and keeping other parameters fixed ($T = 1 \times 10^{-4}$, $\frac{\alpha_L}{\alpha_T} = 10^{4}$, $D_m = 10^{-3}$, $\kappa_fL = 3$). 
	The figures also show the estimated and the true mean concentrations of product $C$ throughout the spatial domain.
	The estimated mean concentrations are a sum of the mean concentrations associated with each temporal feature.
		\label{fig:mean_concentrations_v_o}}
\end{figure}

\begin{figure}
	\centering
	\subfigure[$v_o = 10^{-4}$]
	{\includegraphics[clip=true,width = 0.49\textwidth]{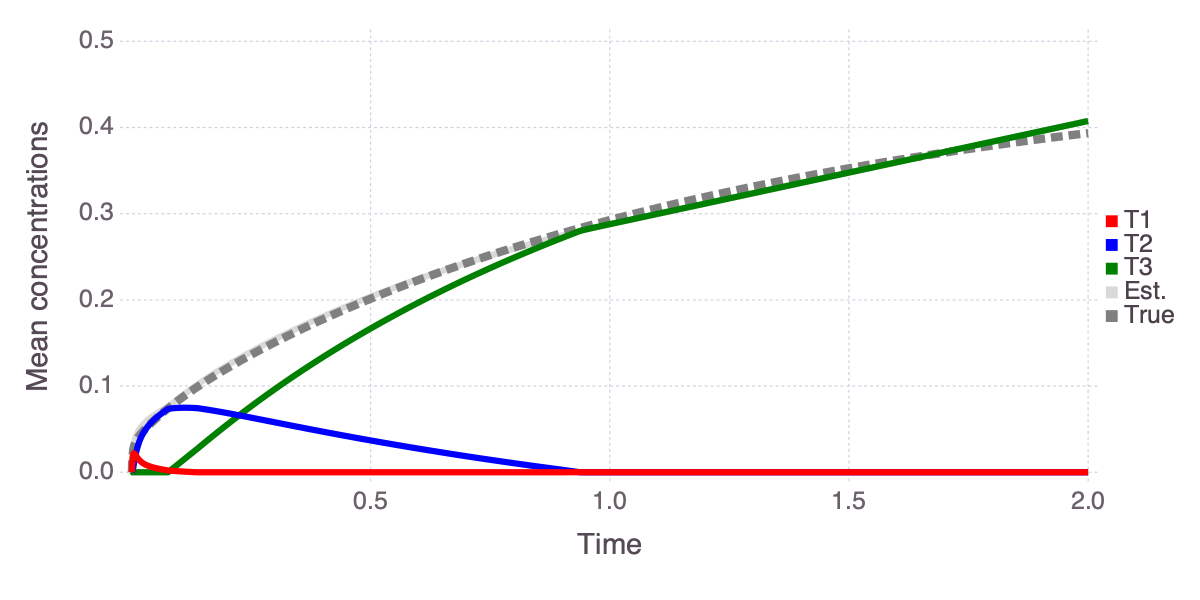}}
	\subfigure[$v_o = 10^{-3}$]
	{\includegraphics[clip=true,width = 0.49\textwidth]{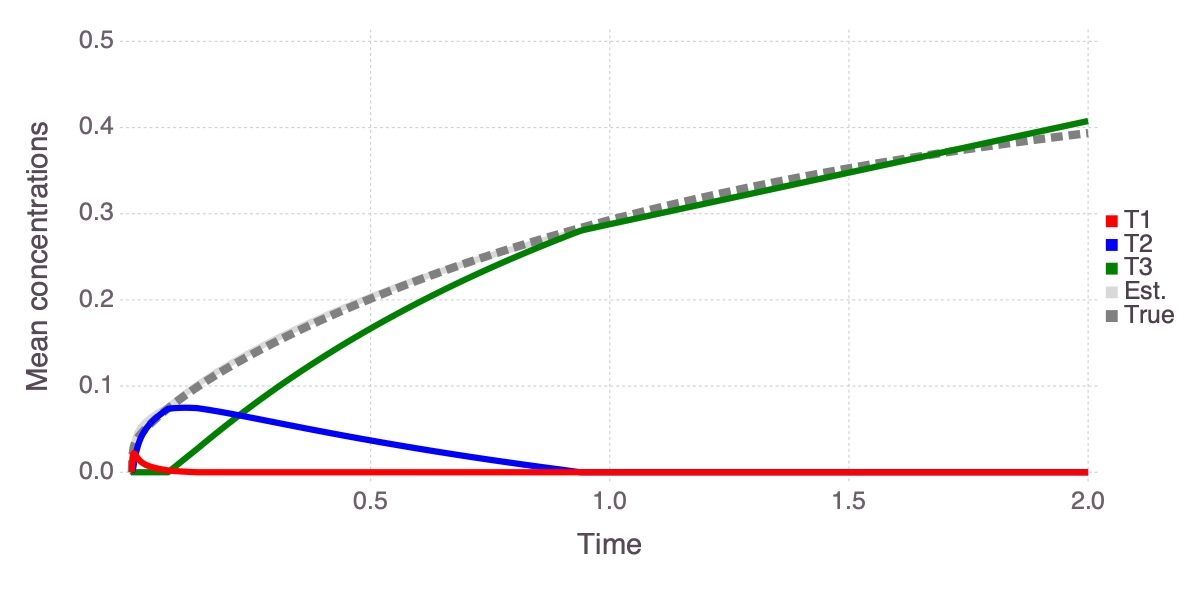}}\\
	\subfigure[$v_o = 10^{-2}$]
	{\includegraphics[clip=true,width = 0.49\textwidth]{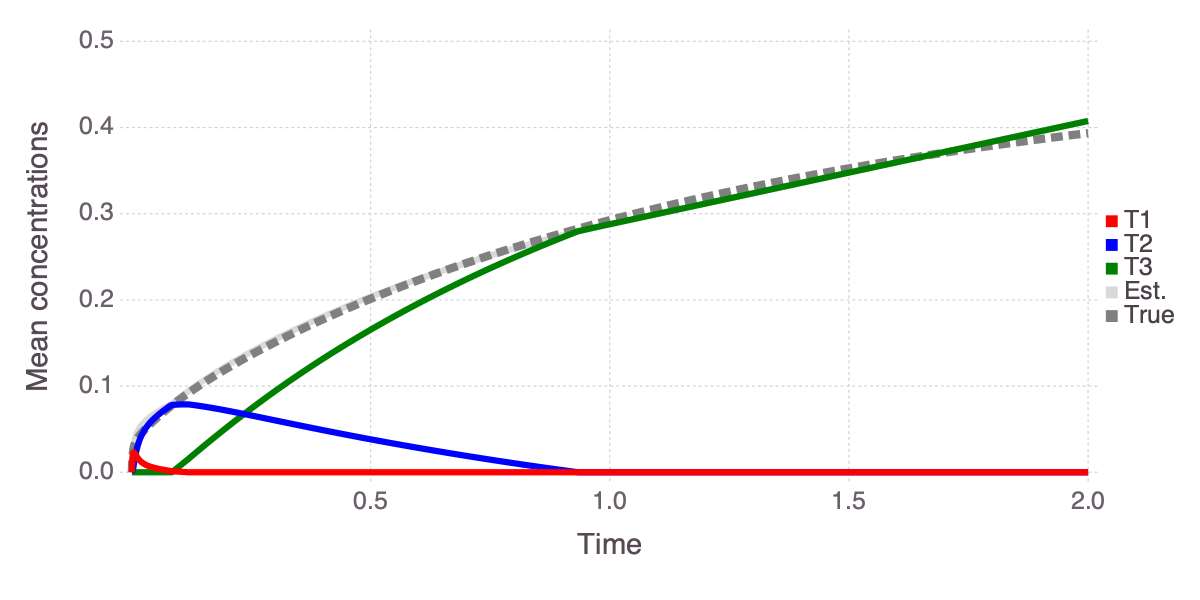}}
	\subfigure[$v_o= 10^{-1}$]
	{\includegraphics[clip=true,width = 0.49\textwidth]{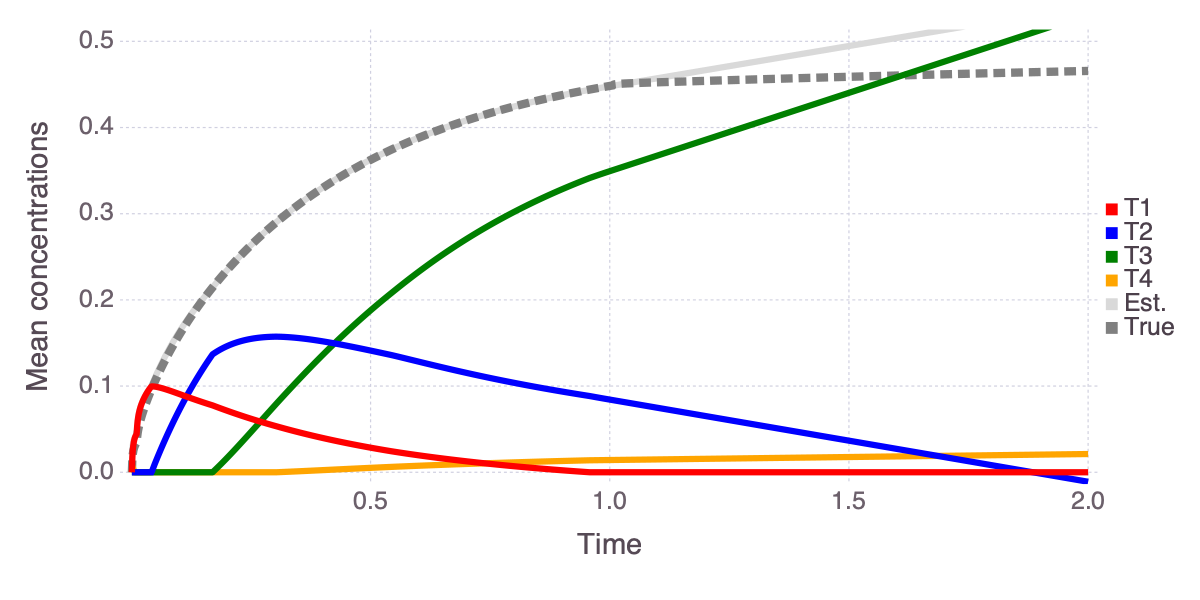}}
	\caption{Comparison of the true and ``blind'' NTF$k$ predictions of the transients in the estimated mean concentrations of product $C$ for times grater than 1.
	The curves in the training period for $t\leq1$ are equivalent to the curves presented in Fig.~\ref{fig:mean_concentrations_v_o}.
	The figures show results for varying $v_0$ and keeping other parameters fixed ($T = 1 \times 10^{-4}$, $\frac{\alpha_L}{\alpha_T} = 10^{4}$, $D_m = 10^{-3}$, $\kappa_fL = 3$).
	The figures also show the estimated and the true mean concentrations of product $C$ throughout the spatial domain.
	The estimated mean concentrations are a sum of the mean concentrations associated with each temporal feature.
	\label{fig:mean_concentrations_v_o_extrapolation}}
\end{figure}

\begin{figure}
	\centering
	{\includegraphics[width = 0.65\textwidth]{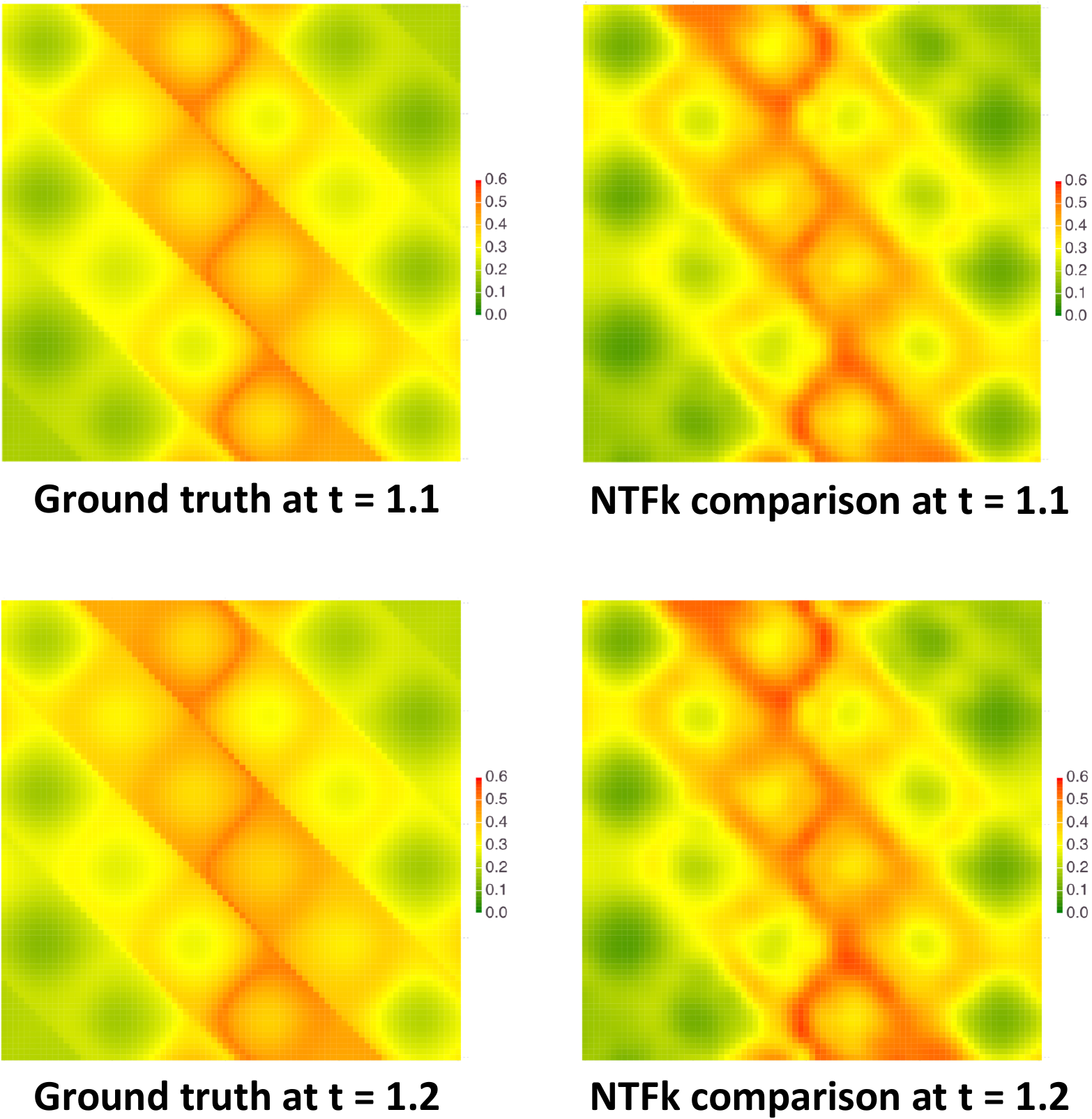}}
	\caption{Comparison of ground truth and NTF$k$ ``blind'' prediction results at times grater than 1 for product $C$.
	The input parameters corresponding to this particular case are equivalent to the ones used in Fig.~\ref{fig:mean_concentrations_v_o} ($v_o = 10^{-3}$, $\frac{\alpha_L}{\alpha_T} = 10^{4}$, $D_m = 10^{-3}$, $\kappa_fL = 3$, and $T = 1 \times 10^{-4}$).
	\label{Fig:Truth_vs_NTFk_Extrapolation}}
\end{figure}

\begin{figure}
	\centering
	\subfigure[$v_o = 10^{-4}$]
	{\includegraphics[clip=true,width = 0.49\textwidth]{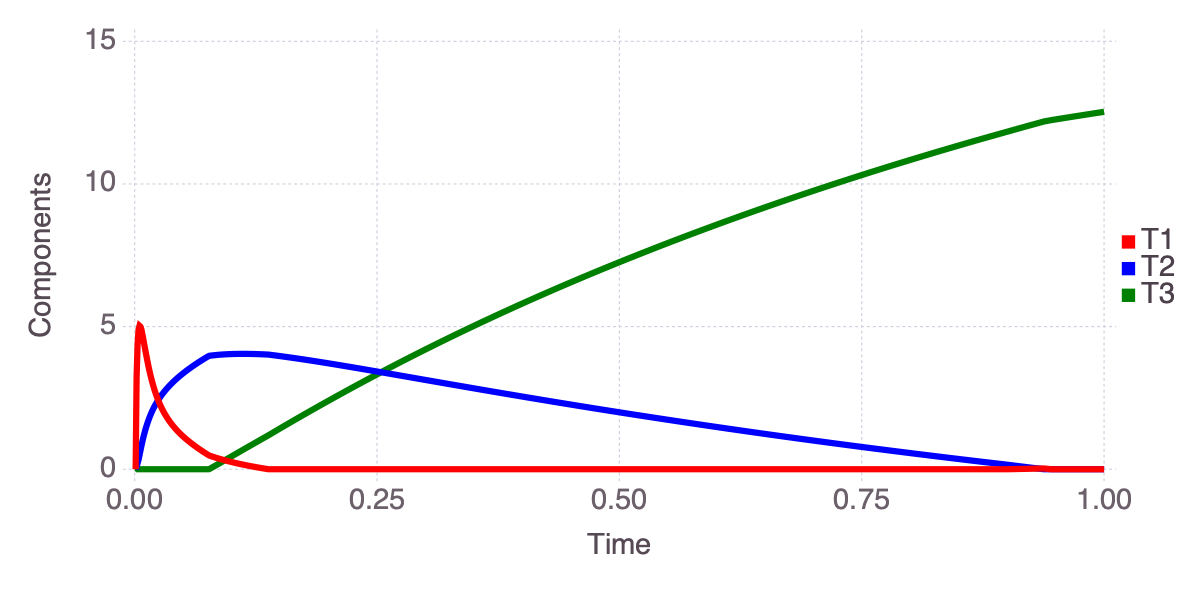}}
	\subfigure[$v_o = 10^{-3}$]
	{\includegraphics[clip=true,width = 0.49\textwidth]{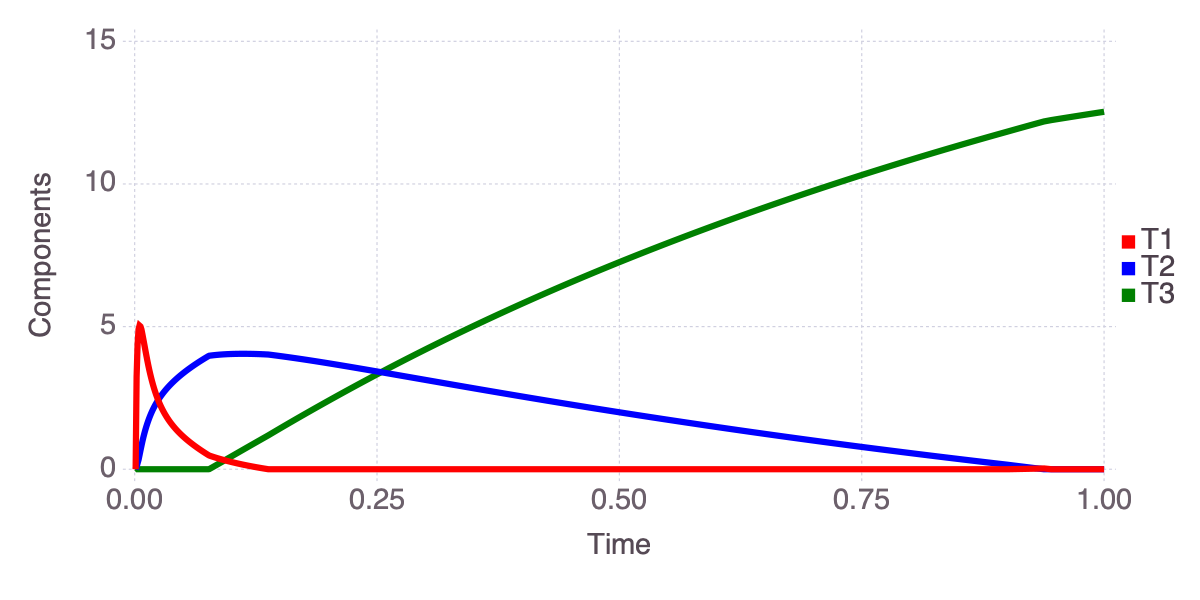}}\\
	\subfigure[$v_o = 10^{-2}$]
	{\includegraphics[clip=true,width = 0.49\textwidth]{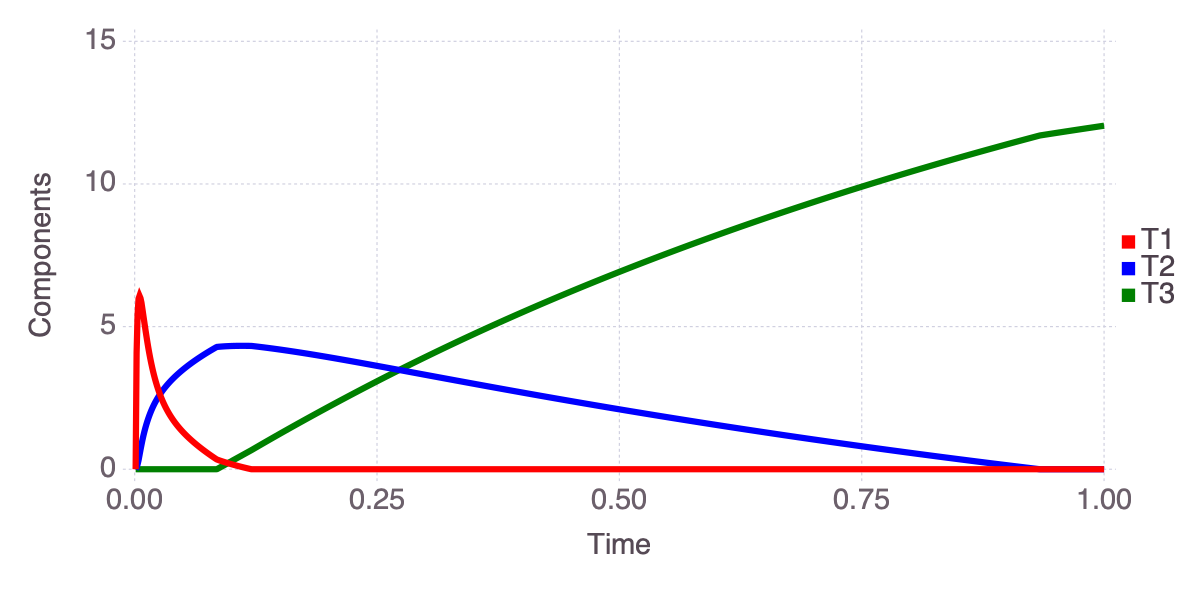}}
	\subfigure[$v_o = 1$]
	{\includegraphics[clip=true,width = 0.49\textwidth]{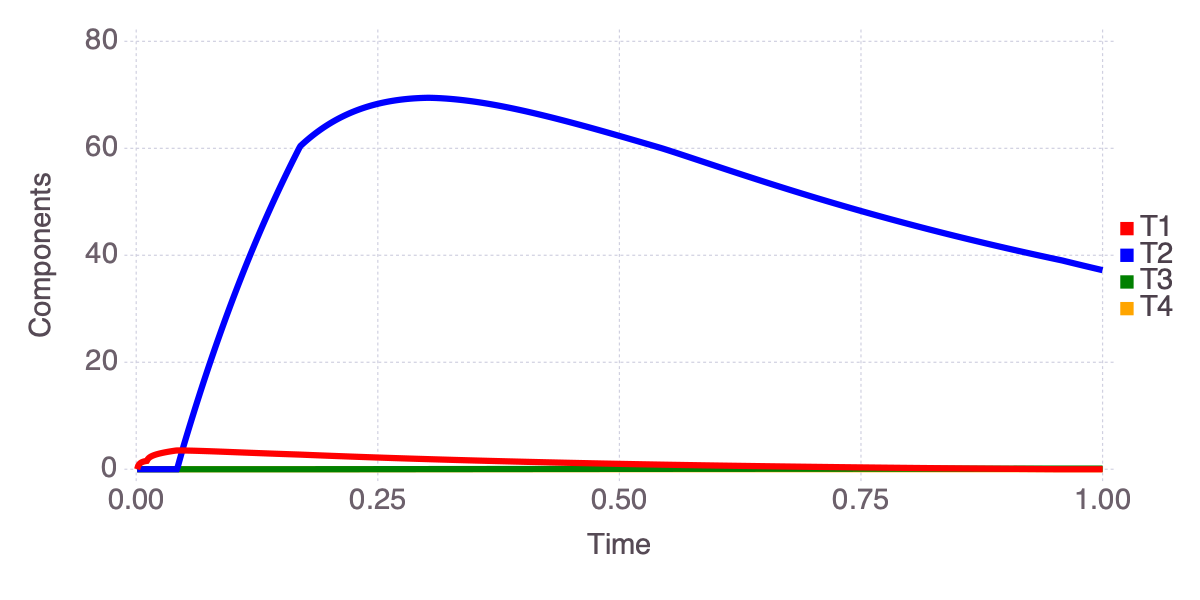}}
	\caption{Components of the temporal matrix $W$ in Eq.\ref{eqn:tucker3}.
	For consistency, the horizontal axis is presented in terms of time where row indices of $W$ are mapped to the respective simulation times. 
	The number of $W$ columns define the number of temporal features.
	The temporal features (T1, T2, T3, and T4) are identified for a series of finite element simulations representing varying $v_0$ and keeping other parameters fixed ($T = 1 \times 10^{-4}$, $\frac{\alpha_L}{\alpha_T} = 10^{4}$, $D_m = 10^{-3}$, $\kappa_fL = 3$). 
		\label{fig:tensor_componenents_v_o}}
\end{figure}

\begin{figure}
	\centering
	\subfigure[$v_o = 10^{-4}$]
	{\includegraphics[clip=true,width = 0.49\textwidth]{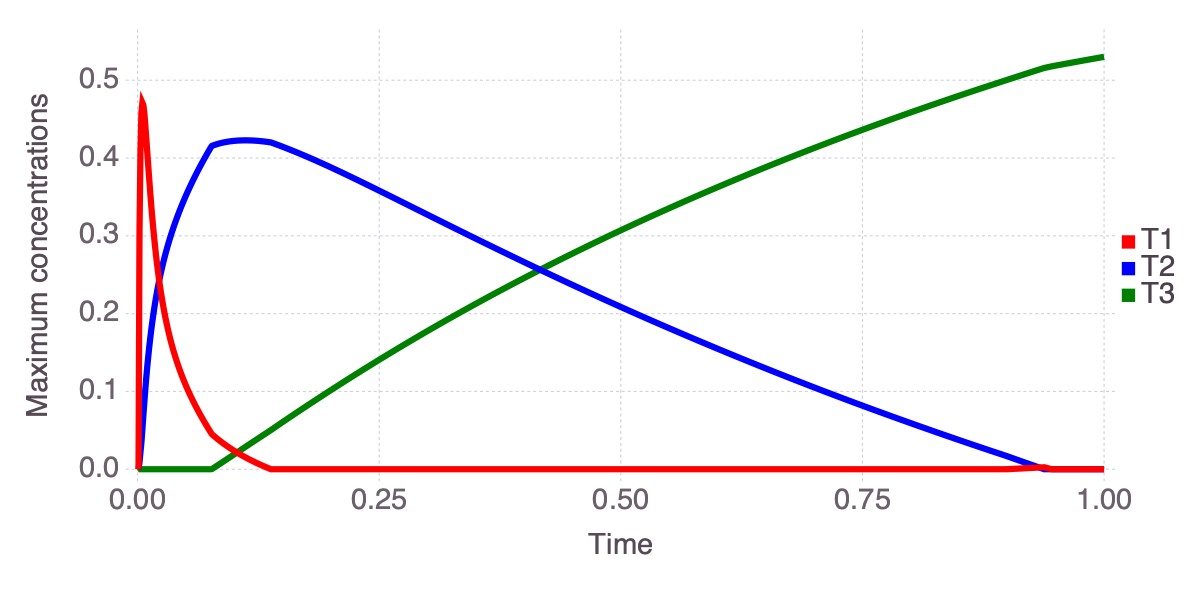}}
	\subfigure[$v_o = 10^{-3}$]
	{\includegraphics[clip=true,width = 0.49\textwidth]{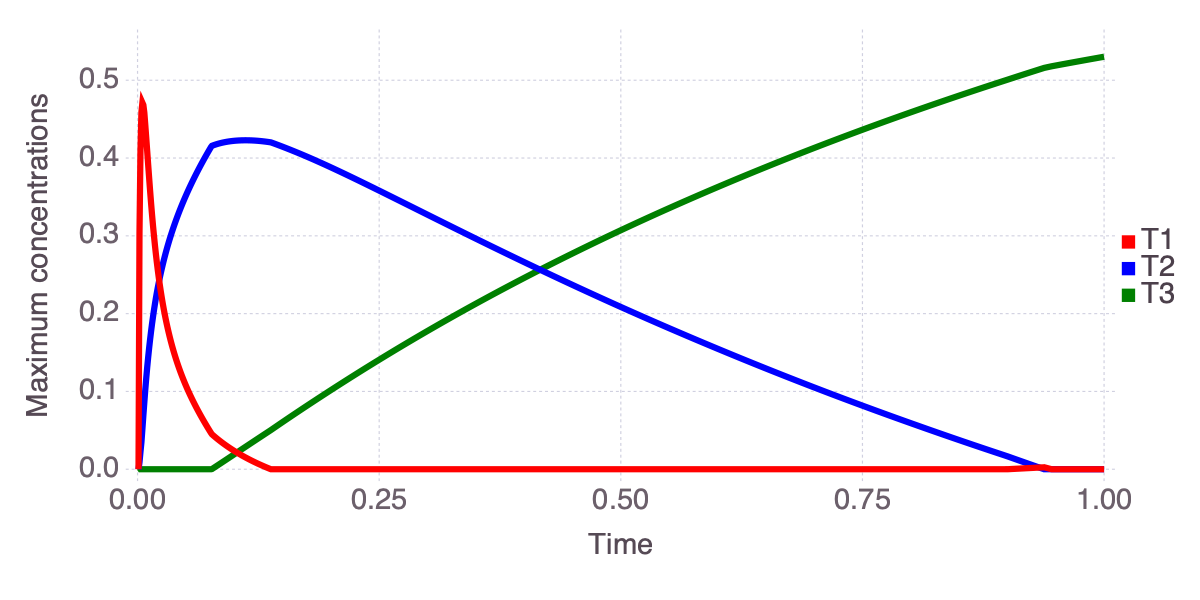}}\\
	\subfigure[$v_o = 10^{-2}$]
	{\includegraphics[clip=true,width = 0.49\textwidth]{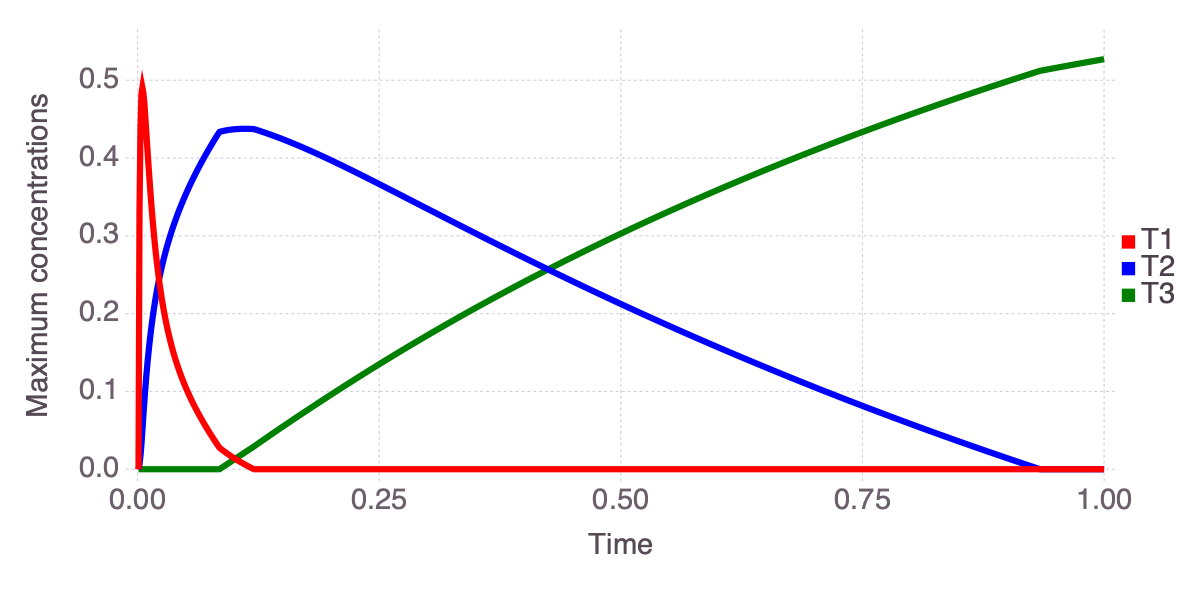}}
	\subfigure[$v_o = 1$]
	{\includegraphics[clip=true,width = 0.49\textwidth]{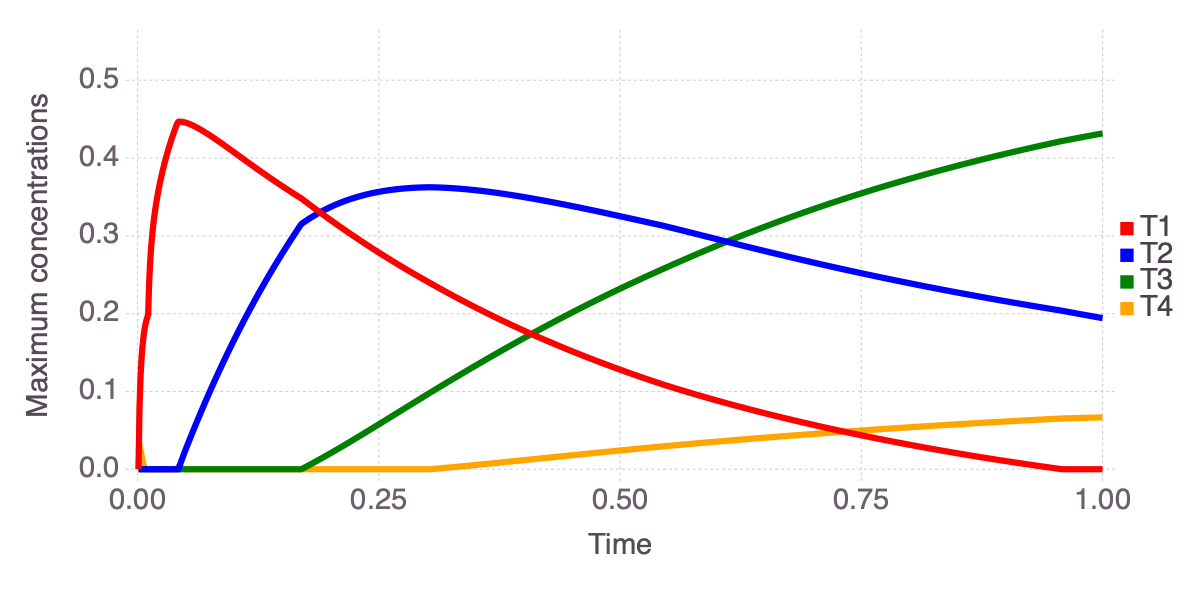}}
	\caption{Transients in the estimated maximum concentrations associated with the identified temporal features (T1, T2, T3, and T4) for a series of simulations representing varying $v_0$ and keeping other parameters fixed ($T = 1 \times 10^{-4}$, $\frac{\alpha_L}{\alpha_T} = 10^{4}$, $D_m = 10^{-3}$, $\kappa_fL = 3$). 
	$T = 1 \times 10^{-4}$ corresponds to fast flipping of vortex structures as given by Eqs.~\eqref{eqn:Vel_x}--\eqref{eqn:Vel_y}.
	$\frac{\alpha_L}{\alpha_T} = 10^{4}$ and $D_m = 10^{-3}$ corresponds to high anisotropic dispersion and low molecular diffusivity.
	$\kappa_fL = 3$ corresponds to medium-scale vortex structures present in the velocity field.
		\label{fig:max_concentrations_v_o}}
\end{figure}

\begin{figure}
	\centering
	\subfigure[$\frac{\alpha_L}{\alpha_T} = 10^{4}$]
	{\includegraphics[clip=true,width = 0.49\textwidth]{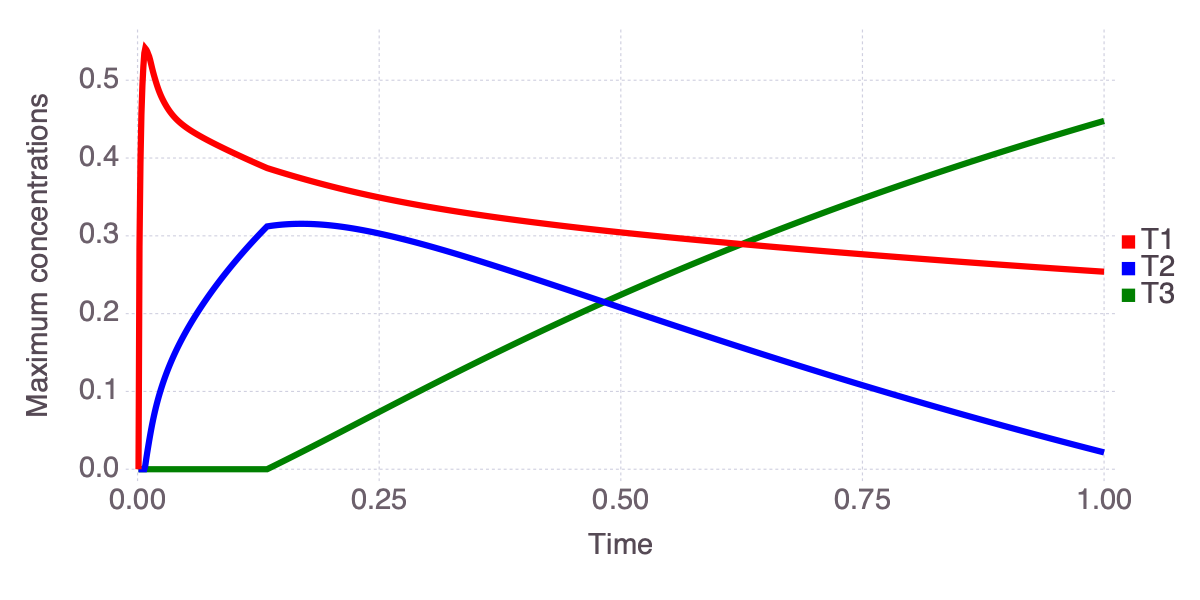}}
	\subfigure[$\frac{\alpha_L}{\alpha_T} = 10^{3}$]
	{\includegraphics[clip=true,width = 0.49\textwidth]{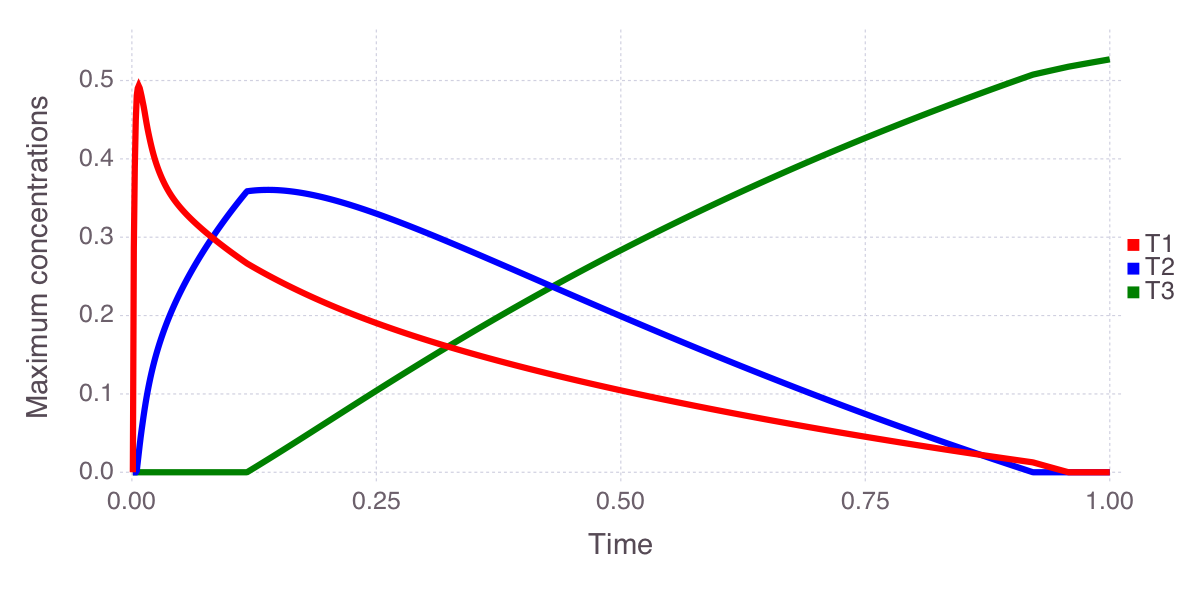}}\\
	\subfigure[$\frac{\alpha_L}{\alpha_T} = 10$]
	{\includegraphics[clip=true,width = 0.49\textwidth]{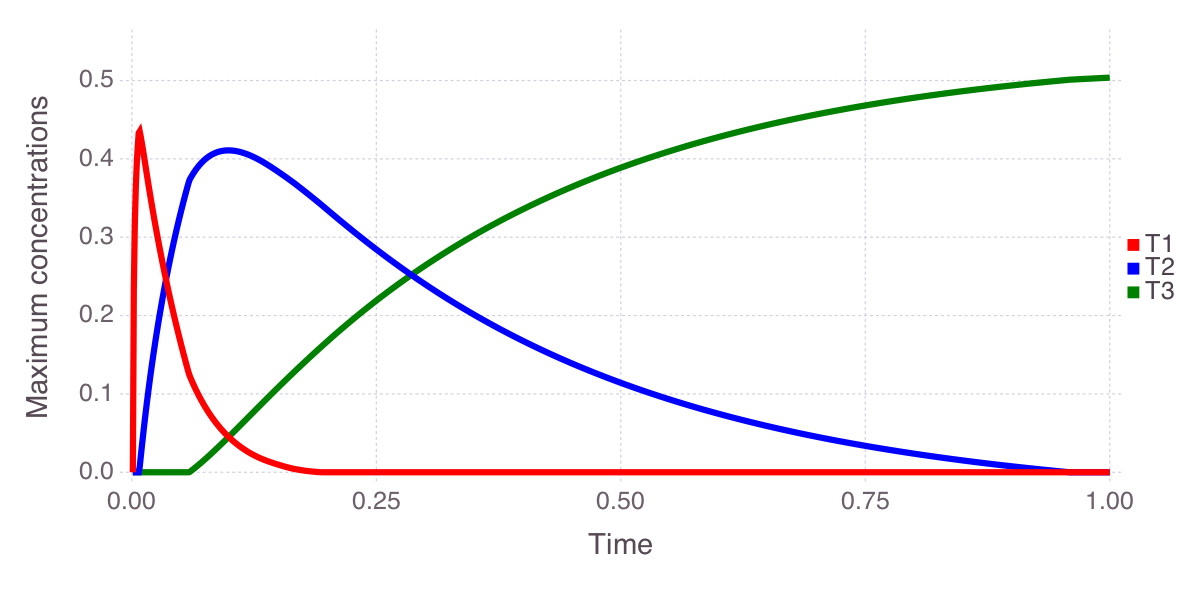}}
	\subfigure[$\frac{\alpha_L}{\alpha_T} = 1$]
	{\includegraphics[clip=true,width = 0.49\textwidth]{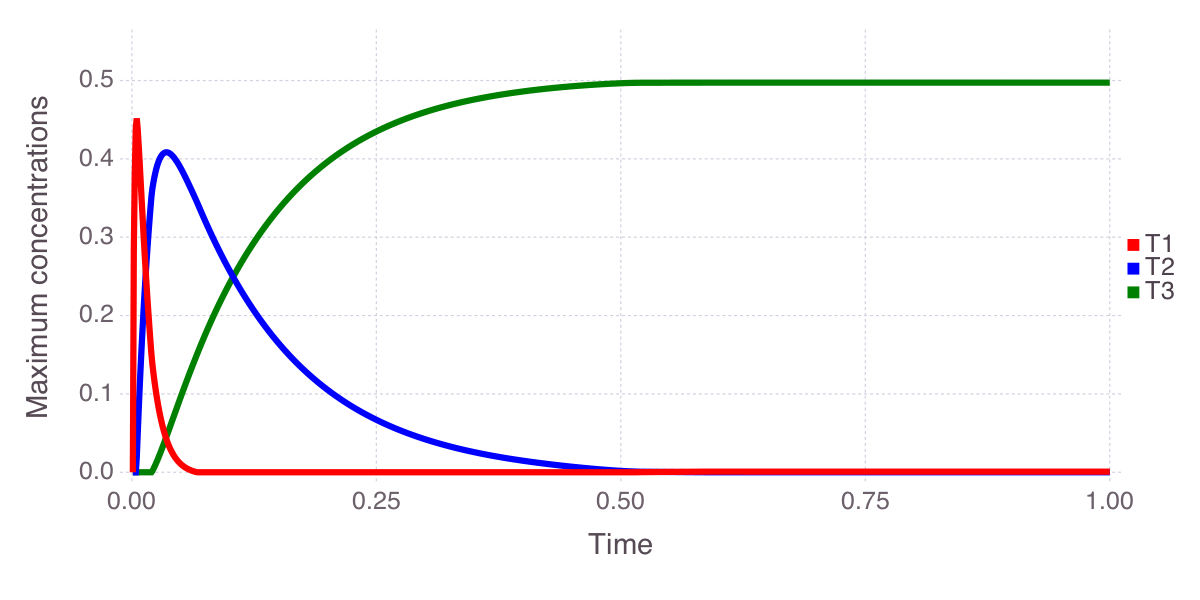}}\\
	\caption{Transients in the estimated maximum concentrations associated with the identified temporal features (T1, T2, and T3) for a series of simulations representing varying anisotropic dispersion contrast $\frac{\alpha_L}{\alpha_T}$ and keeping other parameters fixed ($T = 1 \times 10^{-4}$, $v_o = 10^{-1}$, $D_m = 10^{-3}$, $\kappa_fL = 3$).
$v_o = 10^{-3}$ corresponds to small perturbations in underlying vortex-based velocity field.
$\frac{\alpha_L}{\alpha_T} = 10^{4}$ and $\frac{\alpha_L}{\alpha_T} = 1$ corresponds to high and low anisotropic dispersion.
	\label{fig:max_concentrations_anisotropy}}
\end{figure}

\begin{figure}
	\centering
	\subfigure[$D_m = 10^{-8}$]
	{\includegraphics[clip=true,width = 0.49\textwidth]{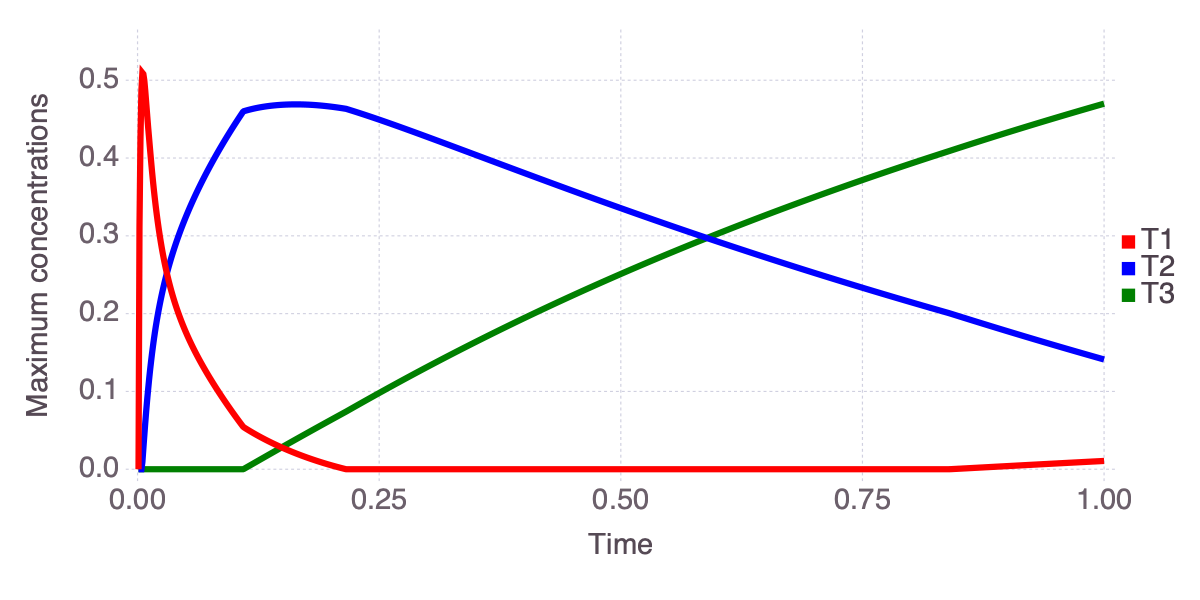}}
    \subfigure[$D_m = 10^{-3}$]
	{\includegraphics[clip=true,width = 0.49\textwidth]{Tensor_Figures/R56C-50000-sa-t2dmax.png}}\\
	\subfigure[$D_m = 10^{-2}$]
	{\includegraphics[clip=true,width = 0.49\textwidth]{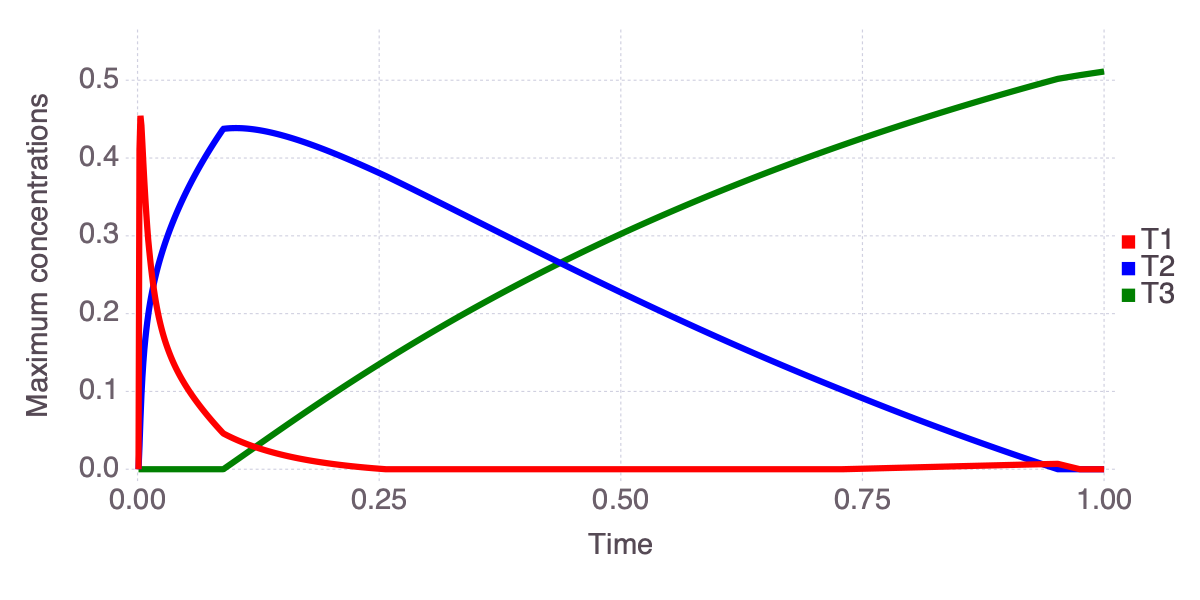}}
	\caption{Transients in the estimated maximum concentrations associated with the identified temporal features (T1, T2, and T3).
Analysis is performed for a series of simulations representing varying molecular diffusivity $D_m$ and keeping other input parameters fixed ($T = 1 \times 10^{-4}$, $\frac{\alpha_L}{\alpha_T} = 10^{4}$, $v_o = 10^{-1}$, $\kappa_fL = 3$).
	\label{fig:max_concentrations_D_m}}
\end{figure}

\begin{figure}
	\centering
	\subfigure[$\kappa_fL = 2$]
	{\includegraphics[clip=true,width = 0.49\textwidth]{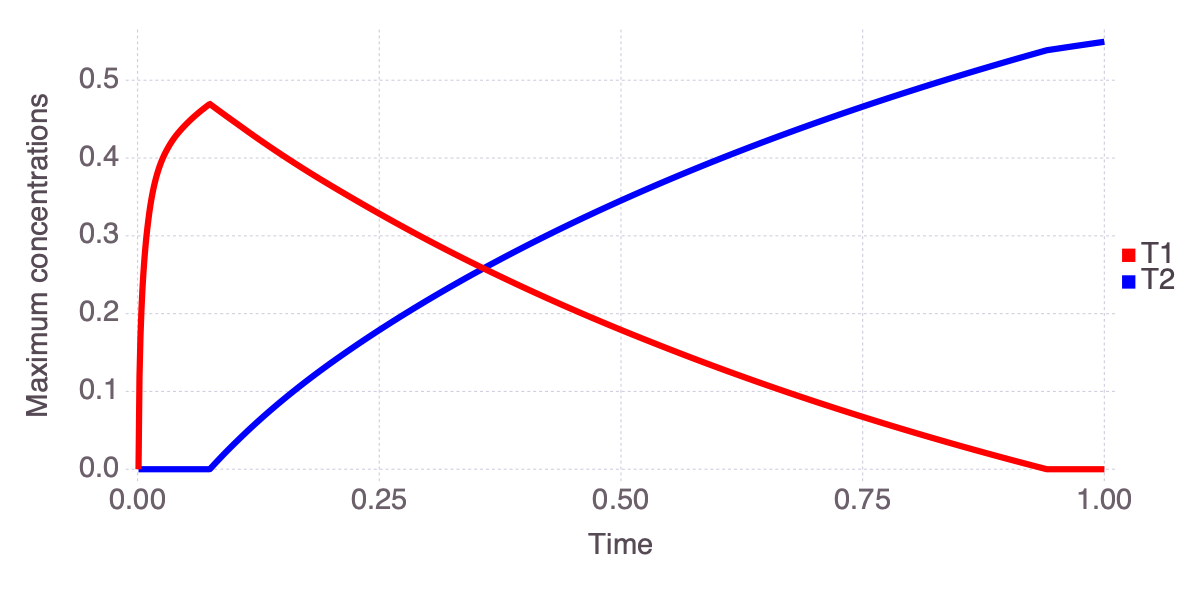}}
	\subfigure[$\kappa_fL = 3$]
	{\includegraphics[clip=true,width = 0.49\textwidth]{Tensor_Figures/R56C-50000-sa-t2dmax.png}}\\
	\subfigure[$\kappa_fL= 5$]
	{\includegraphics[clip=true,width = 0.49\textwidth]{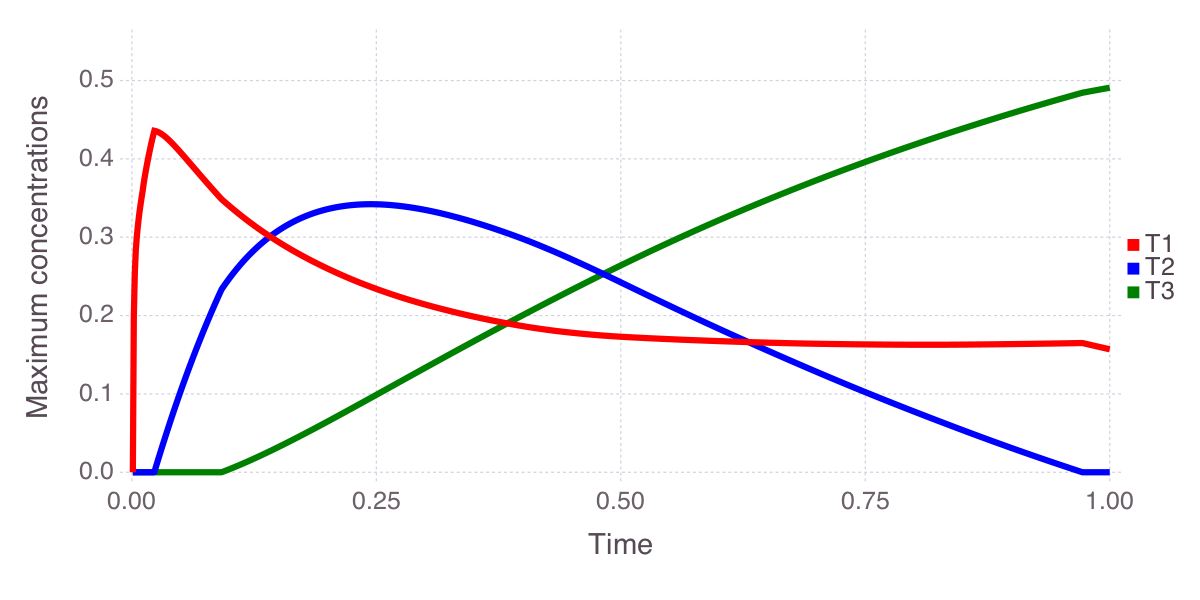}}
	\caption{Transients in the estimated maximum concentrations associated with the identified temporal features (T1, T2, and T3).
Analysis is performed for a series of simulations corresponding to varying spatial scales of underlying vortex-structures $\kappa_fL$ and keeping other parameters fixed ($T = 1 \times 10^{-4}$, $\frac{\alpha_L}{\alpha_T} = 10^{4}$, $v_o = 10^{-1}$, $D_m = 10^{-3}$).
$\kappa_fL = 2$, $\kappa_fL = 3$, and $\kappa_fL = 4$ corresponds to large-scale, medium-scale, and small-scale vortex structures present in the synthetic velocity field.
	\label{fig:max_concentrations_kappa_fL}}
\end{figure}

\begin{figure}
	\centering
	\subfigure[$T = 1 \times 10^{-4}$]
	{\includegraphics[clip=true,width = 0.49\textwidth]{Tensor_Figures/R56C-50000-sa-t2dmax.png}}
	\subfigure[$T = 2 \times 10^{-4}$]
	{\includegraphics[clip=true,width = 0.49\textwidth]{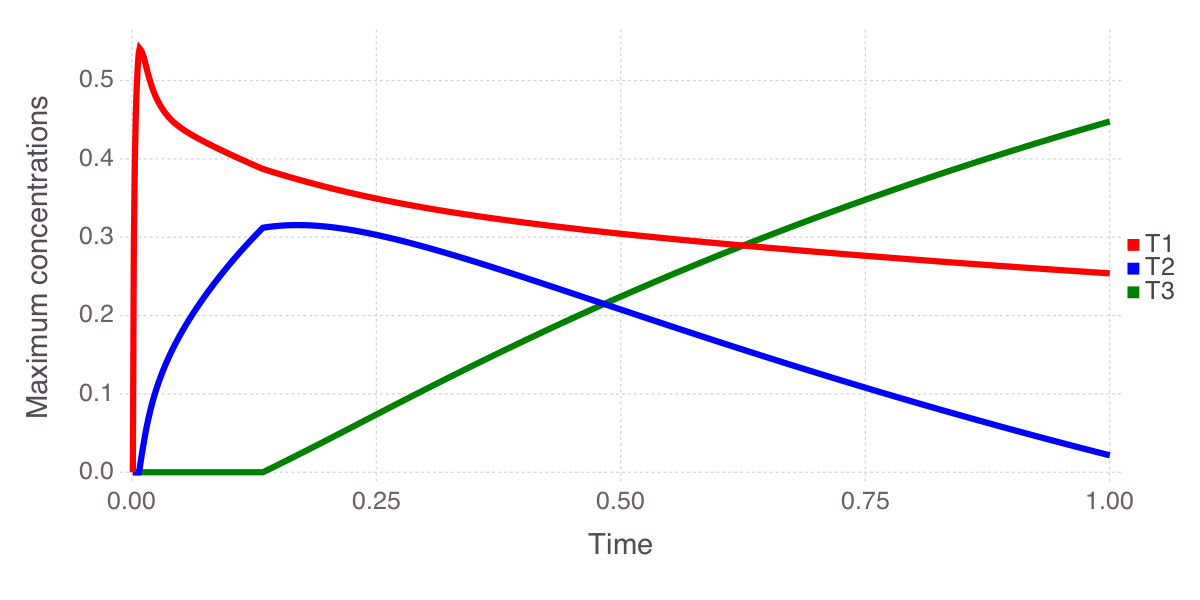}}\\
	\subfigure[$T = 3 \times 10^{-4}$]
	{\includegraphics[clip=true,width = 0.49\textwidth]{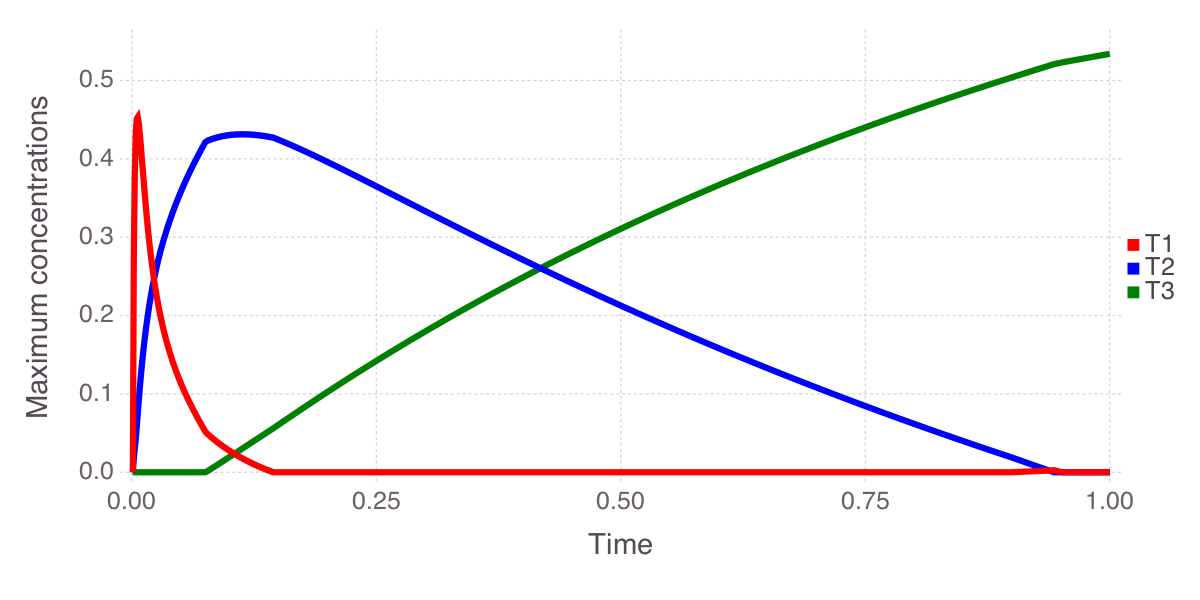}}
	\subfigure[$T = 4 \times 10^{-4}$]
	{\includegraphics[clip=true,width = 0.49\textwidth]{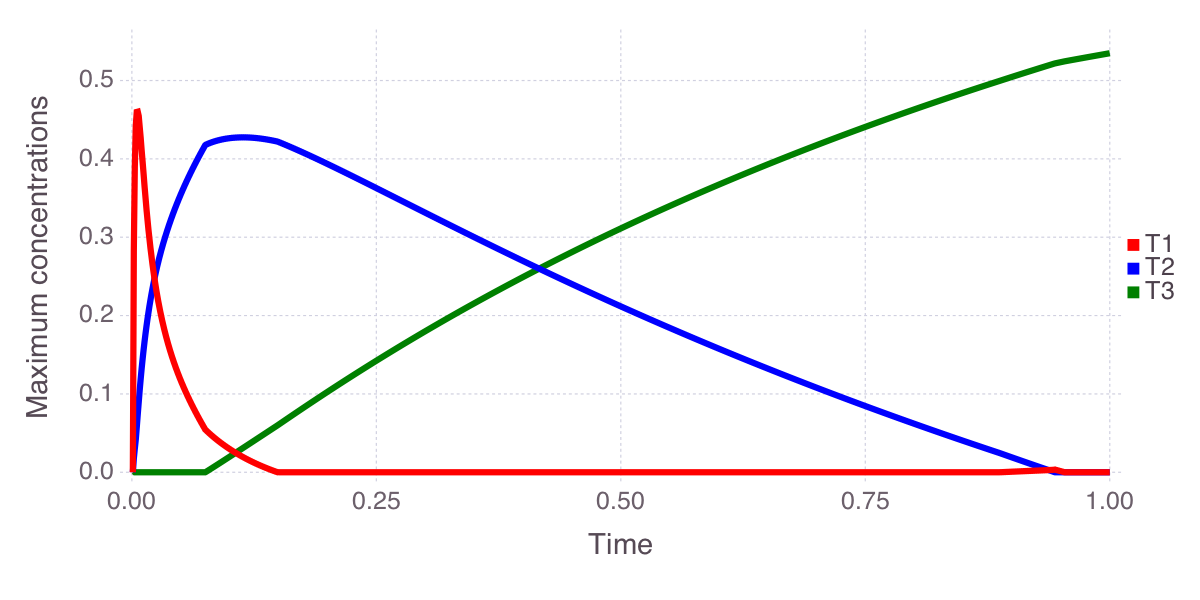}}
	\caption{Transients in the estimated maximum concentrations associated with the identified temporal features for a series of simulations representing varying $T$ and keeping other parameters fixed ($\frac{\alpha_L}{\alpha_T} = 10^{4}$, $v_o = 10^{-1}$, $D_m = 10^{-3}$, $\kappa_fL = 3$).
$T = 1 \times 10^{-4}$ and $T = 4 \times 10^{-4}$ corresponds to fast flipping and slow flipping of vortex structures from clockwise direction to anti-clockwise direction.
	\label{fig:max_concentrations_T}}
\end{figure}

Example results obtained from the NTF$k$ methodology are presented in Figs.~\ref{Fig:Truth_vs_NTFk}--\ref{fig:max_concentrations_T}.
Fig.~\ref{Fig:Truth_vs_NTFk} provides a comparison of the ground truth concentrations with that estimated by the NTF$k$ methodology. 
Ground truth is a high-fidelity numerical simulation with input parameters as:~$v_o = 10^{-3}$, $\frac{\alpha_L}{\alpha_T} = 10^{4}$, $D_m = 10^{-3}$, $\kappa_fL = 3$, and $T = 1 \times 10^{-4}$.
It is clear that the solution obtained from the proposed method compares well with that of the high-fidelity finite element numerical simulation.
The corresponding numerical error ($L_2$-norm) at times $t = 0.1$, $t = 0.5$, and $t = 1.0$ are equal to 1.954, 1.285, and 1.587.
Fig.~\ref{Fig:Truth_vs_NTFk} also shows a linear decomposition of concentration of product $C$ at various stages in time. 
One can see that at earlier stages in time ($t = 0.1$), streamline mixing and mixing due to anisotropic dispersion are dominant. 
Mixing due to molecular diffusion is close to zero, which is expected.
At later stages in time ($t = 0.5$ and $t = 1.0$), one can see that concentration of product $C$ due to streamline mixing and anisotropic dispersion gradually decrease to zero and mixing due to molecular diffusion dominates.
Note that for high anisotropy, longitudinal dispersion enhances reactant mixing along the streamlines during initial times. 
At early times, Fig.~\ref{Fig:Truth_vs_NTFk} shows that that sharp interfaces are formed near the boundary of species.
At t = 0.5, one can see that the spread of the product in the domain due to transverse dispersion.
Transverse dispersion and molecular diffusion enhances reactive-mixing across the streamlines resulting in less sharp interfaces.
As a result, at final times (t = 1.0) we can see that sharp interfaces are being smeared even under high anisotropy because of molecular diffusion.
Based on these observations, we conclude that NTF$k$ features T1, T2 and T3 correspond to the spatial distribution of product $C$ due to longitudinal dispersion, transverse dispersion.and molecular diffusion, respectively.

Fig.~\ref{fig:mean_concentrations_v_o} shows another validation of the proposed NTF$k$.
It compares the true mean concentration of product $C$ obtained from the finite element simulation to that of the estimated mean concentration based on NTF$k$.
From Fig.~\ref{fig:mean_concentrations_v_o}, it is clear that estimated mean agrees well with true mean concentration.
The figure also shows the mean concentrations associated with the NTF$k$ estimated temporal components in the dataset.
The sum of the mean temporal-components concentrations over time reproduces the total NTF$k$ estimated mean concentrations (the brown curves in Fig.~\ref{fig:mean_concentrations_v_o}).

Fig.~\ref{fig:mean_concentrations_v_o_extrapolation} shows ``blind'' prediction of the proposed NTF$k$.
The true mean concentrations for $t > 1$ are calculated using the high-resolution finite element simulations.
``Blind'' predictions of the transients for $t>1$ are obtained through extrapolation of the temporal features identified by NTF$k$ based on analyses of the simulation results for $t\leq1$.
The temporal extrapolation of the features is done using linear basis splines \cite{deBoor1978splines, JuliaMathBsplines}.
From Fig.~\ref{fig:mean_concentrations_v_o_extrapolation}, it is clear that for low values of $v_o$ the extrapolated results agree well with true mean concentrations.
However, for $v_o = 1$ the NTF$k$ ``blind'' predictions are higher than the true mean concentration, which shows the limitations our ML analyses and the applied linear spline extrapolation.
The predictability for the case of $v_o = 1$ can be improved if either higher order splines with different boundary conditions (e.g. ``flat'' boundary condition \cite{deBoor1978splines, JuliaMathBsplines} representing the plateauing of the mean concentration) or the ML analyses were  performed using longer training period (e.g. $t\leq1.5$) which also includes the plateauing  of the concentration curve.
Fig.~\ref{Fig:Truth_vs_NTFk_Extrapolation} presents ``blind'' NTF$k$ predictions for $t > 1$ spatially for the case of $v_o = 10^{-3}$.
This figure is an extension of the results already shown Fig.~\ref{Fig:Truth_vs_NTFk}. 
The ground truth concentrations are obtained by performing high-resolution non-negative finite element simulations for $t > 1$.
From Fig.~\ref{Fig:Truth_vs_NTFk_Extrapolation}, it is clear that NTF$k$ extrapolation results compare well with the ground truth concentrations of product $C$.
The numerical errors ($L_2$-norm) for the ``blind'' predictions at times $t = 1.1$, and $t = 1.2$ are equal to 2.096, and 2.658.
The errors are higher but comparable with the errors for the training period ($t\leq 1$; see Fig.~\ref{Fig:Truth_vs_NTFk}).

As discussed in Section \ref{sec:methodology}, the NTF$k$ predicted concentration are obtained based on products of the core tensor and the matrix factors (Eq.\ref{eqn:tucker3}; Fig.\ref{Fig:Tucker_Models}).
The temporal matrix components of $W$ are visualized in Fig.~\ref{fig:tensor_componenents_v_o}.
The $W$  components drive the transients in the NTF$k$ predicted concentrations.
As expected, the overall shapes of the curves in  Fig.~\ref{fig:tensor_componenents_v_o} are similar to those in Fig.~\ref{fig:mean_concentrations_v_o}.

Figs.~\ref{fig:max_concentrations_v_o}--\ref{fig:max_concentrations_T} show the maximum concentration of product $C$ predicted by each temporal feature identified by our methodology for each time snapshot (there are 1000 snapshots for times between 0.001 to 1.0).
It is important to note that the ML-predicted maximum concentration in some cases exceeds $0.5$ (the true maximum concentration of product $C$ in the input data; for example, see Fig.~\ref{fig:max_concentrations_v_o}a) due to estimation errors.
This estimation error can be reduced by imposing another constraint that the maximum concentration of product $C$ cannot exceed $0.5$. 
In most cases, the proposed NTF$k$ methodology identified three temporal features. 
These are labeled as T1, T2, and T3 in Figs.~\ref{fig:max_concentrations_v_o}--\ref{fig:max_concentrations_T}.
However, in one case, there are only two features (see Fig.~\ref{fig:max_concentrations_kappa_fL}a) while in another there are four features (\ref{fig:max_concentrations_v_o}d).
The first temporal feature T1 is dominated by species mixing due to longitudinal dispersion.
The process of reactive-mixing due to longitudinal dispersion is much faster than mixing of reactants due to transverse dispersion and molecular diffusion.
As a result, we see a sharp increase in the concentration of product $C$ at early times (for example, see Fig.~\ref{fig:max_concentrations_v_o}a).
The second temporal feature T2 corresponds to species mixing due to transverse dispersion.
Finally, the third temporal feature T3 represents species mixing due to molecular diffusion.
The shape of the T3 curve is close to linear when plotted as a square root of time which is consistent with pure isotropic molecular diffusion process.
The above inferred observations are valid for most of the results discussed below.

Fig.~\ref{fig:max_concentrations_v_o} demonstrates how the perturbations in vortex structures (as dictated by $v_0$ values) impacts the identified temporal features.
Figs.~\ref{fig:max_concentrations_v_o}a--\ref{fig:max_concentrations_v_o}d depict changes in the maximum concentration in product $C$ due to varying $v_0$ by keeping other parameters fixed, which are $T = 1 \times 10^{-4}$, $\frac{\alpha_L}{\alpha_T} = 10^{4}$, $D_m = 10^{-3}$, and $\kappa_fL = 3$. 
$\frac{\alpha_L}{\alpha_T} = 10^{4}$ corresponds to high anisotropic dispersion, $D_m = 10^{-3}$ corresponds to low molecular diffusivity, and $\kappa_fL = 3$ corresponds to medium-scale vortex structures present in the velocity field.
As evident from Eqs.~\eqref{eqn:Vel_x} and~\eqref{eqn:Vel_y}, $v_o = 10^{-4}$ corresponds to low deviation and $v_o = 1$ corresponds to high deviation from the symmetric vortex structure of the velocity field. 
For low values of $v_0$ ($10^{-4}$, $10^{-3}$, and $10^{-2}$), one can observe that the maximum concentration of temporal feature T1 rises and decays sharply. 
However, for high values of $v_0$ (for example, $v_o = 1$), maximum concentration of temporal feature T1 decays slowly. 
This is because temporal feature T1 corresponds to formation of product $C$ along the vortex streamlines (due to longitudinal anisotropic dispersion coefficient). 
The non-symmetric vortex structures (caused due to higher values of $v_0$) enhance mixing of reactants resulting in formation of product $C$ even at longer times. 
This is not the case for lower values of $v_0$ because of the symmetric vortex structures present in the velocity field.
For low values of $v_0$ and based on temporal feature T2, one can observe that mixing across the streamlines starts to decay once the maximum concentration of the first temporal feature T1 reaches zero. 
This is due to the fact that the anisotropic contrast is very high, which hinders reactants $A$ and $B$ from dispersing across the streamlines. 
As a result, for temporal feature T2, we see a decrease in the maximum concentration of product $C$ as time progresses.
For higher values of $v_0$, we also observe decay in the maximum concentration of temporal feature T2.
But at larger times (in feature T2), the maximum concentration of product $C$ is still non-zero, which is not the case for lower values of $v_0$.
This is because the non-symmetric nature of the vortex-based velocity field brings the unused reactants near to each other when compared to the symmetric case.
As a result, for longer times (even under high anisotropy) one can see formation of product $C$ for higher values of $v_0$.
The comparison of Figs.~\ref{fig:max_concentrations_v_o}a--\ref{fig:max_concentrations_v_o}c demonstrates that similar temporal features are identified regardless of the differences in $v_0$.
This observation fails only at high values of $v_0$ as shown in Fig.~\ref{fig:max_concentrations_v_o}d.
In this case, the proposed NTF$k$ methodology identifies a fourth temporal feature T4.
This fourth temporal feature (T4) is caused by boundary effects, which are brought on by high values of $v_0$.

Qualitatively, the trends shown in Fig.~\ref{fig:max_concentrations_v_o} and Fig.~\ref{fig:mean_concentrations_v_o} are very similar.
Interesting inference can be drawn from the T1 feature in both the figures.
Even though, the estimated maximum concentration for T1 is high (see Fig.~\ref{fig:max_concentrations_v_o}), the amount of product formed in the domain is low.
For example, see the spatial spread of product concentration in Fig.~\ref{Fig:Truth_vs_NTFk}.
As a result, we can see that mean concentration value corresponding to T1 is low and decreases over time.
Similar inference can be drawn on feature T2.
For temporal features T3, one can see that as the time progresses the maximum and mean concentration have similar trends.

Fig.~\ref{fig:max_concentrations_anisotropy} shows the impact of transverse anisotropic dispersion on the identified temporal features.
On comparison with the results for $v_0$ (see Fig.~\ref{fig:max_concentrations_v_o}), the identified temporal features obtained for different values of $\frac{\alpha_L}{\alpha_T}$ are substantially different.
With the decrease in anisotropic dispersion ratio, it is clear that both the first and second temporal feature (T1 \& T2) become less important at later times, and the third temporal feature (T3) becomes more dominant.
This observation is in accordance with the anisotropic reaction-diffusion process as high anisotropic contrast hinders the transverse dispersion resulting in incomplete mixing of reactants.
At later times, molecular diffusion becomes the major mechanism to enhance mixing between reactants $A$ and $B$ resulting in higher yield of product $C$. 

Fig.~\ref{fig:max_concentrations_D_m} shows the effects of the molecular diffusivity $D_m$ on the identified temporal features.
From this figure, it is clear that the increase in the values of $D_m$ shows an increase in the maximum concentration of T3 at later times.
This is expected because the temporal feature T3 is dominated by the molecular diffusion. 

Fig.~\ref{fig:max_concentrations_kappa_fL} presents the importance of the spatial characteristic scale $\kappa_fL$ on the identified temporal features.
This parameter provides us information on small-scale and large-scale nature of vortex structures present in the velocity field. 
For small values of $\kappa_fL$, we have large-scale vortex structures and for bigger values of $\kappa_fL$, we have small-scale vortex structures.
As $\kappa_fL$ increases, the number of vortices present in the velocity field increases resulting in enhance mixing between reactant $A$ and $B$.
Moreover, it also affects the number of identified temporal features.
In the case for $\kappa_fL = 2$, the NTF$k$ methodology identifies only two temporal features (see Fig.~\ref{fig:max_concentrations_kappa_fL}a).
The proposed method, could not separate the reactive-mixing occurring due to vortex structures and anisotropic dispersion.
For this value of $\kappa_fL$, product formation due to mixing by longitudinal dispersion and transverse dispersion are lumped together as a single feature T1.
However, for values of $\kappa_fL > 2$  (Figs.~\ref{fig:max_concentrations_kappa_fL}b and  \ref{fig:max_concentrations_kappa_fL}c), both of these aspects are distinguished.
It is important to note that as $\kappa_fL$ increases, the first temporal features T1 remains dominant throughout the entire simulation time.
Moreover, T1 does not diminish as compared to other cases presented here (for example, see Figs.~\ref{fig:max_concentrations_D_m}, \ref{fig:max_concentrations_v_o}, and \ref{fig:max_concentrations_anisotropy}) even when the anisotropic contrast is very high.
This is because of the high number of vortex structures present in the velocity field, which increases the reactive-mixing even at the later times. 

Fig.~\ref{fig:max_concentrations_T} shows the impact of flipping the vortex-based velocity field from clockwise direction to anti-clockwise direction (see Eq.~\ref{eqn:Div_Free_Stream_Function}) on the trend observed in formation of product $C$. 
Low values of $T$ correspond to fast flipping of the velocity field from clockwise direction to anti-clockwise direction while high values of $T$ correspond to slow flipping.
For low values of $T$ ($1 \times 10^{-4}$ and $2 \times 10^{-4}$), we observe formation of product $C$ even at later times (see Figs.~\ref{fig:max_concentrations_T}a and \ref{fig:max_concentrations_T}b).
This is due to reactive-mixing along the streamlines due to longitudinal dispersion, which is not observed at higher values of $T$ (see Figs.~\ref{fig:max_concentrations_T}c and \ref{fig:max_concentrations_T}d).
The reason is that fast flipping brings more reactants together resulting in enhanced mixing.
The proposed NTF$k$ methodology is able to identify and delineate the above aspects. 
For example, the identified temporal features between the scenarios when $T\leq 2 \times 10^{-4}$ (see Figs.~\ref{fig:max_concentrations_T}a and \ref{fig:max_concentrations_T}b) and $T > 2 \times 10^{-4}$ (see Figs.~\ref{fig:max_concentrations_T}c and \ref{fig:max_concentrations_T}d) are distinct.
A comparison of the results in Fig.~\ref{fig:max_concentrations_T} demonstrates that for $T > 2 \times 10^{-4}$ the temporal feature T1 is dominant only at very early times where as temporal features T2 \& T3 become more dominant at intermediate and late times. 
This is not the case when $T\leq 2 \times 10^{-4}$ due the fast flipping of vortex-based velocity field.

All the high-resolution numerical simulations performed using the non-negative finite element method are about 200 GB of data.
The tensor decomposition of these datasets using the proposed NTF$k$ methodology has produced a compressed dataset of about 70 MB.
This corresponds to a compression ratio close to $4 \times 10^{-4}$.
In addition to identifying hidden temporal features, the data compression is an added benefit of Tucker-3 NTF$k$ model.
To summarize, the proposed NTF$k$ method substantially reduces the data storage requirements with minimal loss of information and provides accelerated data processing.
%

\section{CONCLUSIONS}
\label{sec:conclusions}
Our results demonstrate the applicability of the NTF$k$ approach to analyze complex large-scale model outputs of anisotropic reaction-diffusion equations.
NTF$k$ is based on a Non-negative Tensor Factorization (NTF) combined with a custom semi-supervised clustering \cite{alexandrov2014blind,vesselinov2018contaminant}.
NTF$k$ allows for the deconstruction of tensor datasets (multi-dimensional arrays) into additive components.
The separation is based on the temporal and spatial variability of the data.
In this paper, we applied NTF$k$ to identify between 2 and 4 temporal features (components) in model simulation outputs representing the impact of species mixing in a fast irreversible bimolecular reaction $A + B \rightarrow C$.
Five model parameters representing physical processes such as anisotropic dispersion contrast, molecular diffusivity, perturbations in velocity field as well as characteristic spatial and time-scales of the velocity field, are varied in our finite element numerical simulations to evaluate the effect of these parameters on formation of product $C$ in space and time.
The features identified by NTF$k$ have physical meaning and enable the interpretation of the governing process impacting the model outputs.
The identified features appear to be dominated by vortex structure, anisotropic dispersion, molecular diffusion, and boundary effects.
The analyzed species mixing problem is challenging but important for various real world applications (e.g., contaminant remediation, material design, etc.).
These ML results can play a critical role in improving our conceptual understanding of these reaction-diffusion processes.
Our NTF$k$ work also acts as data compression of the numerical simulations and could streamline the development of reduced-order models that can be applied to predict the system behavior in a more computationally efficient manner.

The application of the proposed NTF$k$ approach is not limited to reactive-mixing.
NTF$k$ can readily be applied to any observed or simulated datasets that can be represented as a tensor (multi-dimensional array) and have separable latent signatures or features.
Frequently, these signatures/features are related to underlying physical processes governing the spatial and temporal variability in the data.
Extracting these signatures and structure-preserving features can help the development of conceptual models, reduced-order models, and simplified closed-form mathematical expressions, which can then be used to predict the system behavior.
Furthermore, NTF$k$ analyses can be applied to detect anomalies or disruptions in the data, which might be caused by additional phenomena not detected in past.
Lastly, unsupervised machine learning methods such as NTF$k$ can be coupled with supervised machine learning methods to perform deep learning. 

\section*{ACKNOWLEDGEMENTS}
This research was funded by the Environmental Programs Directorate of the Los Alamos National Laboratory.
In addition, Velimir V. Vesselinov and Daniel O'Malley were supported by the DiaMonD project (An Integrated Multifaceted Approach to Mathematics at the Interfaces of Data, Models, and Decisions, U.S. Department of Energy Office of Science, Grant \#11145687).
Maruti Kumar Mudunuru and Satish Karra also thank the support of the LANL Laboratory Directed Research and Development Early Career Award 20150693ECR.
Maruti Kumar Mudunuru also thanks the support of LANL Chick-Keller Postdoctoral Fellowship through Center for Space and Earth Sciences (CSES).
Velimir V. Vesselinov, Boian S. Alexandrov, and Daniel O'Malley were also supported by LANL LDRD grants 20180060DR and 20190020DR.
This research used resources provided by the Los Alamos National Laboratory Institutional Computing Program, which is supported by the U.S. Department of Energy National Nuclear Security Administration under Contract No. DE-AC52-06NA25396.

\bibliographystyle{unsrt}
\bibliography{Master_References/extracted}
%


\end{document}